\RequirePackage[l2tabu, orthodox]{nag}
\documentclass[12pt,a4paper]{article}

\usepackage{fixltx2e}

\usepackage{comment}

\usepackage{hyperref}

\usepackage[font=small]{caption}
\usepackage[text={450pt,650pt},headheight={14.5pt},centering]{geometry}

\usepackage{lmodern}
\usepackage{graphicx}
\usepackage{subfig}
\usepackage{booktabs}
\usepackage{dsfont}
\usepackage{collref}
\usepackage{tikz}
\usetikzlibrary{plotmarks,calc}
\usetikzlibrary{arrows}

\tikzstyle arrow=[thick,rounded corners=18pt,-latex]
\tikzstyle box=[draw,rounded corners,outer sep=4pt]
\tikzstyle B node=[outer sep=0pt]
\tikzstyle Q node=[inner sep=1pt,outer sep=0pt]

\usepackage[sort,compress]{cite}

\usepackage{amsmath,amssymb}
\usepackage{mathrsfs}
\usepackage{mathtools}
\usepackage{bbm}
\usepackage{slashed}
\usepackage{braket}

\usepackage{hyperref}

\usepackage{xspace}

\usepackage{accents}

\DeclareMathAlphabet{\mathsfit}{\encodingdefault}{\sfdefault}{m}{sl}

\numberwithin{equation}{section}

\makeatletter
 \let\old@startsection=\@startsection
 \let\oldl@section=\l@section
 \renewcommand{\@startsection}[6]{\old@startsection{#1}{#2}{#3}{#4}{#5}{#6\mathversion{bold}}}
 \renewcommand{\l@section}[2]{\oldl@section{\mathversion{bold}#1}{#2}}
\makeatother

\newcommand{\fl}{q}
\newcommand{\flt}{\tilde{q}}

\renewcommand{\leq}{\leqslant}

\def\XXint#1#2#3{{\setbox0=\hbox{$#1{#2#3}{\int}$}
    \vcenter{\hbox{$#2#3$}}\kern-.5\wd0}}

\newcommand{\ce}{\text{c.e.}}

\newcommand{\stL}{\tilde{\text{\tiny L}}}
\newcommand{\stR}{\tilde{\text{\tiny R}}}

\newcommand{\AdS}{\text{AdS}}
\newcommand{\CFT}{\text{CFT}}
\newcommand{\Sphere}{\mathrm{S}}
\newcommand{\Torus}{\mathrm{T}}
\newcommand{\CP}{\mathrm{CP}}

\newcommand{\Smat}{\mathcal{S}}

\newcommand{\action}{S}
\newcommand{\lagr}{L}

\newcommand{\de}{\textup{d}}

\newcommand{\acommPB}[2]{\{#1,#2\}_{\scriptscriptstyle\text{PB\strut}}}

\newcommand{\alg}[1]{\mathfrak{#1}}

\newcommand{\suA}{\bullet}
\newcommand{\suB}{\circ}

\newcommand{\so}{\alg{so}}

\newcommand{\su}{\alg{su}}
\newcommand{\psu}{\alg{psu}}

\newcommand{\gen}[1]{\mathbf{#1}}

\newcommand{\smallL}{\sL}
\newcommand{\smallR}{\sR}

\newcommand{\Integers}{\mathbb{Z}}

\newcommand{\1}{\mathbf{1}}

\newcommand{\superN}{\mathcal{N}}

\newcommand{\ie}{\textit{i.e.}\xspace}
\newcommand{\eg}{\textit{e.g.}\xspace}
\newcommand{\cf}{\textit{cf.}\xspace}

\newcommand{\pri}[1]{\accentset{\prime}{#1}}

\newcommand{\sL}{\mbox{\tiny L}}
\newcommand{\sR}{\mbox{\tiny R}}

\newcommand{\hslashslash}{%
  \raisebox{.9ex}{%
    \scalebox{.7}{%
      \rotatebox[origin=c]{17}{$-$}%
    }%
  }%
}
\renewcommand{\k}{%
  {%
   \vphantom{k}%
   \ooalign{\kern-.05em\smash{\hslashslash}\hidewidth\cr$k$\cr}%
   \kern-.025em
  }%
}

\newcommand{\h}{h}

\begin{document}
\thispagestyle{empty}

\begin{flushright}\footnotesize\ttfamily
Imperial-TP-OOS-2014-04\\
HU-Mathematik-2014-21\\
HU-EP-14/34
\end{flushright}
\vspace{1em}

\begin{center}
\textbf{\Large\mathversion{bold} The complete worldsheet S matrix of superstrings on $\AdS_3\times \Sphere^3\times \Torus^4 $ with mixed three-form flux}

\vspace{2em}

\textrm{\large Thomas Lloyd${}^1$, Olof Ohlsson Sax${}^2$, Alessandro Sfondrini${}^{3}$\\
and Bogdan Stefa\'nski, jr.${}^1$ } 

\vspace{2em}

\begingroup\itshape

1. Centre for Mathematical Science, City University London,\\ Northampton Square, EC1V 0HB, London, U.K.\\[0.2cm]

2. The Blackett Laboratory, Imperial College,\\  SW7 2AZ, London,
U.K.\\[0.2cm]

3. Institut f\"ur Mathematik und Institut f\"ur Physik, Humboldt-Universit\"at zu Berlin\\
IRIS Geb\"aude, Zum Grossen Windkanal 6, 12489 Berlin, Germany\par\endgroup

\vspace{1em}

\texttt{Tom.Lloyd.1@city.ac.uk, o.olsson-sax@imperial.ac.uk, Alessandro.Sfondrini@physik.hu-berlin.de, Bogdan.Stefanski.1@city.ac.uk}

%%%%%%%%

\end{center}

\vspace{3em}

\begin{abstract}\noindent
We determine the off-shell symmetry algebra and  representations of Type IIB superstring theory on $\AdS_3\times\Sphere^3\times\Torus^4 $ with mixed R-R and NS-NS three-form flux. We use these to derive the non-perturbative worldsheet S~matrix of fundamental excitations of the superstring theory. Our analysis includes both massive and massless modes and shows how turning on mixed three-form flux results in an integrable deformation of the S~matrix of the pure R-R theory.
\end{abstract}

%%%%%%%%%%%%%%%%%%%%%%%%%%%%%%%%%%%%%%%%%%%%%%%%%%%%%%%%%%%%%%%%%%%%%%%%%%%
\newpage

\tableofcontents

\section{Introduction}
\label{sec:introduction}
The investigation of the $\AdS/\CFT$ correspondence~\cite{Maldacena:1997re,Witten:1998qj,Gubser:1998bc} using integrability techniques has led to a remarkably successful quantitative description of the 't~Hooft, or planar, limit~\cite{'tHooft:1973jz} of certain classes of dual theories. The two best known examples are type IIB strings on~$\AdS_5\times \Sphere^5$ and the dual $\mathcal{N}=4$ Supersymmetric Yang-Mills (SYM) theory, and type IIA string theory on $\AdS_4\times \CP^3$ and the dual ABJM Chern-Simons theory~\cite{Aharony:2008ug}.\footnote{See~\cite{Arutyunov:2009ga,Beisert:2010jr,Klose:2010ki} for reviews and a more complete list of references.} The integrability methods used to understand these dual pairs can also be extended to their deformations, orbifolds and orientifolds~\cite{Zoubos:2010kh,vanTongeren:2013gva}, suggesting that other classes of examples of the $\AdS/\CFT$ correspondence may also be amenable to this approach.

Superstrings on~$\AdS_3\times \mathcal{M}_7$ backgrounds with 16 real supersymmetries~\cite{Pesando:1998wm, Rahmfeld:1998zn,Park:1998un, Metsaev:2000mv,Babichenko:2009dk} have been shown to be classically integrable~\cite{Babichenko:2009dk}, opening the  possibility of understanding  such $\AdS_3/\CFT_2$ dualities with integrability methods. However, unlike the more supersymmetric cases mentioned above, such backgrounds contain massive and massless worldsheet excitations. While the massive excitations could be investigated \cite{David:2008yk, Babichenko:2009dk,David:2010yg, OhlssonSax:2011ms, Borsato:2012ud,Borsato:2012ss,Abbott:2012dd,Beccaria:2012kb,Beccaria:2012pm,Borsato:2013qpa,Borsato:2013hoa,Abbott:2013ixa} using more conventional integrability methods developed in the context of $\AdS_5/\CFT_4$ and $\AdS_4/\CFT_3$,%
\footnote{%
See also \cite{Sfondrini:2014via} for a review and a  more extensive list of references.
} massless modes appeared initially to be rather different and difficult to incorporate fully into the holographic integrability approach. For example, these massless excitations made it difficult to apply directly some of the integrability methods, such as the finite-gap techniques, to the non-perturbative theory~\cite{Babichenko:2009dk,OhlssonSax:2011ms}. Initially, progress was made by considering the massless modes at the weakly-coupled spin-chain point~\cite{Sax:2012jv} and in the finite-gap equations of classical  string theory~\cite{Lloyd:2013wza}. 

Recently, through the analysis of the off-shell symmetry algebra of the theory and its representations, a complete non-perturbative worldsheet S matrix of type IIB superstring theory~$\AdS_3\times \Sphere^3\times \Torus^4$ supported by R-R flux was constructed~\cite{Borsato:2014exa,Borsato:2014hja}. This provided a unified description of massive and massless worldsheet excitations in an integrable framework, where all worldsheet excitations are \emph{non-relativistic} and so massless scattering can take place. This allows one to circumvent the more abstract constructions of massless \emph{relativistic} S matrices found in the integrability literature~\cite{Zamolodchikov:1992zr, Fendley:1993wq,Fendley:1993xa}.

Unlike the higher-dimensional $\AdS$ backgrounds, Type IIB string theory on $\AdS_3\times \Sphere^3\times \Torus^4$ has a large moduli space of parameters and can be supported by a mixture of Neveu-Schwarz-Neveu-Schwarz (NS-NS) and Ramond-Ramond (R-R) three-form fluxes.
The relations between these backgrounds are governed by U-duality transformations and were analysed extensively in ref.~\cite{Larsen:1999uk}. In particular, type IIB S-duality relates $\AdS_3\times \Sphere^3\times \Torus^4$ backgrounds supported by different three-form fluxes: the pure R-R flux background can be obtained from the near-horizon limit of D1- and D5-branes, while backgrounds supported by mixed three-form fluxes involve the near-horizon limit of NS5-branes and fundamental strings in addition to the D1- and D5-branes.

In the bosonic non-linear sigma model, turning on the NS-NS three form flux yields a Wess-Zumino-Witten (WZW) term in the action~\cite{Novikov:1982ei,Witten:1983tw, Witten:1983ar}.
In units where the $\AdS$ radius is one, the R-R three form $F$ and the NS-NS three form $H$ are given by
\begin{equation}
  F = \flt\, \bigl( \operatorname{Vol}(\AdS_3) + \operatorname{Vol}(\Sphere^3) \bigr) , \qquad
  H = \fl\, \bigl( \operatorname{Vol}(\AdS_3) + \operatorname{Vol}(\Sphere^3) \bigr) ,
\end{equation}
where the coefficients $\fl$ and $\flt$ satisfy 
\begin{equation}
  \fl^2 + \flt^2 = 1.
\end{equation}
The parameter $\fl$ is related to the quantised coupling $k$ of the WZW model
\begin{equation}
  k = \fl \sqrt{\lambda}\,.
\end{equation}
where~$\lambda$ is the 't~Hooft coupling, which in turn parameterises the string tension $\sqrt{\lambda}/2\pi$.
Note that since we have%
\footnote{The theory is well defined and supersymmetric when $|\fl|\leq 1$. For simplicity, we restrict to positive~$\fl$. A parity transformation  on the worldsheet amounts to~$\fl\to-\fl$, and can be used to consider $-1\leq\fl<0$.}
 $0 \leq \fl \leq 1$, the coupling $\sqrt{\lambda}$ satisfies $\sqrt{\lambda} \ge k$.

If the supersymmetric completions of such $\AdS_3$ backgrounds are tractable by integrability, we will have the exciting possibility of studying families of integrable models with deformation parameters related to (some of) the string moduli. Indeed, in ref.~\cite{Cagnazzo:2012se} the classical superstring action for the mixed-flux $\AdS_3\times \Sphere^3\times \Torus^4$ backgrounds was shown to be integrable. This led to rapid progress in understanding the role integrability plays in the massive sector of the theory~\cite{Hoare:2013pma,Hoare:2013ida,Hoare:2013lja,Ahn:2014tua,Babichenko:2014yaa}. 

In this paper we derive the complete non-perturbative asymptotic worldsheet S~matrix, including the massless modes, for all mixed-flux $\AdS_3\times \Sphere^3\times \Torus^4$ backgrounds. We do this by computing the off-shell symmetry algebra of the theory and using it to determine the two-body S~matrix, which satisfies the Yang-Baxter equation. This allows us to treat the massive and massless modes on the same footing and shows that the approach used in refs.~\cite{Borsato:2014exa,Borsato:2014hja} to tackle massless modes is likely to be applicable to more general $\AdS/\CFT$ integrability settings where such modes frequently appear~\cite{Babichenko:2009dk,Sorokin:2011rr,Wulff:2014kja}. When restricted to the massive sector, our S~matrix reduces to the one presented in ref.~\cite{Hoare:2013ida,Hoare:2013lja}.

This paper is structured as follows. In section~\ref{sec:off-shell-algebra} we derive the off-shell symmetry algebra of the theory from the type IIB superstring action for $\AdS_3\times\Sphere^3\times\Torus^4$ with mixed flux in light-cone gauge, including the exact form of the off-shell central charges. In section~\ref{sec:algebra} we present the representations of the symmetry algebra~$\mathcal{A}$ that enter the S~matrix construction and then deform these representations in a way that produces the shortening condition for these exact charges. In section~\ref{sec:smat} we use these representations to construct an invariant S~matrix for the all worldsheet  excitations of the mixed-flux theories, up to a number of dressing factors which we constrain by crossing symmetry. We conclude in section~\ref{sec:conclusion}. We relegate some more technical details to the appendices.

\section{Superstrings on \texorpdfstring{$\AdS_3\times \Sphere^3\times \Torus^4$}{AdS3 x S3 x T4} with mixed three-form flux and their off-shell symmetry algebra}
\label{sec:off-shell-algebra}

In this section we write down the fully gauge-fixed action for  type IIB superstring theory on $\AdS_3\times \Sphere^3\times \Torus^4$ with mixed flux, determine the classical charges of this theory and compute the off-shell Poisson-bracket algebra $\mathcal{A}$ that the charges satisfy.

The coset formulation~\cite{Cagnazzo:2012se} of type IIB superstring theory on $\AdS_3\times \Sphere^3\times \Torus^4$ with mixed flux is useful in investigating classical integrability of these theories. However, the coset action which can be obtained from a Green-Schwarz action~\cite{Grisaru:1985fv} by fully fixing the kappa symmetry to the so-called coset kappa gauge~\cite{Babichenko:2009dk}, does not allow for a straightforward computation of the Poisson brackets of the symmetries. This is because the massless fermions have non-standard kinetic terms in the bosonic light-cone gauge. Instead, one needs to work directly with the Green-Schwarz action~\cite{Grisaru:1985fv} in the BMN light-cone kappa gauge. While expressions for this action are known explicitly up to quartic order in fermions~\cite{Wulff:2013kga}, we will only work up to quadratic order in fermions and so can use the actions written down in ref.~\cite{Cvetic:1999zs}.

We begin by writing down expressions for the Killing spinors of this background in section~\ref{sec:Killing-spinors}.
% in the metric~\eqref{eq:metric}. 
In section~\ref{sec:action} we gauge fix the bosonic action of the theory and write down the explicit expressions for the non-dynamical fields.
In section~\ref{sec:GS-action} we use the Killing spinors to write down explicitly the superstring action in a mixed-flux background and impose the BMN light-cone kappa gauge. In section~\ref{sec:currents-and-algebra} we write down the Noether currents for the charges that generate the algebra $\mathcal{A}$ and compute the Poisson brackets of these charges off shell in order to determine the classical algebra.

\subsection{Killing spinors of IIB supergravity on \texorpdfstring{$\AdS_3\times \Sphere^3\times \Torus^4$}{AdS3 x S3 x T4} with mixed flux}
\label{sec:Killing-spinors}

Expressions for Killing spinors on $\Sphere^n$ and $\AdS_n$ can be found in ref.~\cite{Lu:1996rhb,Lu:1998nu} in a particular coordinate system. Throughout this paper we will find it useful to work in a different coordinate system---one that is well suited for expansion around the BMN ground state---and so present the expressions for Killing spinors in this coordinate system below. Explicitly, we take the $\AdS_3\times \Sphere^3\times \Torus^4$ metric to be
\begin{equation}
\label{eq:metric}
  ds^2 = ds_{\AdS_3}^2 + ds_{\Sphere^3}^2 + dX_i dX_i  ,
\end{equation}
where
\begin{equation}
\label{eq:s3metric}
  ds^2_{\Sphere^3} = +\Bigl(\frac{1 - \frac{y_3^2 + y_4^2}{4}}{1 + \frac{y_3^2 + y_4^2}{4}}\Bigr)^2 d\phi^2 + \Bigl(\frac{1}{1 + \frac{y_3^2 + y_4^2}{4}}\Bigr)^2 ( dy_3^2 + dy_4^2 )
\end{equation}
and
\begin{equation}
\label{eq:ads3metric}
  ds^2_{\AdS^3} = -\Bigl(\frac{1 + \frac{z_1^2 + z_2^2}{4}}{1 - \frac{z_1^2 + z_2^2}{4}}\Bigr)^2 dt^2 + \Bigl(\frac{1}{1 - \frac{z_1^2 + z_2^2}{4}}\Bigr)^2 ( dz_1^2 + dz_2^2 )  ,
\end{equation}
 In these coordinates the NS-NS $B$ field is given by
\begin{equation}
  B = 
  \frac{\fl}{\bigl( 1 - \frac{z^2}{4}\bigr)^2} ( z_1 \, dz_2 \wedge dt + z_2 \, dt \wedge dz_1 )
  + \frac{\fl}{\bigl( 1 + \frac{y^2}{4}\bigr)^2} ( y_3 \, dy_4 \wedge d\phi + y_4 \, d\phi \wedge dy_3) .
\end{equation}
This leads to the NS-NS three form
\begin{equation}
  H = dB = 
  2 \fl \frac{\hphantom{\bigl(}1 + \frac{z^2}{4}\hphantom{\bigl)^3}}{\bigl( 1 - \frac{z^2}{4}\bigr)^3} dt \wedge dz_1 \wedge dz_2
  + 2 \fl \frac{\hphantom{\bigl(}1 - \frac{y^2}{4}\hphantom{\bigl)^3}}{\bigl( 1 + \frac{y^2}{4}\bigr)^3} dy_3 \wedge dy_4 \wedge d\phi .
\end{equation}
Hence, the R-R and NS-NS three forms have tangent space components
\begin{equation}
  F_{012} = F_{345} = 2 \flt , \qquad H_{012} = H_{345} = 2 \fl.
\end{equation}

The Killing spinor equations take the form
\begin{equation}
\label{eq:AdS3-S3-T4-Killing-spinor-eq}
    \bigl( \delta_{IJ}(\partial_m + \frac{1}{4} \slashed{\omega}_m) + \frac{1}{48}\sigma^3_{IJ} \slashed{F} \slashed{E}_m + \frac{1}{8} \sigma^1_{IJ}\slashed{H}_m\bigr) \tilde{\varepsilon}_J  = 0  , 
\end{equation}
%where the covariant derivative is given by
%\begin{equation}
%  D_m\varepsilon^I = (\partial_m + \frac{1}{4} \slashed{\omega}_m)\varepsilon^I ,
%\end{equation}
%The contraction of $H$ is given by
%\begin{equation}
%  H_{MAB} \Gamma^{AB} = +2q E_M^A \epsilon_{ABC} \Gamma^{BC}
%  =
%  \begin{cases}
 %   + 4 \fl E_M^A \Gamma_A \Gamma^{012} , & \text{for $M = t, z_1, z_2$} , \\
%    + 4 \fl E_M^A \Gamma_A \Gamma^{345} , & \text{for $M = y_1, y_2, \phi$} ,
%  \end{cases}
%\end{equation}
where $\omega_m$ is the spin-connection, whose explicit components can be found in ref.~\cite{Borsato:2014hja} and the fluxes can be written as
\begin{equation}
\label{eq:Hslash}
 \slashed{H}_m\equiv  H_{mAB} \Gamma^{AB} = 2 \fl \bigl( \slashed{E}_m ( \Gamma^{012} + \Gamma^{345} ) + ( \Gamma^{012} + \Gamma^{345} ) \slashed{E}_m \bigr) ,
\end{equation}
and
\begin{equation}
\label{eq:Fslash}
\slashed{F}\equiv  F_{ABC} \Gamma^{ABC} = 12\flt ( \Gamma^{012} + \Gamma^{345} ) .
\end{equation}

As shown in appendix~\ref{app:killing-spinors}, the solutions of the Killing spinor equations~\eqref{eq:AdS3-S3-T4-Killing-spinor-eq} are
\begin{equation}\label{eq:killing-spinor-lin-combs}
  \tilde{\varepsilon}_1 = \sqrt{\frac{1+\flt}{2}} \, \varepsilon_1 - \sqrt{\frac{1-\flt}{2}} \, \varepsilon_2 , \qquad
  \tilde{\varepsilon}_2 = \sqrt{\frac{1+\flt}{2}} \, \varepsilon_2 + \sqrt{\frac{1-\flt}{2}} \, \varepsilon_1 ,
\end{equation}
where $\varepsilon_I$ are the pure R-R background Killing spinors found in ref.~\cite{Borsato:2014hja}. Recall that these latter spinors can be written as
\begin{equation}
  \varepsilon^1
  =
  \hat{M} \varepsilon_0^1,
  \qquad\qquad
  \varepsilon^2
  =
  \check{M} \varepsilon_0^2,
\end{equation}
where $\varepsilon_0^I$ are constant 9+1 dimensional Majorana-Weyl spinors, which further satisfy
\begin{equation}
  \frac{1}{2} ( 1 + \Gamma^{012345} ) \varepsilon^I = 
  \frac{1}{2} ( 1 + \Gamma^{012345} ) \varepsilon_0^I  = 0.
\end{equation}
Explicit expressions for the matrices  $\hat{M}$ and $\check{M}$ are given in equations~\eqref{eq:M0-def} and~\eqref{eq:Mt-def}. We will find it useful to separate the dependence of these matrices on the transverse and light-cone coordinates of $\AdS_3\times \Sphere^3$
\begin{equation}
  \hat{M}(z_{\underline{i}},y_{\underline{i}},t,\phi) = M_0(z_{\underline{i}},y_{\underline{i}}) M_t(t,\phi) , \qquad
  \check{M}(z_{\underline{i}},y_{\underline{i}},t,\phi) = M_0^{-1}(z_{\underline{i}},y_{\underline{i}}) M_t^{-1}(t,\phi) .
\end{equation}
Below, so as not to over-crowd the notation, we will drop the explicit coordinate dependence and simply write $\hat{M}$, $\check{M}$, $M_t^{\pm 1}$ and $M_0^{\pm 1}$.

Before ending this sub-section we would like to use the matrices $\hat{M}$ and $\check{M}$ to define tangent-space rotations $ \hat{\mathcal{M}}_B{}^A$ and $\check{\mathcal{M}}_B{}^A$ which will be useful in the following sub-sections
\begin{equation}\label{eq:MN-identities}
  \hat{M}^{-1} \Gamma^A \hat{M} = \Gamma^B \hat{\mathcal{M}}_B{}^A ,
  \qquad
  \check{M}^{-1} \Gamma^A \check{M} = \Gamma^B \check{\mathcal{M}}_B{}^A .
\end{equation}
These matrices are block diagonal,
\begin{equation}
\label{eq:orth-rot-ads3-s3}
  \hat{\mathcal{M}} = \hat{\mathcal{M}}_{\AdS_3} \oplus \hat{\mathcal{M}}_{\Sphere^3} \oplus \1_4 ,
  \qquad
  \check{\mathcal{M}} = \check{\mathcal{M}}_{\AdS_3} \oplus \check{\mathcal{M}}_{\Sphere^3} \oplus \1_4  ,
\end{equation}
and explicit expressions for them can be found in Appendix C of~\cite{Borsato:2014hja}.

\subsection{The mixed-flux bosonic action and gauge-fixing}
\label{sec:action}

In this sub-section we write down the bosonic action of the mixed-flux background and impose uniform light-cone gauge. The gauge-fixing determines the non-dynamical fields  ($x^{\pm}$ and $\gamma^{\alpha\beta}$) in terms of the physical degrees of freedom. Below, when computing the symmetry algebra $\mathcal{A}$, we will work to quartic order in transverse bosons and quadratic order in transverse fermions. As a result, we will only need explicit expressions for the non-dynamical fields up to zeroth order in fermions and quadratic order in transverse bosons. The bosonic action is\footnote{%
  We suppress the overall string tension $\sqrt{\lambda}/2\pi$ in the worldsheet action, and only reinsert it in the final result for the central charge.%
} %
\begin{equation}
  \action_B = -\frac{1}{2} \int_{-r}^{+r} d\sigma \bigl(
  \gamma^{\alpha\beta} G_{mn} \partial_\alpha X^m \partial_\beta X^n + \epsilon^{\alpha\beta} B_{mn} \partial_\alpha X^m \partial_\beta X^n
  \bigr),
\end{equation}
where the range of the worldsheet coordinate $\sigma$ is given by $-r < \sigma < +r$.

%The bosonic equations of motion can be written as
%\begin{equation}
%  \hat{K}_m^A \eta_{AB} \partial_{\alpha} ( \gamma^{\alpha\beta} \hat{K}_n^B \partial_\beta X^n ) - \frac{1}{4} %\epsilon^{\alpha\beta} H_{mnk} \partial_\alpha X^n \partial_\beta X^k = 0 ,
%\end{equation}
%where the $\hat{K}_m^A$ are a prefered set of vielbeins defined by an orthogonal rotation of the ``diagonal'' vielbein %$E_m^A$
%\begin{equation}
%\label{eq:def-of-hat-check-vielbeins}
%  \hat{K}_m^A = \hat{\mathcal{M}}^A{}_B E_m^B ,
%  \qquad
%  \check{K}_m^A = \hat{\mathcal{M}}^A{}_B E_m^B .
%\end{equation}
%The matrices $\hat{\mathcal{M}}$ and $\check{\mathcal{M}}$ are defined in equation~\eqref{eq:MN-identities}. They follow %from considering bilinears formed out of the Killing spinors $\varepsilon^I$, and hence they satisfy the Killing vector %equations of the background and generate the $\so(2,2) \oplus \so(4) = \algSL(2) \oplus \algSL(2) \oplus \su(2) %\oplus \su(2)$ isometry algebra of $\AdS_3\times \Sphere^3$.
%
Introducing the canonical momenta
\begin{equation}
  p_m = - \gamma^{00} G_{MN} \dot{X}^N - \gamma^{01} G_{MN} \pri{X}^N - B_{MN} \pri{X}^N
\end{equation}
the action can be written in the first order form
\begin{equation}
  \action_B = \int_{-r}^{+r} d\sigma \bigl( p_m \dot{X}^m + \frac{\gamma^{01}}{\gamma^{00}} C_1 + \frac{1}{2\gamma^{00}} C_2 \bigr) ,
\end{equation}
with
\begin{equation}
  C_1 = p_m \pri{X}^m
\end{equation}
and
\begin{equation}
  C_2 = G^{mn} p_m p_n + G_{mn} \pri{X}^m\pri{X}^n  + 2 G^{mn} B_{nk} p_m \pri{X}^K + G^{mn} B_{mk} B_{nl} \pri{X}^k \pri{X}^l .
\end{equation}
The constraints $C_1=0$ and $C_2=0$ are equivalent to the Virasoro constraints
\begin{equation}
  \gamma^{11} G_{mn} \dot{X}^m \pri{X}^n + \gamma^{01} G_{mn} \dot{X}^m \dot{X}^n = 0 , \qquad
  \gamma^{00} G_{mn} \dot{X}^m \dot{X}^n - \gamma^{11} G_{mn} \pri{X}^m \pri{X}^n = 0.
\end{equation}
We want to fix uniform light-cone gauge in which $x^+ = \tau$ and $p_-$ is constant.\footnote{%
  We set $x^{\pm} = \frac{1}{2} ( \phi \pm t )$.
} %
Solving the Virasoro constraints we find\footnote{We have checked that these equations are consistent using the equations of motion for the transverse bosons.}
\begin{equation}
  \begin{aligned}
    \dot{x}^- &= - \frac{1}{4} (\dot{z}^2 + \dot{y}^2 +\dot{x}^2+\pri{z}^2 + \pri{y}^2 + \pri{x}^2 -z^2 -y^2), \\
    \pri{x}^- &= -\frac{1}{2}(\dot{z} \cdot \pri{z} + \dot{y} \cdot \pri{y} + \dot{x} \cdot \pri{x}).
  \end{aligned}
\end{equation}
Using the $x^{\pm}$ equations of motion  we further find that to quadratic order in fields the worldsheet metric is
\begin{equation}
  \begin{aligned}
    \gamma^{00} &= -1 + \frac{z^2-y^2}{2} - \frac{q}{2} \epsilon^{\underline{i}\underline{j}}(z_{\underline{i}}\pri{z}_{\underline{j}}-y_{\underline{i}}\pri{y}_{\underline{j}}) , \\
    \gamma^{01} &= \frac{q}{2}\epsilon^{\underline{i}\underline{j}}(z_{\underline{i}}\pri{z}_{\underline{j}}-y_{\underline{i}}\pri{y}_{\underline{j}}) , \\
    \gamma^{11} &=+1 + \frac{z^2-y^2}{2} - \frac{q}{2} \epsilon^{\underline{i}\underline{j}}(z_{\underline{i}}\pri{z}_{\underline{j}}-y_{\underline{i}}\pri{y}_{\underline{j}}) .
  \end{aligned}
\end{equation}
Note in particular that for $\fl \neq 0$, the worldsheet metric is non-diagonal already at quadratic order in fields.

In the limit where the light-cone momentum
\begin{equation}
  P_- = \int_{-r}^{+r} d\sigma \, p_- = 4r
\end{equation}
is infinite, the worldsheet becomes decompactified and we are effectively working on a plane rather than a cylinder.
In the transverse directions we impose periodic boundary conditions $x^i(-r) = x^i(+r)$ and $x^{\underline{i}}(-r) = x^{\underline{i}}(+r)$. Physical closed string states should further be periodic in the light-cone direction $x^-$. This leads to the condition
\begin{equation}
  \Delta x^- = x^-(+r) - x^-(-r) = \int_{-r}^{+r} d\sigma \, \pri{x}^- = 0 .
\end{equation}
The quantity $\Delta x^-$ is directly related to the worldsheet momentum
\begin{equation}
  p_{\text{ws}} = - \int_{-r}^{+r} d\sigma \, ( p_{\underline{i}} \pri{x}^{\underline{i}} + p_i \pri{x}^i ) = 2 \Delta x^- .
\end{equation}
Here we have assumed that there is no winding along the $\phi$ direction. For non-zero winding number $w \in \Integers$, the level-matching condition takes the form
\begin{equation}
  \label{eq:level-matching}
  p_{\text{ws}} = 2\pi w .
\end{equation}
However, in the rest of this section we will work at zero winding. Moreover, we are mainly interested in studying the symmetries of the worldsheet theory when we go \emph{off shell} by allowing the worldsheet momentum to take arbitrary values.

\subsection{The Green-Schwarz action}
\label{sec:GS-action}
In this sub-section we describe the gauge-fixing of the Green-Schwarz action in a form that will be particularly suited to computing $\mathcal{A}$. This computation requires the action to quadratic order in fermions, and we give explicit expressions for the action to this order in appendix~\ref{app:lagrangian}. The Green-Schwarz action for type IIB superstrings in a general supergravity background was written down in terms of superfields in ref.~\cite{Grisaru:1985fv} and explicit expressions in an expansion of fermions up to quadratic~\cite{Cvetic:1999zs} and quartic~\cite{Wulff:2013kga} order are known. 
 The Lagrangian can be written as
\begin{equation}
  \lagr = \lagr_B + \lagr_{\text{\scriptsize kin}} + \lagr_{\text{WZ}}  ,
\end{equation}
with the bosonic part, $\lagr_B$, discussed in the previous sub-section. The remaining part of the Lagrangian is split into two terms: a term dependent on the worldsheet metric, $\lagr_{\text{\scriptsize kin}}$, and the Wess-Zumino term $\lagr_{\text{WZ}}$. Up to quadratic order in fermions these are given by ~\cite{Cvetic:1999zs}
\begin{align}
  \lagr_{\text{kin}} &= 
  -i\gamma^{\alpha\beta} \bar{\tilde{\theta}}_I \slashed{E}_\alpha \bigl( \delta^{IJ} D_\beta + \frac{1}{48} \sigma_3^{IJ} \slashed{F} \slashed{E}_\beta + \frac{1}{8} \sigma_1^{IJ} \slashed{H}_\beta \bigr) \tilde{\theta}_J  ,\\
  \lagr_{\text{WZ}} &= 
  +i\epsilon^{\alpha\beta} \bar{\tilde{\theta}}_I \sigma_1^{IJ} \slashed{E}_\alpha \bigl( \delta^{JK} D_\beta + \frac{1}{48} \sigma_3^{JK} \slashed{F} \slashed{E}_\beta + \frac{1}{8} \sigma_1^{JK} \slashed{H}_\beta \bigr) \tilde{\theta}_K .
\end{align}
It is helpful to perform a field re-definition of the fermions so as to end up with fermionic coordinates that are best adapted to the underlying integrable structure. It is easiest to understand this field redefinition as a combination of two transformations of the fermionic coordinates $\tilde{\theta}_I$ appearing above. Initially we ``rotate'' the fermions along the $I-J$ index
\begin{equation}
\label{eq:fermion-IJ-rot}
  \tilde{\theta}_1 = \sqrt{\frac{1+\flt}{2}} \, \theta_1 - \sqrt{\frac{1-\flt}{2}} \, \theta_2  , \qquad
  \tilde{\theta}_2 = \sqrt{\frac{1+\flt}{2}} \, \theta_2 + \sqrt{\frac{1-\flt}{2}} \, \theta_1 .
\end{equation}
This ensures that the kinetic term in the Lagrangian is diagonal in terms of the $\theta_I$. There are now two different field redefinitions that are useful to consider for different purposes. The first is to redefine
\begin{equation}\label{eq:theta-vartheta-def}
  \begin{aligned}
    \theta_1 &= \hat{M} \frac{1-\Gamma^{012345}}{2} \vartheta_1^- + \hat{M} \frac{1+\Gamma^{012345}}{2} \vartheta^+_1  , \\
    \theta_2 &= \check{M} \frac{1-\Gamma^{012345}}{2} \vartheta^-_2 + \check{M} \frac{1+\Gamma^{012345}}{2} \vartheta^+_2 .
  \end{aligned}
\end{equation}
In the resulting action, before kappa-gauge fixing, supersymmetry is realised as a shift on the fermions $\vartheta^-_I$. The expression for the resulting Lagrangian is written down in equations~\eqref{eq:pre-gauge-action-kin} and~\eqref{eq:pre-gauge-action-WZ}. However, we will be interested in a (suitably) kappa gauge fixed action, and so we will need to perform a different field redefinition to the one above. It turns out that a particular kappa gauge simplifies the computation of the algebra $\mathcal{A}$. This kappa gauge is the so-called BMN light-cone kappa gauge for fermions that are neutral with respect to the two $U(1)$'s associated to shifts along $t$ and $\phi$~\cite{Alday:2005ww}. As a result, in addition to~\eqref{eq:fermion-IJ-rot}, the second redefinition of the fermions we perform is
\begin{equation}\label{eq:theta-chi-eta-def-t}
  \begin{aligned}
    \theta_1 &= 
    \frac{1}{2} ( 1 + \Gamma^{012345} ) \mathrlap{M_0 \chi_1}\hphantom{M_0^{-1} \chi_2} 
    + \frac{1}{2} ( 1 - \Gamma^{012345} ) M_0 \eta_1 ,
    \\
    \theta_2 &= \frac{1}{2} ( 1 + \Gamma^{012345} ) M_0^{-1} \chi_2 + \frac{1}{2} ( 1 - \Gamma^{012345} ) M_0^{-1} \eta_2 .
  \end{aligned}
\end{equation}
where the matrix $M_0$ and its inverse are defined in equation~\eqref{eq:M0-def}. The fermions $\eta_I$ and $\chi_I$ correspond to the massive and massless fermions, respectively, of the integrable S matrix. 

% We also introduce light-cone coordinates
% \begin{equation}
%   E^{\pm} = \frac{1}{2} ( E^5 \pm E^0 ) , \qquad
%   x^{\pm} = \frac{1}{2} ( \phi \pm t ) ,
% \end{equation}
% and so have
% \begin{equation}
%   E_{x^+}^{+} = E_{x^-}^{-} = \frac{1}{2} \bigl( E_{\phi}^5 + E_t^0 \bigr) , \qquad
%   E_{x^+}^{-} = E_{x^-}^{+} = \frac{1}{2} \bigl( E_{\phi}^5 - E_t^0 \bigr) .
% \end{equation}
% %The light-cone components of the tangent space metric are given by
% %\begin{equation}
% %  \eta^{+-} = \eta^{-+} = + \frac{1}{2} , \qquad
% %  \eta_{+-} = \eta_{-+} = + 2 .
% %\end{equation}
% In terms of these the bosonic Lagrangian becomes
% \begin{equation}
%   \lagr_B = - \frac{1}{2} \gamma^{\alpha\beta} \bigl(
%   4 E^+_\alpha E^-_\beta + E^{\underline{i}}_\alpha E^{\underline{i}}_\beta + 
%   E^i_\alpha E^i_\beta 
%   \bigr) .
% \end{equation}

Having defined fermions that are neutral under shifts in $t$ and $\phi$, we impose the BMN light-cone kappa gauge
\begin{equation}\label{eq:bmn-lc-kappa-gauge}
  \Gamma^+ \eta_I = 0 , \qquad
  \Gamma^+ \chi_I = 0 , \qquad
  \Gamma^{\pm} = \frac{1}{2} \bigl( \Gamma^5 \pm \Gamma^0 \bigr) .
\end{equation}
The resulting light-cone kappa gauged-fixed action is written down in equations~\eqref{eq:post-gauge-action-kin} and~\eqref{eq:post-gauge-action-WZ}.

\subsection{The algebra \texorpdfstring{$\mathcal{A}$}{A}}
\label{sec:currents-and-algebra}

In this section we give the algebra $\mathcal{A}$ of (super)charges which commute with the Hamiltonian. As in the case of pure R-R flux the algebra itself is given by\footnote{%
  We use the direct sum to denote the sum of the subalgebras as vector spaces. This does not imply that they commute with each other as will be clear when we write down the full commutation relations in section~\ref{sec:algebra}.%
}%
\begin{equation}
\label{eq:Aalgebra}
  \mathcal{A} = \psu(1|1)^4_{\ce} \oplus \so(4) ,
\end{equation}
where the subscript $\ce$ denotes a four-fold central extension. In section~\ref{sec:currents} we first give the supercurrents $j_I$ that generate the algebra. In section~\ref{sec:supercurrent-algebra} we consider the Poisson brackets of these supercurrents,  see that we produce the expected Hamiltonian and compute the central charges of the algebra. In particular we find that the off-shell central charges are simple rescalings of the ones of the pure R-R theory. Throughout this section we give explicit results in the main text to quadratic order in fields, with higher order results given in the appendices. We use a ``hybrid'' expansion~\cite{Arutyunov:2006ak} in which we expand order by order in fermions and transverse bosons, but keep all factors of the light-cone coordinate $x^-$ exact. This allows us to compute the central charges exactly in momentum.

\subsubsection{Supercurrents}
\label{sec:currents}
The supercurrents of the algebra $\mathcal{A}$  to quadratic order are given by
\begin{align}
  \label{eq:quadratic-currents}
  j^{\tau}_1 =
  i e^{+x^- \gamma^{34}} \bigl(
  & ( \dot{z}^{\underline{i}} - \dot{y}^{\underline{i}} ) \gamma_{\underline{i}} \eta_1
  + ( z^{\underline{i}} + y^{\underline{i}} ) \gamma^{34} \gamma_{\underline{i}} \eta_1
  - ( \pri{z}^{\underline{i}} - \pri{y}^{\underline{i}} ) \gamma_{\underline{i}} ( \flt \eta_2 + \fl \eta_1 ) \\\nonumber
  + & \dot{x}^i \gamma^{34} \tilde{\tau}_i \chi_1
  - \pri{x}^i \gamma^{34} \tilde{\tau}_i ( \flt \chi_2 + \fl \chi_1 )
  \bigr) ,
  \\\nonumber
  j^{\tau}_2 =
  i e^{-x^- \gamma^{34}} \bigl(
  & ( \dot{z}^{\underline{i}} - \dot{y}^{\underline{i}} ) \gamma_{\underline{i}} \eta_2
  - ( z^{\underline{i}} + y^{\underline{i}} ) \gamma^{34} \gamma_{\underline{i}} \eta_2 
  - ( \pri{z}^{\underline{i}} - \pri{y}^{\underline{i}} ) \gamma_{\underline{i}} ( \flt \eta_1 - \fl \eta_2 ) \\\nonumber
  + & \dot{x}^i \gamma^{34} \tilde{\tau}_i \chi_2
  - \pri{x}^i \gamma^{34} \tilde{\tau}_i ( \flt \chi_1 - \fl \chi_2 )
  \bigr) ,
  \\\nonumber
  j^{\sigma}_1 =
  i e^{+x^- \gamma^{34}} \bigl(
  & ( \dot{z}^{\underline{i}} - \dot{y}^{\underline{i}} ) \gamma_{\underline{i}} ( \flt \eta_2 + \fl \eta_1 )
  + ( z^{\underline{i}} + y^{\underline{i}} ) \gamma^{34} \gamma_{\underline{i}} ( \flt \eta_2 + \fl \eta_1 )
  - ( \pri{z}^{\underline{i}} - \pri{y}^{\underline{i}} ) \gamma_{\underline{i}} \eta_1 \\\nonumber
  + & \dot{x}^i \gamma^{34} \tilde{\tau}_i ( \flt \chi_2 + \fl \chi_1 ) 
  - \pri{x}^i \gamma^{34} \tilde{\tau}_i ) \chi_1
  \bigr) ,
  \\\nonumber
  j^{\sigma}_2 =
  i e^{-x^- \gamma^{34}} \bigl(
  & ( \dot{z}^{\underline{i}} - \dot{y}^{\underline{i}} ) \gamma_{\underline{i}} ( \flt \eta_1 - \fl \eta_2 )
  - ( z^{\underline{i}} + y^{\underline{i}} ) \gamma^{34} \gamma_{\underline{i}} ( \flt \eta_1 - \fl \eta_2 )
  - ( \pri{z}^{\underline{i}} - \pri{y}^{\underline{i}} ) \gamma_{\underline{i}} \eta_2 \\\nonumber
  + & \dot{x}^i \gamma^{34} \tilde{\tau}_i ( \flt \chi_1 - \fl \chi_2 ) 
  - \pri{x}^i \gamma^{34} \tilde{\tau}_i \chi_2
  \bigr) .
\end{align}
In appendix~\ref{app:quartic-currents} we give expressions for the supercurrents to cubic order in transverse bosons and leading order in fermions. These expressions for the supercurrents are given in terms of fermions written as bispinors of $\so(4)_1 \oplus \so(4)_2 \in \so(1,9)$,\footnote{We have suppressed the corresponding spinor indices. These can easily be put back in; as defined in~\ref{app:conventions} the fermions carry spinor indices $(\eta_I)^{\underline{\dot{a}}\dot{b}}$ and $(\chi_I)^{\underline{a}b}$, while the $\so(4)$ gamma matrices carry indices $(\gamma_{\underline{i}})^{\underline{a}}{}_{\underline{\dot{b}}}$, $(\tilde{\tau}_i)^{\dot{a}}{}_b$ and $(\gamma^{34})^{\underline{a}}{}_{\underline{b}}$.} corresponding to rotations of $(z^{\underline{i}}, y^{\underline{i}})$ and $x^i$, as defined in ref.~\cite{Borsato:2014hja}.
The Lagrangian of the theory does not preserve~$\so(4)_1$, which is in fact broken to $\so(2)^2$ corresponding to separate rotations of~$z^{\underline{i}}$ and~$y^{\underline{i}}$. Nevertheless we will find it useful to write expression in these notation. On the other hand, $\so(4)_2$ is unbroken, and is part of~$\mathcal{A}$ in equation~\eqref{eq:Aalgebra}.
 Our conventions for the gamma matrices in these expressions are given in equation~\eqref{eq:so4-gamma}.
 
We have checked that the currents given in equation~\eqref{eq:quadratic-currents} and appendix~\ref{app:quartic-currents} satisfy the conservation equation
\begin{equation}
\partial_\tau j_I^\tau + \partial_\sigma j_I^\sigma = 0 ,
\end{equation}
to the required order using the equations of motion arising from the Lagrangian computed in the previous section.

\subsubsection{The algebra from the supercurrents}
\label{sec:supercurrent-algebra}

To find the off-shell symmetry algebra we need to compute Poisson brackets of the supercurrents, which in turn requires the Poisson brackets of the fermions. Explicit expressions for these are given in appendix~\ref{app:poisson}. For the Poisson brackets of two charges with the same index we find\footnote{Here $\epsilon$ symbols carry appropriate spinor indices, which we have suppressed for brevity.}
\begin{equation}
  \begin{aligned}
    \int d\sigma \, d\sigma' \acommPB{ j_1^{\tau}(\sigma) }{ j_1^{\tau}(\sigma') }
    &= + \frac{i}{2} \int d\sigma ( \mathcal{H} + \mathcal{M} ) \, \epsilon \, \epsilon , 
    \\
    \int d\sigma \, d\sigma' \acommPB{ j_2^{\tau}(\sigma) }{ j_2^{\tau}(\sigma') }
    &= + \frac{i}{2} \int d\sigma ( \mathcal{H} - \mathcal{M} ) \, \epsilon \, \epsilon .
  \end{aligned}
\end{equation}
The bosonic Hamiltonian density $\mathcal{H}$ is given to quadratic order by
\begin{equation}
  \mathcal{H} = \frac{1}{2} \bigl(
  p_{\underline{i}} p^{\underline{i}} + p_i p^i
  + \pri{z}_{\underline{i}} \pri{z}^{\underline{i}} + \pri{y}_{\underline{i}} \pri{y}^{\underline{i}} + \pri{x}_i \pri{x}^i
  + z_{\underline{i}} z^{\underline{i}} + y_{\underline{i}} y^{\underline{i}}
  - 2 \fl \epsilon^{\underline{ij}} ( z_{\underline{i}} \pri{z}_{\underline{j}} + y_{\underline{i}} \pri{y}_{\underline{j}} )
  \bigr) .
\end{equation}
The full quartic bosonic Hamiltonian can be found in equation~\eqref{eq:quartic-hamiltonian}.
The ``mass'' term $\mathcal{M}$ is given by
\begin{equation}
  \mathcal{M} = 
  - \epsilon^{\underline{ij}} ( p_{\underline{i}} z_{\underline{j}} + p_{\underline{i}} y_{\underline{j}} )
  - \fl ( p_{\underline{i}} \pri{z}^{\underline{i}} + p_{\underline{i}} \pri{y}^{\underline{i}} + p_i \pri{x}^i ) .
\end{equation}
It is important to note that this expression does not receive any corrections at quartic order.

Calculating the Poisson bracket between the two charges with different index we find
\begin{align}
  \int d\sigma \, d\sigma' \acommPB{ j_1^{\tau}(\sigma) }{ j_2^{\tau}(\sigma') }
  &= - \frac{i\flt}{2} \int d\sigma \Bigl[
  \partial_{\sigma} \bigl( e^{2\gamma^{34} x^-} \bigr)
  - \frac{1}{8} e^{2\gamma^{34} x^-} \partial_\sigma \bigl[(z^2 - y^2)^2\bigr]
  \\ \nonumber &\qquad
   - \partial_\sigma \bigl( e^{2\gamma^{34} x^-} \frac{z^2 - y^2}{2}
   - z^{\underline{i}} y^{\underline{j}} \bigl( 1 + \frac{z^2 - y^2}{4} \bigr) \gamma_{\underline{ij}} \bigr)
  \Bigr] \gamma^{34} \epsilon \, \epsilon.
\end{align}
The total derivative on the second line integrates to zero. The second term in the first line can be integrated by parts. The result is of higher order in  transverse bosons and can therefore be dropped. Hence, we are left with
\begin{equation}
  \int d\sigma \, d\sigma' \acommPB{ j_1^{\tau}(\sigma) }{ j_2^{\tau}(\sigma') } = 
  - \frac{i \flt}{2} e^{+2\gamma^{34} x^-(-\infty)} \bigl( e^{+\gamma^{34} p_{\text{ws}}} - 1 \bigr) \gamma^{34} \epsilon \, \epsilon
\end{equation}
Hence, we find the central charge
\begin{equation}
C=\frac{i\zeta}{2} \frac{\flt \sqrt{\lambda}}{2\pi} \left( e^{ip_{\text{ws}}} - 1 \right)
\end{equation}
where we have reintroduced the string tension, and where $\zeta=\mathrm{exp}(2ix^-(-\infty))$. This is related to the central charge of the pure R-R theory by a rescaling by $\flt$. It is precisely the fact that $C$ depends non-linearly on the momentum which imposes a non-local coproduct on the symmetry algebra, which we will discuss in the section~\ref{sec:coproduct}.

% Note that we find here a form for the charge $M$ which after putting back the coupling constant is
% \begin{equation}
%   M = m \pm \fl h p .
% \end{equation}
% Therefore we have the dispersion relation
% \begin{equation}
%     H^2 = M^2 + 4 \bar{C} C = (\fl h p \pm m)^2 + 4 \flt^2 h^2 \sin^2\frac{p}{2} .
% \end{equation}
% For integer $m$ this is just the giant magnon dispersion relation discussed in~\cite{Hoare:2013lja}.

% Note that the exponential $x^-$-dependent factor cancels out in the $\acomm{j_1}{j_1}$ bracket. Hence, there seems to be no way of getting a charge $M$ of the form
% \begin{equation}
%   M = 2qh\sin\frac{p}{2} \pm m
% \end{equation}
% which was proposed in~\cite{Hoare:2013ida}.

\section{Symmetry algebra and representations}
\label{sec:algebra}
We have seen that the off-shell symmetry algebra~$\mathcal{A}$ takes the same form in the mixed-flux case as it did in the pure-R-R one. Furthermore, in the limit where the NS-NS flux vanishes, we expect to recover precisely the same representations that were described in detail in ref.~\cite{Borsato:2014hja}. For this reason, we begin in section~\ref{sec:RRreprs} by briefly recalling the representations arising in the pure-R-R case. Then in sections~\ref{sec:near-BMN-representations} and~\ref{sec:exact-repr-mixed}, we describe how these are deformed, first in the near-plane-wave limit, and then in the full theory. We will see that the deformation can be completely understood by suitably altering the representation parameters of the pure-R-R case. As we proceed we will encounter, and comment on, several new features of the mixed-flux background.

\subsection{Overview of the pure-R-R symmetries and representations}
\label{sec:RRreprs}
The off-shell symmetry algebra $\mathcal{A}$ for type IIB superstrings on the pure-R-R $\AdS_3\times\Sphere^3\times\Torus^4$ background has been found in ref.~\cite{Borsato:2014hja}. There it was found that  $\mathcal{A}$ is a central extension of $\psu(1|1)^4\oplus\so(4)_{2}$. Writing down the supercharges in components we have
\begin{equation}
\label{eq:cealgebra}
\begin{aligned}
&\{\gen{Q}_{\smallL}^{\ \dot{a}},\overline{\gen{Q}}{}_{\smallL \dot{b}}\} =  \frac{1}{2}\delta^{\dot{a}}_{\ \dot{b}}\,(\gen{H}+\gen{M}),
&\qquad &\{\gen{Q}_{\smallL}^{\ \dot{a}},{\gen{Q}}{}_{\smallR \dot{b}}\} = \delta^{\dot{a}}_{\ \dot{b}}\,\gen{C},\\
&\{\gen{Q}_{\smallR  \dot{a}},\overline{\gen{Q}}{}_{\smallR}^{\ \dot{b}}\} = \frac{1}{2}\delta^{\ \dot{b}}_{\dot{a}}\,(\gen{H}-\gen{M}),
&\qquad &\{\overline{\gen{Q}}{}_{\smallL \dot{a}},\overline{\gen{Q}}{}_{\smallR}^{\ \dot{b}}\} = \delta^{\ \dot{b}}_{\dot{a}}\,\overline{\gen{C}}.
\end{aligned}
\end{equation}
The supercharges carry labels ``L'' and ``R'' corresponding to the left and right labels in the superisometry algebra $\su(1,1|2)_{\sL} \oplus \su(1,1|2)_{\sR}$. We also decompose
\begin{equation}
\so(4)_2=\su(2)_{\bullet}\oplus\su(2)_{\circ},
\end{equation}
so that the massive fermions are charged only under $\su(2)_{\bullet}$ .
The lower and upper dotted indices correspond to the fundamental and anti-fundamental representation of $\su(2)_{\bullet}$ respectively. Finally, the central charges on the one-particle representation are
\begin{equation}
\label{eq:allloop-centralcharges-RR}
\begin{aligned}
&\gen{C}=+\frac{i h}{2}(e^{+i p}-1),
\qquad\qquad &&
\overline{\gen{C}}=-\frac{i h}{2}(e^{-i p}-1),\\
&\gen{H}=\sqrt{m^2+4h^2\sin\bigl(\frac{p}{2}\bigr)^2},
\qquad\qquad &&
\gen{M}=m,
\end{aligned}
\end{equation}
where $p$ is the momentum, $m$ is an angular momentum taking values $\pm 1,0$ and $h$ is the coupling constant, which is expected to be a so-far undetermined function of the 't~Hooft coupling, $h=h(\lambda)$.

\subsubsection{Exact representations for the pure-R-R theory}
\label{sec:RR-exactrepr}
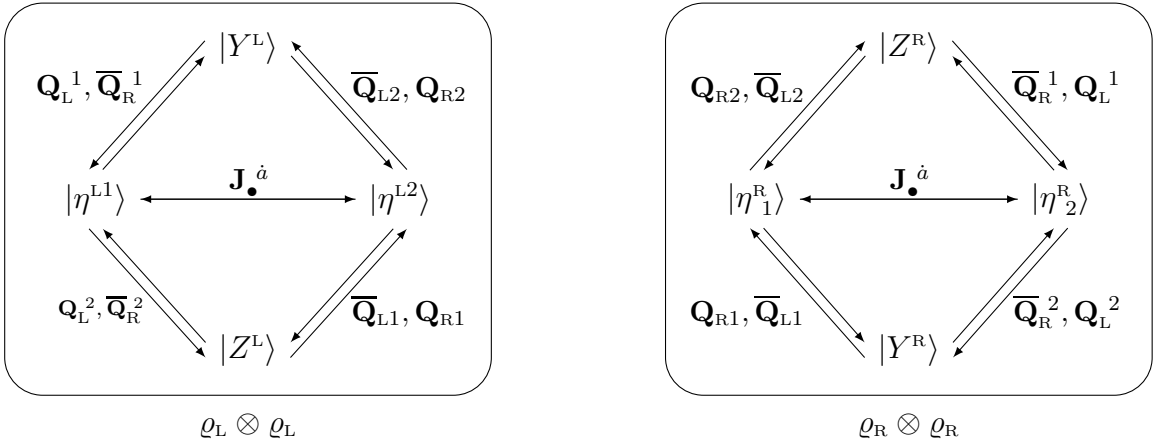
\begin{figure}[t]
  \centering
  \begin{tikzpicture}[%
    box/.style={outer sep=1pt},
    Q node/.style={inner sep=1pt,outer sep=0pt},
    arrow/.style={-latex}
    ]%

    \node [box] (PhiM) at ( 0  , 2cm) { $\ket{Y^{\sL}}$};
    \node [box] (PsiP) at (-2cm, 0cm) { $\ket{\eta^{\sL 1}}$};
    \node [box] (PsiM) at (+2cm, 0cm) { $\ket{\eta^{\sL 2}}$};
    \node [box] (PhiP) at ( 0  ,-2cm) { $\ket{Z^{\sL}}$};

    \newcommand{\horshift}{0.09cm,0cm}
    \newcommand{\vershift}{0cm,0.10cm}
 
    \draw [arrow] ($(PhiM.west) +(\vershift)$) -- ($(PsiP.north)-(\horshift)$) node [pos=0.5,anchor=south east,Q node] {\small $\gen{Q}^{\ 1}_{\sL},\overline{\gen{Q}}{}^{\ 1}_{\sR}$};
    \draw [arrow] ($(PsiP.north)+(\horshift)$) -- ($(PhiM.west) -(\vershift)$) node [pos=0.5,anchor=north west,Q node] {};

    \draw [arrow] ($(PsiM.south)-(\horshift)$) -- ($(PhiP.east) +(\vershift)$) node [pos=0.5,anchor=south east,Q node] {};
    \draw [arrow] ($(PhiP.east) -(\vershift)$) -- ($(PsiM.south)+(\horshift)$) node [pos=0.5,anchor=north west,Q node] {\small $\overline{\gen{Q}}{}_{\sL 1},\gen{Q}_{\sR 1}$};

    \draw [arrow] ($(PhiM.east) -(\vershift)$) -- ($(PsiM.north)-(\horshift)$) node [pos=0.5,anchor=north east,Q node] {};
    \draw [arrow] ($(PsiM.north)+(\horshift)$) -- ($(PhiM.east) +(\vershift)$) node [pos=0.5,anchor=south west,Q node] {\small $\overline{\gen{Q}}{}_{\sL 2},{\gen{Q}}{}_{\sR 2}$};

    \draw [arrow] ($(PsiP.south)-(\horshift)$) -- ($(PhiP.west) -(\vershift)$) node [pos=0.5,anchor=north east,Q node] {\scriptsize $
%    -
    \gen{Q}^{\ 2}_{\sL},\overline{\gen{Q}}{}^{\ 2}_{\sR}$};
    \draw [arrow] ($(PhiP.west) +(\vershift)$) -- ($(PsiP.south)+(\horshift)$) node [pos=0.5,anchor=south west,Q node] {};

    \draw [arrow] (PsiM) -- (PsiP) node [pos=0.6,anchor=south west,Q node] {\small $\gen{J}^{\ \dot{a}}_{\suA}$};
    \draw [arrow] (PsiP) -- (PsiM);

\draw[rounded corners=5mm] (-3.2cm,-2.6cm)rectangle (3.2cm,2.6cm);

\node (reprlabel) at ( 0  , -3cm) { $\varrho_{\sL}\otimes\varrho_{\sL}$};
  \end{tikzpicture}
\hspace{2cm}
  \begin{tikzpicture}[%
    box/.style={outer sep=1pt},
    Q node/.style={inner sep=1pt,outer sep=0pt},
    arrow/.style={-latex}
    ]%

    \node [box] (PhiM) at ( 0  , 2cm) { $\ket{Z^{\sR}}$};
    \node [box] (PsiP) at (-2cm, 0cm) { $\ket{\eta^{\sR}_{\  1}}$};
    \node [box] (PsiM) at (+2cm, 0cm) { $\ket{\eta^{\sR}_{\  2}}$};
    \node [box] (PhiP) at ( 0  ,-2cm) { $\ket{Y^{\sR}}$};

    \newcommand{\horshift}{0.09cm,0cm}
    \newcommand{\vershift}{0cm,0.10cm}
 
    \draw [arrow] ($(PsiP.north)-(\horshift)$) -- ($(PhiM.west) +(\vershift)$) node [pos=0.5,anchor=south east,Q node] {\small $\gen{Q}_{\sR 2},\overline{\gen{Q}}{}_{\sL 2}$};
    \draw [arrow] ($(PhiM.west) -(\vershift)$) -- ($(PsiP.north)+(\horshift)$) node [pos=0.5,anchor=north west,Q node] {};

    \draw [arrow] ($(PhiP.east) +(\vershift)$) -- ($(PsiM.south)-(\horshift)$) node [pos=0.5,anchor=south east,Q node] {};
    \draw [arrow] ($(PsiM.south)+(\horshift)$) -- ($(PhiP.east) -(\vershift)$) node [pos=0.5,anchor=north west,Q node] {\small $\overline{\gen{Q}}{}^{\ 2}_{\sR},\gen{Q}_{\sL}^{\ 2}$};

    \draw [arrow] ($(PsiM.north)-(\horshift)$) -- ($(PhiM.east) -(\vershift)$) node [pos=0.5,anchor=north east,Q node] {};
    \draw [arrow] ($(PhiM.east) +(\vershift)$) -- ($(PsiM.north)+(\horshift)$) node [pos=0.5,anchor=south west,Q node] {\small $\overline{\gen{Q}}{}^{\ 1}_{\sR}, \gen{Q}_{\sL}^{\ 1}$};

    \draw [arrow] ($(PhiP.west) -(\vershift)$) -- ($(PsiP.south)-(\horshift)$) node [pos=0.5,anchor=north east,Q node] {\small $\gen{Q}_{\sR 1}, \overline{\gen{Q}}{}_{\sL 1}$};
    \draw [arrow] ($(PsiP.south)+(\horshift)$) -- ($(PhiP.west) +(\vershift)$) node [pos=0.5,anchor=south west,Q node] {};

    \draw [arrow] (PsiM) -- (PsiP) node [pos=0.6,anchor=south west,Q node] {\small $\gen{J}^{\ \dot{a}}_{\suA}$};
    \draw [arrow] (PsiP) -- (PsiM);
    
\draw[rounded corners=5mm] (-3.2cm,-2.6cm)rectangle (3.2cm,2.6cm);

\node (reprlabel) at ( 0  , -3cm) { $\varrho_{\sR}\otimes\varrho_{\sR}$};
  \end{tikzpicture}
\caption{%
The massive excitations and their transformation properties under~$\mathcal{A}$.
The left and right panel depict the left and right representations respectively. The bosons $Z^{\sL,\sR}$ are excitations on $\AdS_3$ while $Y^{\sL,\sR}$ are excitations on $\Sphere^3$. Note that the massive fermions $\eta^{\sL \dot{a}},\eta^{\sR \dot{a}}$ are charged under~$\su(2)_{\bullet}$. The tensor products below each diagram indicate how each representation can be obtained from the short fundamental representations of $\su(1|1)^2_{\ce}$ introduced in section~\ref{sec:repr-smallalgebra}.
}
\label{fig:repr:massive}
  \end{figure}
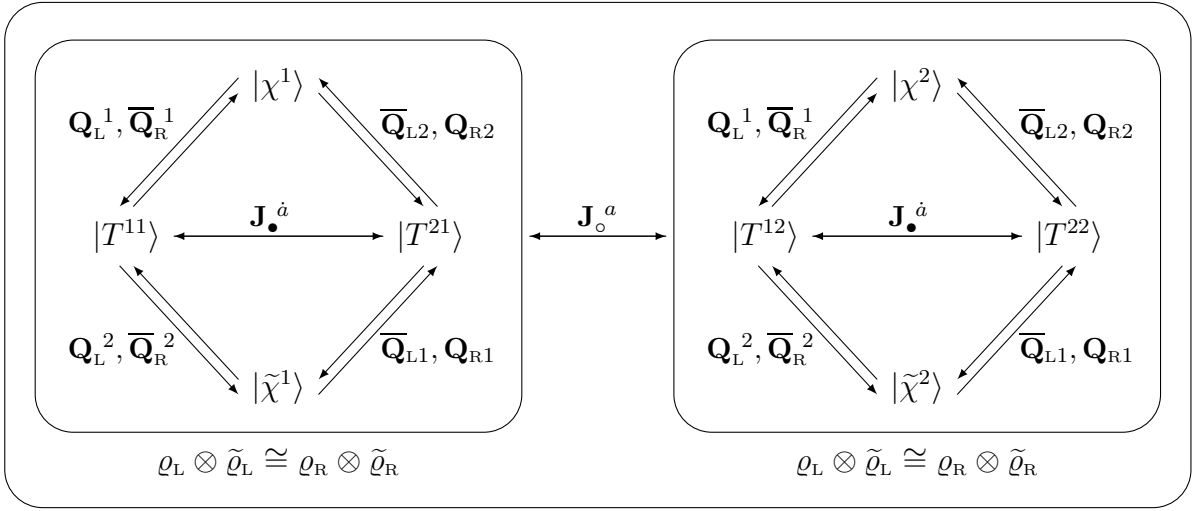
\begin{figure}
  \begin{tikzpicture}[%
    box/.style={outer sep=1pt},
    Q node/.style={inner sep=1pt,outer sep=0pt},
    arrow/.style={-latex}
    ]%
\begin{scope}[xshift=-4.2cm]
    \node [box] (PhiM) at ( 0  , 2cm) { $\ket{\chi^{1}}$};
    \node [box] (PsiP) at (-2cm, 0cm) { $\ket{T^{11}}$};
    \node [box] (PsiM) at (+2cm, 0cm) { $\ket{T^{21}}$};
    \node [box] (PhiP) at ( 0  ,-2cm) { $\ket{\widetilde{\chi}^{1}}$};

    \newcommand{\horshift}{0.09cm,0cm}
    \newcommand{\vershift}{0cm,0.10cm}
 
    \draw [arrow] ($(PhiM.west) +(\vershift)$) -- ($(PsiP.north)-(\horshift)$) node [pos=0.5,anchor=south east,Q node] {\small $\gen{Q}^{\ 1}_{\sL},\overline{\gen{Q}}{}_{\sR}^{\ 1}$};
    \draw [arrow] ($(PsiP.north)+(\horshift)$) -- ($(PhiM.west) -(\vershift)$) node [pos=0.5,anchor=north west,Q node] {};

    \draw [arrow] ($(PsiM.south)-(\horshift)$) -- ($(PhiP.east) +(\vershift)$) node [pos=0.5,anchor=south east,Q node] {};
    \draw [arrow] ($(PhiP.east) -(\vershift)$) -- ($(PsiM.south)+(\horshift)$) node [pos=0.5,anchor=north west,Q node] {\small $\overline{\gen{Q}}{}_{\sL 1}, \gen{Q}_{\sR 1}$};

    \draw [arrow] ($(PhiM.east) -(\vershift)$) -- ($(PsiM.north)-(\horshift)$) node [pos=0.5,anchor=north east,Q node] {};
    \draw [arrow] ($(PsiM.north)+(\horshift)$) -- ($(PhiM.east) +(\vershift)$) node [pos=0.5,anchor=south west,Q node] {\small $\overline{\gen{Q}}{}_{\sL 2}, \gen{Q}_{\sR 2}$};

    \draw [arrow] ($(PsiP.south)-(\horshift)$) -- ($(PhiP.west) -(\vershift)$) node [pos=0.5,anchor=north east,Q node] {\small $\gen{Q}^{\ 2}_{\sL},\overline{\gen{Q}}{}_{\sR}^{\ 2}$};
    \draw [arrow] ($(PhiP.west) +(\vershift)$) -- ($(PsiP.south)+(\horshift)$) node [pos=0.5,anchor=south west,Q node] {};

    \draw [arrow] (PsiM) -- (PsiP) node [pos=0.65,anchor=south west,Q node] {\small $\gen{J}^{\ \dot{a}}_{\suA}$};
    \draw [arrow] (PsiP) -- (PsiM);

\draw[rounded corners=5mm] (-3.2cm,-2.6cm)rectangle (3.2cm,2.6cm);

\node (reprlabel) at ( 0  , -3cm) { $\varrho_{\sL}\otimes\widetilde{\varrho}_{\sL} \cong\varrho_{\sR}\otimes\widetilde{\varrho}_{\sR}$};
\end{scope}
\begin{scope}[xshift=0cm]
    \draw [arrow] (-0.9cm,0cm) -- (0.9cm,0cm) node [Q node] at (0cm,0.22cm) {\small $\gen{J}^{\ a}_{\suB}$};
    \draw [arrow] (0.9cm,0cm) -- (-0.9cm,0cm);
%    node [Q node] at (0cm,0.6cm) {\scriptsize $\gen{J}^{\alpha}_{\ \suB}$};
    %
\end{scope}
\begin{scope}[xshift=4.2cm]

    \node [box] (PhiM) at ( 0  , 2cm) { $\ket{\chi^{2}}$};
    \node [box] (PsiP) at (-2cm, 0cm) { $\ket{T^{12}}$};
    \node [box] (PsiM) at (+2cm, 0cm) { $\ket{T^{22}}$};
    \node [box] (PhiP) at ( 0  ,-2cm) { $\ket{\widetilde{\chi}^{2}}$};

    \newcommand{\horshift}{0.09cm,0cm}
    \newcommand{\vershift}{0cm,0.10cm}
 
    \draw [arrow] ($(PhiM.west) +(\vershift)$) -- ($(PsiP.north)-(\horshift)$) node [pos=0.5,anchor=south east,Q node] {\small $\gen{Q}^{\ 1}_{\sL},\overline{\gen{Q}}{}_{\sR}^{\ 1}$};
    \draw [arrow] ($(PsiP.north)+(\horshift)$) -- ($(PhiM.west) -(\vershift)$) node [pos=0.5,anchor=north west,Q node] {};

    \draw [arrow] ($(PsiM.south)-(\horshift)$) -- ($(PhiP.east) +(\vershift)$) node [pos=0.5,anchor=south east,Q node] {};
    \draw [arrow] ($(PhiP.east) -(\vershift)$) -- ($(PsiM.south)+(\horshift)$) node [pos=0.5,anchor=north west,Q node] {\small $\overline{\gen{Q}}{}_{\sL 1}, \gen{Q}_{\sR 1}$};

    \draw [arrow] ($(PhiM.east) -(\vershift)$) -- ($(PsiM.north)-(\horshift)$) node [pos=0.5,anchor=north east,Q node] {};
    \draw [arrow] ($(PsiM.north)+(\horshift)$) -- ($(PhiM.east) +(\vershift)$) node [pos=0.5,anchor=south west,Q node] {\small $\overline{\gen{Q}}{}_{\sL 2}, \gen{Q}_{\sR 2}$};

    \draw [arrow] ($(PsiP.south)-(\horshift)$) -- ($(PhiP.west) -(\vershift)$) node [pos=0.5,anchor=north east,Q node] {\small $\gen{Q}^{\ 2}_{\sL}, \overline{\gen{Q}}{}_{\sR}^{\ 2}$};
    \draw [arrow] ($(PhiP.west) +(\vershift)$) -- ($(PsiP.south)+(\horshift)$) node [pos=0.5,anchor=south west,Q node] {};

    \draw [arrow] (PsiM) -- (PsiP) node [pos=0.65,anchor=south west,Q node] {\small $\gen{J}^{\ \dot{a}}_{\suA}$};
    \draw [arrow] (PsiP) -- (PsiM);

\draw[rounded corners=5mm] (-3.2cm,-2.6cm)rectangle (3.2cm,2.6cm);

\node (reprlabel) at ( 0  , -3cm) { $\varrho_{\sL}\otimes\widetilde{\varrho}_{\sL} \cong\varrho_{\sR}\otimes\widetilde{\varrho}_{\sR}$};
\end{scope}
\draw[rounded corners=5mm] (-7.8cm,-3.6cm)rectangle (7.8cm,3.1cm);
  \end{tikzpicture}
\caption{%
The massless excitations transform in two irreducible representations of~$\psu(1|1)^4_{\ce}$, which form a doublet under~$\su(2)_{\circ}$. Each of these representation can equivalently be taken to be the massless limit of a left or a right representation with a fermionic highest-weight state. For definiteness, here we take both of them to be given by left representations.
Below each $\psu(1|1)^4_{\ce}$ diagram we indicate how each representation can be obtained from one of two (left or right) isomorphic tensor products of fundamental $\su(1|1)^2_{\ce}$ representations, see  section~\ref{sec:repr-smallalgebra}.
}
\label{fig:repr:massless}
\end{figure}

The fundamental excitations of the theory are 8 bosons and 8 fermions, which arrange themselves into three irreducible representations of~$\mathcal{A}$. The $4+4$ massive excitations transform in two irreducible representations, that we call ``left'' and ``right'' and depict in figure~\ref{fig:repr:massive}. The remaining modes transform in the ``massless'' representation of~$\mathcal{A}$, depicted in figure~\ref{fig:repr:massless}. All these are short  representations of $\psu(1|1)^4_{\ce}$, \ie they satisfy the shortening condition
\begin{equation}
\label{eq:shortening}
\gen{H}^2=\gen{M}^2+4\,\gen{C}\,\overline{\gen{C}}.
\end{equation}
The left representation is four-dimensional and has $m=+1$. It is an irreducible representation of  $\psu(1|1)^4_{\ce}$ and it owes its name to the fact that on shell only the left supercharges act non-trivially on its module. We can write it as
\begin{equation}\label{eq:repr-massive-L}
  \begin{aligned}
    \gen{Q}_{\sL}^{\ \dot{a}} \ket{Y_p^{\sL}} &= a^{\sL}_p \ket{\eta^{\sL \dot{a}}_p},
    \qquad
    &\gen{Q}_{\sL}^{\ \dot{a}} \ket{\eta^{\sL \dot{b}}_p} &= \epsilon^{\dot{a}\dot{b}} \, a^{\sL}_p \ket{Z_p^{\sL}}, \\
    %%%% 
    \overline{\gen{Q}}{}_{\sL \dot{a}} \ket{Z_p^{\sL}} &=  - \epsilon_{\dot{a}\dot{b}}  \, \bar{a}^{\sL}_p \ket{\eta^{\sL \dot{b}}_p},
    \qquad
    &\overline{\gen{Q}}{}_{\sL \dot{a}} \ket{\eta^{\sL \dot{b}}_p}& =  \delta_{\dot{a}}^{\ \dot{b}}  \, \bar{a}^{\sL}_p \ket{Y_p^{\sL}}, \\[4pt]
    %%% 
    %%% 
    \gen{Q}_{\sR \dot{a}} \ket{Z^{\sL}_p} &= - \epsilon_{\dot{a}\dot{b}} \,  b^{\sL}_p \ket{\eta^{\sL \dot{b}}_p},
    \qquad
    &\gen{Q}_{\sR \dot{a}} \ket{\eta^{\sL \dot{b}}_p} &= \delta_{\dot{a}}^{\ \dot{b}} \, b^{\sL}_p \ket{Y^{\sL}_p},\\
    %%% 
    \overline{\gen{Q}}{}_{\sR}^{\ \dot{a}} \ket{Y^{\sL}_p} &= \bar{b}^{\sL}_p \ket{\eta^{\sL \dot{a}}_p},
    \qquad
    &\overline{\gen{Q}}{}_{\sR}^{\ \dot{a}} \ket{\eta^{\sL \dot{b}}_p} &= \epsilon^{\dot{a}\dot{b}} \,  \bar{b}^{\sL}_p \ket{Z^{\sL}_p}.
  \end{aligned}
\end{equation}
The representation coefficients $a^{\sL}_p$ and $b^{\sL}_p$ are such as to reproduce the central charges~\eqref{eq:allloop-centralcharges-RR} and $b^{\sL}_p$ vanishes on shell, \ie, $b^{\sL}_{p=0}=0$. We will comment more on the $a^{\sL}_p$ and $b^{\sL}_p$ in the next subsection.

The right representation is also four-dimensional and has $m=-1$. It is given by
\begin{equation}\label{eq:repr-massive-R}
  \begin{aligned}
    \gen{Q}_{\sL}^{\ \dot{a}} \ket{Z_p^{\sR}} &=  b^{\sR}_p \ket{\eta^{\sR \dot{a}}_p},
    \qquad
    &\gen{Q}_{\sL}^{\ \dot{a}} \ket{\eta^{\sR \dot{b}}_p} &=- \epsilon^{\dot{a}\dot{b}} \,  b^{\sR}_p \ket{Y_p^{\sR}},\\
        \overline{\gen{Q}}{}_{\sL \dot{a}} \ket{Y_p^{\sR}} &= \epsilon_{\dot{a}\dot{b}} \,  \bar{b}^{\sR}_p \ket{\eta^{\sR \dot{b}}_p},
    \qquad
    &\overline{\gen{Q}}{}_{\sL \dot{a}} \ket{\eta^{\sR \dot{b}}_p} &= \delta_{\dot{a}}^{\ \dot{b}} \,  \bar{b}^{\sR}_p \ket{Z_p^{\sR}},\\[4pt]
   \gen{Q}_{\sR \dot{a}} \ket{Y_p^{\sR}} &=  \epsilon_{\dot{a}\dot{b}} \,  a^{\sR}_p \ket{\eta^{\sR \dot{b}}_p},
    \qquad
    &\gen{Q}_{\sR \dot{a}} \ket{\eta^{\sR \dot{b}}_p} &= \delta_{\dot{a}}^{\ \dot{b}} \,  a^{\sR}_p \ket{Z_p^{\sR}}, \\
    %%% 
    \overline{\gen{Q}}{}_{\sR}^{\ \dot{a}} \ket{Z_p^{\sR}} &= \bar{a}^{\sR}_p \ket{\eta^{\sR \dot{a}}_p},
    \qquad
    &\overline{\gen{Q}}{}_{\sR}^{\ \dot{a}} \ket{\eta^{\sR \dot{b}}_p} &= - \epsilon^{\dot{a}\dot{b}}  \, \bar{a}^{\sR}_p \ket{Y_p^{\sR}}.
  \end{aligned}
\end{equation}
Note that this right representation follows from the previous one by relabelling everywhere~L$\leftrightarrow$R. We will refer this $\mathbb{Z}_2$ symmetry as \emph{left-right symmetry} (LR symmetry).

Finally, the massless representation is eight-dimensional with four bosons $T^{\dot{a}a}$ and four fermions $\chi^{a},\widetilde{\chi}^{a}$. This representation has $m=0$ and is given by two irreducible representations of~$\psu(1|1)^4_{\ce}$ that form a doublet under~$\su(2)_{\circ}\subset\so(4)_2$. Each of these $\psu(1|1)^4_{\ce}$ representations can equivalently be obtained by using the left or right representations above and taking a massless limit of the coefficients $a^{\sL}_p,b^{\sL}_p$ or $a^{\sR}_p,b^{\sR}_p$. This is due to the fact that the left and right representations become isomorphic in the massless limit. For definiteness, let us take the representation coefficients to be inherited from the left representation. Then we have
\begin{equation}\label{eq:repr-massless}
  \begin{aligned}
    \gen{Q}_{\sL}^{\ \dot{a}} \ket{T^{\dot{b}a}_p}& = \epsilon^{\dot{a}\dot{b}} a^{\sL}_p \ket{\widetilde{\chi}^a_p},
    \qquad
    &\gen{Q}_{\sL}^{\ \dot{a}} \ket{\chi^{a}_p} \;&=  a^{\sL}_p \ket{T^{\dot{a}a}_p}, \\
    %%% 
    \overline{\gen{Q}}{}_{\sL \dot{a}} \ket{\widetilde{\chi}^{a}_p}\;& = -\epsilon_{\dot{a}\dot{b}} \bar{a}^{\sL}_p \ket{T^{\dot{b}a}_p},
    \qquad
    &\overline{\gen{Q}}{}_{\sL \dot{a}} \ket{T^{\dot{b}a}_p} &= \delta_{\dot{a}}^{\ \dot{b}} \bar{a}^{\sL}_p \ket{\chi^a_p}, \\[4pt]
    %%% 
    %%%
    \gen{Q}_{\sR \dot{a}} \ket{T^{\dot{b}a}_p} &= \delta_{\dot{a}}^{\ \dot{b}} b^{\sL}_p \ket{\chi^a_p},
    \qquad
    &\gen{Q}_{\sR \dot{a}} \ket{\widetilde{\chi}^a_p} \;&= -\epsilon_{\dot{a}\dot{b}} b^{\sL}_p \ket{T^{\dot{b}a}_p}, \\
    %%% 
    \overline{\gen{Q}}{}_{\sR}^{\ \dot{a}} \ket{\chi^a_p}\;& = \bar{b}^{\sL}_p \ket{T^{\dot{a}a}_p},
    \qquad
    &\overline{\gen{Q}}{}_{\sR}^{\ \dot{a}} \ket{T^{\dot{b}a}_p} &= \epsilon^{\dot{a}\dot{b}} \bar{b}^{\sL}_p \ket{\widetilde{\chi}^a_p}.
  \end{aligned}
\end{equation}
Note also that the highest weight state of the massless representations are fermionic, namely $\ket{\chi^{a}}$, in contrast with the ones of the left and right representations, that are $\ket{Y^{\sL}}$ and $\ket{Z^{\sR}}$ respectively.

\subsubsection{Representation coefficients for the pure R-R theory}
\label{sec:RR-repr-coeff}
In absence of NS-NS fluxes, the representations coefficients are
\begin{equation}
\begin{aligned}
\label{eq:RR-repr-coeff}
&a^{\sL}_p = a^{\sR}_p = \eta_p e^{i\xi}, \qquad&& \bar{a}^{\sL}_p = \bar{a}^{\sR}_p = \eta_p e^{-ip/2} e^{-i\xi},\\
&b^{\sL}_p = b^{\sR}_p = -\frac{\eta_p}{x^-_p} e^{-ip/2} e^{i\xi}, \qquad\qquad &&\bar{b}^{\sL}_p = \bar{b}^{\sR}_p = -\frac{\eta_p}{x^+_p} e^{-i\xi},
\end{aligned}
\end{equation}
with
\begin{equation}
\eta_p = e^{ip/4}  \sqrt{\frac{ih}{2}(x^-_p - x^+_p)}.
\end{equation}
The equality of the left- and right-representation coefficients in equation~\eqref{eq:RR-repr-coeff} indicates that left-right symmetry is particularly simple in the pure-R-R case.
The Zhukovski variables~$x^{\pm}_p$ are mass-dependent:
\begin{equation}
\label{eq:zhukovski}
\frac{x^+_p}{x^-_p}=e^{ip},
\qquad
x^+_p +\frac{1}{x^+_p} -x^-_p -\frac{1}{x^-_p} = \frac{2i \, |m|}{h}.
\end{equation}
In fact, the dependence of the representation parameters on $m=\pm1,0$ is entirely encoded in $x^{\pm}_p$.
The phase~$\xi$ is irrelevant for the one-particle representation, but is instrumental for defining the two-particle representation, \ie in order to define a non-trivial coproduct~\cite{Borsato:2014hja}, similarly to what happens for $\AdS_5\times\Sphere^5$ strings~\cite{Arutyunov:2006yd}.

Let us note that while the excitations in the left and right modules have kinematic properties similar to the one of $\AdS_5\times\Sphere^5$ excitations, new features emerge in the massless case. When $m=0$, the Zhukovski variables satisfy the additional constraint
\begin{equation}
x^{+}_p=\frac{1}{x^{-}_p}.
\end{equation}
Moreover, the vanishing of $\gen{M}$ imposes that the representation coefficients satisfy
\begin{equation}
\label{eq:massless-rep-shortening1}
|a^{\sL,\sR}_p|^2=|b^{\sL,\sR}_p|^2,\qquad
\text{at}\quad m=0,
\end{equation}
and the dispersion relation becomes non-analytic,
\begin{equation}
E(p)=2h\,\Big|\sin\bigl(\frac{p}{2}\bigr)\Big|.
\end{equation}
This last property can be physically interpreted as an indication that left- and right-movers on the worldsheet should be treated as two different species of particles, similarly to what is done in the relativistic case.

\subsubsection{Representations of \texorpdfstring{$\su(1|1)^2_{\ce}$}{su(1|1)**2 c.e.}}
\label{sec:repr-smallalgebra}
It is useful to introduce the $\su(1|1)^2_{\ce}$ algebra, whose anticommutation relations are
\begin{equation}
\label{eq:cealgebra-small}
\begin{aligned}
&\{\gen{Q}_{\smallL},\overline{\gen{Q}}{}_{\smallL}\} =  \frac{1}{2}\,(\gen{H}+\gen{M}),
&\qquad &\{\gen{Q}_{\smallL},{\gen{Q}}{}_{\smallR}\} = \gen{C},\\
&\{\gen{Q}_{\smallR},\overline{\gen{Q}}{}_{\smallR} \}= \frac{1}{2}\,(\gen{H}-\gen{M}),
&\qquad &\{\overline{\gen{Q}}{}_{\smallL},\overline{\gen{Q}}{}_{\smallR}\} = \overline{\gen{C}}.
\end{aligned}
\end{equation}
The short representations of this algebra are two-dimensional, and have been studied in ref.~\cite{Borsato:2012ud}. Once again, we have a left representation $\varrho_{\sL}$
\begin{equation}\label{eq:su(1|1)2-repr1}
  \begin{aligned}
    \gen{Q}_{\smallL} \ket{\phi^{\sL}_p} &= a^{\sL}_p \ket{\psi^{\sL}_p} , \qquad &
    \overline{\gen{Q}}{}_{\smallL} \ket{\psi^{\sL}_p} &= \bar{a}^{\sL}_p \ket{\phi^{\sL}_p} , \\
    \gen{Q}_{\smallR} \ket{\psi^{\sL}_p} &= b^{\sL}_p \ket{\phi^{\sL}_p} , \qquad &    \overline{\gen{Q}}{}_{\smallR} \ket{\phi^{\sL}_p} &= \bar{b}^{\sL}_p \ket{\psi^{\sL}_p}
 .
  \end{aligned}
\end{equation}
Similarly, we can consider a right representation~$\varrho_{\sR} $
\begin{equation}\label{eq:su(1|1)2-reprR}
  \begin{aligned}
 \gen{Q}_{\smallL} \ket{\psi_p^{\sR}} &= b^{\sR}_p \ket{\phi_p^{\sR}} , \qquad &
    \overline{\gen{Q}}{}_{\smallL} \ket{\phi_p^{\sR}} &= \bar{b}^{\sR}_p \ket{\psi_p^{\sR}} , \\
    \gen{Q}_{\smallR} \ket{\phi_p^{\sR}} &= a^{\sR}_p \ket{\psi_p^{\sR}} , \qquad &
    \overline{\gen{Q}}{}_{\smallR} \ket{\psi_p^{\sR}} &= \bar{a}^{\sR}_p \ket{\phi_p^{\sR}}    .
  \end{aligned}
\end{equation}
Two more representations, which we denote by $\widetilde{\varrho}_{\sL}$ and $\widetilde{\varrho}_{\sR}$, can be obtained from the ones above by exchanging bosons with fermions.

As discussed in detail in ref.~\cite{Borsato:2013qpa}, appropriate tensor products of pairs of these representations are isomorphic to the~$\psu(1|1)^4_{\ce}$ representations discussed above. In fact, a similar structure will be present also in the mixed-flux case, and we will exploit it to write down the S~matrix.

Let us sketch these isomorphisms. Firstly, note that we can obtain~$\psu(1|1)^4_{\ce}$ supercharges from those of the tensor products of~$\su(1|1)^2_{\ce}$ by setting%
\footnote{This tensor-product structure is similar to the one of $\psu(2|2)^2_{\ce}$, which is the off-shell symmetry algebra of $\AdS_5\times\Sphere^5$ superstrings~\cite{Arutyunov:2009ga,Beisert:2010jr}.}
\begin{equation}
  \begin{aligned}
    \gen{Q}_{\smallL}^{\ 1} = \gen{Q}_{\smallL} \otimes \1 , \qquad
    \gen{Q}_{\smallL}^{\ 2} = \1 \otimes \gen{Q}_{\smallL}, \qquad
    \gen{Q}_{\smallR 1} = \gen{Q}_{\smallR} \otimes \1 , \qquad
    \gen{Q}_{\smallR 2} = \1 \otimes \gen{Q}_{\smallR} ,
  \end{aligned}
\end{equation}
and similarly for their conjugates. Clearly then ${\varrho}_{\sL}\otimes {\varrho}_{\sL}$, ${\varrho}_{\sR}\otimes {\varrho}_{\sR}$, etc.\@ are representations of $\psu(1|1)^4_{\ce}$. What is more, one can check that the left representation given in equation~\eqref{eq:repr-massive-L} is isomorphic to~${\varrho}_{\sL}\otimes {\varrho}_{\sL}$, while the right one~\eqref{eq:repr-massive-R} is isomorphic to~${\varrho}_{\sR}\otimes {\varrho}_{\sR}$. As for the two~$\psu(1|1)^4_{\ce}$ modules that constitute the massless $\mathcal{A}$ module, each of them can be given either by ${\varrho}_{\sL}\otimes \widetilde{\varrho}_{\sL}$ or by ${\varrho}_{\sR}\otimes \widetilde{\varrho}_{\sR}$. This is consistent with the equivalence of left and right representations when $m=0$, and with the fact that the massless modules have fermionic highest-weight states.

The details of the isomorphisms outlined above are reviewed in appendix~\ref{app:tensorprod}.

\subsection{Representations in the near-plane-wave limit for the mixed-flux theory}
\label{sec:near-BMN-representations}
The near-plane-wave limit of the symmetry algebra~\cite{Gava:2002xb,Berenstein:2002jq} can be read off the explicit expression of the supercurrents~\eqref{eq:quadratic-currents}, truncated at quadratic order in the fields. This yields a representation on the fields and conjugate momenta $X, P, \eta, \bar{\eta}$, and so on. For our purposes it is more useful to work in terms of the excitations. To this end, we introduce creation and annihilation operators in the usual way. For the bosons, we schematically write
\begin{equation}
\begin{aligned}
a^{\dagger}(p)\approx \int\frac{\de\sigma}{\sqrt{\omega(p,m,\fl)}} \bigl(\omega(p,m,\fl)\,X-i P\bigr) e^{+ip\sigma},\\
a(p)\approx \int\frac{\de\sigma}{\sqrt{\omega(p,m,\fl)}} \bigl(\omega(p,m,\fl)\,X+i P\bigr) e^{-ip\sigma}.
\end{aligned}
\end{equation}
This representation depends on the energy~$\omega(p,m,\fl)$, which is function of the mass~$m$ and on the flux parameter~$\fl$. For the fermions we write
\begin{equation}
\begin{aligned}
d^{\dagger}(p)\approx \int\frac{\de\sigma}{\sqrt{\omega(p,m,\fl)}} \bigl(f(p,m,\fl)\,\eta-i g(p,m,\fl)\, \bar{\eta}\bigr) e^{+ip\sigma},\\
d(p)\approx \int\frac{\de\sigma}{\sqrt{\omega(p,m,\fl)}} \bigl(f(p,m,\fl)\,\eta+i g(p,m,\fl)\, \bar{\eta}\bigr) e^{-ip\sigma},
\end{aligned}
\end{equation}
where we introduced the wave-function parameters $f(p,m,\fl)$ and~$g(p,m,\fl)$. The creation operators generate the space of fundamental excitations, which as we reviewed consists of sixteen particles
\begin{equation}
\begin{gathered}
\ket{Z^{\sL,\sR}}=a^\dagger_{\sL,\sR\, z}\ket{0},\quad
\ket{Y^{\sL,\sR}}=a^\dagger_{\sL,\sR\, y}\ket{0},\quad
\ket{\eta^{\sL \dot{a}}}=d^{\ \dot{a} \dagger}_{\sL}\ket{0},\quad
\ket{\eta^{\sR}_{\ \dot{a}}}=d^{\dagger}_{\sR \dot{a}}\ket{0},\\
\ket{T^{\dot{a}a}}=a^{\dot{a} a\dagger}\ket{0},\qquad
\ket{\chi^{a}}=d^{a\,\dagger}\ket{0},\qquad
\ket{\widetilde{\chi}^{a}}=\tilde{d}^{a\,\dagger}\ket{0}.
\end{gathered}
\end{equation}

In appendix~\ref{app:charges:osc} we extract the supercharges from the supercurrents constructed in section~\ref{sec:off-shell-algebra}, which indeed gives an algebra of the form~\eqref{eq:cealgebra}. Furthermore, we rewrite them in terms of oscillators, obtaining
\begin{equation}
\label{eq:supercharges-leading-order-rep}
\begin{aligned}
&\gen{Q}_{\smallL}^{\ {\dot{a}}}= \int \de p \ \Bigl[
 (d_{\sL}^{\ {\dot{a}}\,\dagger} a_{\sL y} + \epsilon^{{\dot{a}\dot{b}}}\, a_{\sL z}^\dagger  d_{\sL {\dot{b}}})f_p^{\sL}
+ (a_{\sR y}^\dagger  d_{\sR}^{\ {\dot{a}}} +\epsilon^{{\dot{a}\dot{b}}}\, d_{\sR {\dot{b}}}^\dagger  a_{\sR z})\,g_p^{\sR} \\
&\qquad\qquad\qquad\qquad\qquad\qquad\qquad\qquad\qquad\qquad
+ \left( \epsilon^{{\dot{a}\dot{b}}}\, \tilde{d}^{{a}\,\dagger}a_{{\dot{b}a}}+a^{{\dot{a}a}\,\dagger}d_{{a}}\right)\tilde{f}_p
\Bigr],\\
%%%%
&\gen{Q}_{\smallR {\dot{a}}}=\int \de p \ \Bigl[
 (d_{\sR {\dot{a}}}^\dagger  a_{\sR y} -\epsilon_{{\dot{a}\dot{b}}}\, a_{\sR z}^\dagger d_{\sR}^{\ {\dot{b}}})f_p^{\sR}
+ (a_{\sL y}^\dagger  d_{\sL {\dot{a}}} -\epsilon_{{\dot{a}\dot{b}}}\,  d_{\sL}^{\ {\dot{b}}\,\dagger} a_{\sL z})\,g_p^{\sL}\\
&\qquad\qquad\qquad\qquad\qquad\qquad\qquad\qquad\qquad\qquad
 + \left( d^{{a}\,\dagger}a_{{\dot{a}a}}-\epsilon_{{\dot{a}\dot{b}}}\, a^{{\dot{b}a}\,\dagger}\tilde{d}_{{a}}\right)\tilde{g}_p
 \Bigr],
%%%%
\end{aligned}
\end{equation}
and similarly for their conjugates. Note that we suppressed the dependence of $f_p^{\sL,\sR}, \tilde{f}_p$ and $g_p^{\sL,\sR}, \tilde{g}_p$ on $m$ and $\fl$ for ease of notation.

This representation is indeed of the form~(\ref{eq:repr-massive-L}--\ref{eq:repr-massless}) up to fixing the representation coefficients, and closely resembles near-plane wave limit of the pure-R-R one discussed in ref.~\cite{Borsato:2014hja}. Note however that wave-function parameters are $f_p^{\sL}\neq f_p^{\sR}$ and $g_p^{\sL}\neq g_p^{\sR}$, unlike what happened in the pure-R-R case. This is reflected by the values of the central charges
\begin{equation}
\label{eq:MH-near-BMN}
\gen{M}=\begin{cases}
\fl p+1 &\text{left}, \\
\fl p-1 &\text{right}, \\
\fl p &\text{massless},
\end{cases}
\qquad\qquad
\gen{H}=\begin{cases}
\sqrt{\flt^2+(p+\fl)^2} &\text{left}, \\
\sqrt{\flt^2+(p-\fl)^2}  &\text{right}, \\
\sqrt{p^2} &\text{massless},
\end{cases}
\end{equation}
while the off-shell central charges take the same form for all representations, and are both real
\begin{equation}
\gen{C}=\overline{\gen{C}}=-\frac{\flt}{2}\, p.
\end{equation}
This is consistent with the tree-level analysis of symmetries in the massive sector~\cite{Hoare:2013ida} and with the leading-order massless dispersion relation~\cite{Berenstein:2002jq}.

\subsubsection{Representation coefficients}
\label{sec:repr-paraman-nfs}
The wave-function parameters play the role of representation coefficients. In fact, comparing~\eqref{eq:supercharges-leading-order-rep} with the pure-R-R representations of section~\ref{sec:RRreprs}, we see that the real parameter $f_p$ should be the near-plane-wave limit of~$a_p$ and~$\bar{a}_p$, while~$g_p$ should be the limit of~$b_p$ and~$\bar{b}_p$. Therefore, the precise form of $f_p$ and $g_p$ will be important in order to fix $a_p, \bar{a}_p, b_p$ and $\bar{b}_p$ in the full theory.

Let us begin from the massive representations. We have
\begin{equation}
\begin{aligned}
\omega^{\sL}_p=\sqrt{\flt^2+(p+\fl)^2},
\qquad
f^{\sL}_p=\sqrt{\frac{1+\fl p+\omega^{\sL}_p}{2}},
\qquad
g^{\sL}_p=-\frac{\flt\,p}{2\,f^{\sL}_p},\\
\omega^{\sR}_p=\sqrt{\flt^2+(p-\fl)^2},
\qquad
f^{\sR}_p=\sqrt{\frac{1-\fl p+\omega^{\sR}_p}{2}},
\qquad
g^{\sR}_p=-\frac{\flt\,p}{2\,f^{\sR}_p}.
\end{aligned}
\end{equation}
where the energy is related to the representation parameters by~$\omega_p=f_p^2+g_p^2$.
The different sign in front of the $p$-linear terms is explained by the necessity of reproducing~\eqref{eq:MH-near-BMN}, and ultimately is a consequence of the fact that the NS-NS flux breaks parity invariance. In particular, this implies that LR symmetry will require a non-trivial map of the representation coefficients too.

For the massless representation we have
\begin{equation}
\tilde{\omega}_p\equiv \tilde{\omega}^{\sL}_p=\sqrt{p^2},
\qquad
\tilde{f}_p\equiv \tilde{f}^{\sL}_p=\sqrt{\frac{\fl p+\tilde{\omega}_p}{2}}
\qquad
\tilde{g}_p\equiv \tilde{g}^{\sL}_p=-\frac{\flt\,p}{2\,\tilde{f}_p}.
\end{equation}
This may appear troubling: we have argued in section~\ref{sec:RRreprs} that at least at~$\fl=0$ massless modes can be equivalently obtained from the left or right representation, and indeed this seems to be the case looking at~\eqref{eq:MH-near-BMN}. However, the values of $\tilde{f}_p$ and~$\tilde{g}_p$ come from a massless limit of $f^{\sL}_p$ and $g^{\sL}_p$. To see how this is inessential, let us define new massless parameters, now as limit of $f^{\sR}_p$ and $g^{\sR}_p$:
\begin{equation}
\tilde{\omega}^{\sR}_p=\sqrt{p^2},
\qquad
\tilde{f}^{\sR}_p=\sqrt{\frac{-\fl p+\tilde{\omega}^{\sR}_p}{2}}
\qquad
\tilde{g}^{\sR}_p=-\frac{\flt\,p}{2\,\tilde{f}^{\sR}_p}.
\end{equation}
Let us rescale \eg the massless fermion creation operators in \eqref{eq:supercharges-leading-order-rep} as
\begin{equation}
d_{a}\to \frac{\tilde{f}_p^{\sL}}{\tilde{g}^{\sR}_p}\, d_a,\quad 
d^{a\,\dagger}\to \frac{\tilde{f}_p^{\sL}}{\tilde{g}^{\sR}_p}\,d^{a\,\dagger},\qquad
\tilde{d}_{a}\to -\frac{\tilde{g}^{\sR}_p}{\tilde{f}^{\sL}_p}\,\tilde{d}_a,\quad 
\tilde{d}^{a\,\dagger}\to -\frac{\tilde{g}^{\sR}_p}{\tilde{f}^{\sL}_p}\,\tilde{d}^{a\,\dagger},
\end{equation}
and note the identities
\begin{equation}
(\tilde{f}_p^{\sL})^2=(\tilde{g}^{\sR}_p)^2,
\qquad
(\tilde{g}_p^{\sL})^2=(\tilde{f}^{\sR}_p)^2,
\qquad
\tilde{f}^{\sL}_p\,\tilde{f}^{\sR}_p= \tilde{g}^{\sL}_p\,\tilde{g}^{\sR}_p.
\end{equation}
In this way we replace everywhere the parameters $\tilde{f}_p\equiv\tilde{f}_p^{\sL}$ and $\tilde{g}_p\equiv\tilde{g}_p^{\sL}$ with $\tilde{f}_p^{\sR}$ and $\tilde{g}_p^{\sR}$, and in fact obtain the massless limit of a right representation. In summary, also in the mixed-flux case the massless representation can equivalently be described as a left or a right one, at least in the near-plane-wave limit. For definiteness, we will adopt the first choice.

\subsection{Exact representations for the mixed-flux theory}
\label{sec:exact-repr-mixed}
When we go beyond the near-plane-wave limit, we expect the representations discussed in the previous subsection to be deformed. In particular, as we have computed in section~\ref{sec:supercurrent-algebra}, the off-shell central charges will be non-linear functions of the worldsheet momentum,
\begin{equation}
\label{eq:Cexact}
\gen{C}=+\frac{i\h}{2}(e^{+i\,\gen{P}}-1),
\qquad
\overline{\gen{C}}=-\frac{i\h}{2}(e^{-i\,\gen{P}}-1).
\end{equation}
Here we introduce the mixed-flux coupling constant $\h=\h(\lambda,\flt)$, which enters as an overall normalisation of the central charge. In the worldsheet calculation we found that for large $\sqrt{\lambda}$
\begin{equation}
  \h(\lambda,\flt) \approx \frac{\flt \sqrt{\lambda}}{2\pi}\, .
\end{equation}
However, this relation might receive perturbative and non-perturbative corrections in $1/\sqrt{\lambda}$, analogously to what happens for string theory in $\AdS_4\times\CP^3$~\cite{Nishioka:2008gz,Gaiotto:2008cg}.\footnote{%
  Recently a proposal has been made for the all-loop $\lambda$-dependence of the function $h(\lambda)$ in $\AdS_4 \times \CP^3$~\cite{Gromov:2014eha}.
} %
Note that we have absorbed a factor $\flt$ into the definition of $\h$. This makes~\eqref{eq:Cexact} take the same form as in the pure R-R case, but differs from the conventions of previous literature.

In our discussion of  pure-R-R representations at the start of this section the eigenvalue of~$\gen{M}$ was a real number. However, from the string theory computations of section~\ref{sec:currents} we see that it should be a function of the total worldsheet momentum~$\gen{P}$. This may appear surprising as we expect~$\gen{M}$ to be a \emph{quantised} angular momentum on a physical state. As we detail in section~\ref{sec:mass-comments}, this can be achieved if we take
\begin{equation}
\label{eq:Mexact}
\gen{M}=m+\k\,\gen{P} ,
\end{equation}
where $m=\pm1,0$ depending on which representation we are considering, as in equation~\eqref{eq:MH-near-BMN}.
The constant $\k$ is related to the WZW level $k$ by
\begin{equation}
  \k = \frac{k}{2\pi} = \frac{q\sqrt{\lambda}}{2\pi} \,.
\end{equation}
In this way, using the shortening condition~\eqref{eq:shortening}, we conclude that the all-loop dispersion relation is
\begin{equation}
\label{eq:all-loop-dispersion}
E(p)=\sqrt{(m+\k\,p)^2+4\h^2\,\sin^2\bigl(\frac{p}{2}\bigr)} \,,
\end{equation}
which for massive particles confirms what was found by the analysis of giant magnons~\cite{Hoare:2013lja, Ahn:2014tua, Babichenko:2014yaa}. It is interesting to note that in the massless case the dispersion relation is non-analytic also at $\k \neq 0$. To make this evident, we write
\begin{equation}
\label{eq:all-loop-mless-dispersion}
E(p) = \k |p| \sqrt{ 1 + \frac{4 \h^2 \sin^2\bigl(\frac{p}{2}\bigr)}{\k^2 p^2}} , \qquad
\text{at }m=0 ,
\end{equation}
where the square root is analytic when $p$ is in the vicinity of the real line.

\subsubsection{Exact representation parameters}
\label{sec:exact-repr-mixed:params}
We now want to construct three irreducible representations of $\mathcal{A}$ that in the limit $\fl\to0$ coincide with the pure-R-R ones which we recalled in section~\ref{sec:RRreprs}, and whose near-plane-wave limit is the one we computed in section~\ref{sec:near-BMN-representations}. To this end, it will be sufficient to suitably deform the representation coefficients of equations \eqref{eq:repr-massive-L}--\eqref{eq:repr-massless}. In particular, we define
\begin{equation}\label{eq:abparam}
\begin{aligned}
a^{\sL}_p &= \eta_p^{\sL}\, e^{i\xi},
&\quad
\bar{a}^{\sL}_p &= \eta_p^{\sL}\, e^{-ip/2} e^{-i\xi},
&\quad
b^{\sL}_p &= -\frac{\eta_p^{\sL}}{x^-_{\sL\, p}} e^{-ip/2} e^{i\xi},
&\quad
\bar{b}^{\sL}_p &= -\frac{\eta^{\sL}_p}{x^+_{\sL\,p}} e^{-i\xi},\\
a^{\sR}_p &= \eta_p^{\sR}\, e^{i\xi},
&\quad
\bar{a}^{\sR}_p &= \eta_p^{\sR}\, e^{-ip/2} e^{-i\xi},
&\quad
b^{\sR}_p &= -\frac{\eta_p^{\sR}}{x^-_{\sR\, p}} e^{-ip/2} e^{i\xi},
&\quad
\bar{b}^{\sR}_p &= -\frac{\eta^{\sR}_p}{x^+_{\sR\,p}} e^{-i\xi},
\end{aligned}
\end{equation}
with
\begin{equation}
\eta_p^{\sL} = e^{ip/4}\sqrt{\frac{i\h}{2}(x^-_{\sL\, p} - x^+_{\sL\, p})} ,
\qquad
\eta_p^{\sR} = e^{ip/4}\sqrt{\frac{i\h}{2}(x^-_{\sR\, p} - x^+_{\sR\, p})}.
\end{equation}
Here we have introduced two sets of Zhukovski variables~$x^{\pm}_{\sL\,p}$ and $x^{\pm}_{\sR\,p}$, which satisfy
\begin{equation}
\label{eq:zhukovski2}
\begin{gathered}
\frac{x^+_{\sL\,p}}{x^-_{\sL\,p}}=e^{ip},
\qquad
x^+_{\sL\,p} +\frac{1}{x^+_{\sL\,p}} -x^-_{\sL\,p} -\frac{1}{x^-_{\sL\,p}} = \frac{2i \, (|m|+\k \, p)}{\h},\\
\frac{x^+_{\sR\,p}}{x^-_{\sR\,p}}=e^{ip},
\qquad
x^+_{\sR\,p} +\frac{1}{x^+_{\sR\,p}} -x^-_{\sR\,p} -\frac{1}{x^-_{\sR\,p}} = \frac{2i \, (|m|-\k\, p)}{\h}.
\end{gathered}
\end{equation}
These equations can be solved by setting
\begin{equation}
\begin{gathered}
x^{\pm}_{\sL\,p}=\frac{(|m| + \k p)+\sqrt{(|m| + \k p)^2 + 4\h^2 \sin^2(\frac{p}{2})}}{2\h\sin(\frac{p}{2})}e^{\pm\frac{i}{2}p} ,\\
x^{\pm}_{\sR\,p}=\frac{(|m| - \k p)+\sqrt{(|m| - \k p)^2 + 4\h^2 \sin^2(\frac{p}{2})}}{2\h\sin(\frac{p}{2})}e^{\pm\frac{i}{2}p} ,
\end{gathered}
\end{equation}
as usual with~$m=\pm1,0$. 
In this way, we reproduce the central charges~(\ref{eq:Cexact}--\ref{eq:Mexact}) and the dispersion relation~\eqref{eq:all-loop-dispersion}.
It is interesting to note that for the massless modes it is no longer true that $x^+=1/x^-$. This identity is replaced by
\begin{equation}
\label{eq:magic-relation}
x^{\pm}_{\sL\,p}=\frac{1}{x^{\mp}_{\sR\,p}}\qquad
\text{at}\quad m=0.
\end{equation}
Owing to this equality, we can check the following identities for the representation coefficients of the massless representation
\begin{equation}
\label{eq:massless-rep-shortening2}
a^{\sL}_p\,\bar{a}^{\sR}_p=b^{\sL}_p\,\bar{b}^{\sR}_p=a^{\sR}_p\,\bar{a}^{\sL}_p=b^{\sR}_p\,\bar{b}^{\sL}_p,\qquad
|a^{\sL}_p|^2=|b^{\sR}_p|^2,
\qquad
|a^{\sR}_p|^2=|b^{\sL}_p|^2,
\end{equation}
that generalise~\eqref{eq:massless-rep-shortening1}. Note that it is not true that \eg $|a^{\sL}_p|^2=|b^{\sL}_p|^2$ at $\k\neq 0$.

\subsubsection{Equivalent representations for massless modes}
\label{eq:equiv-reprs-mless}
In equation~\eqref{eq:repr-massless} we chose to describe all massless modes by the massless limit of left representations. In terms of $\su(1|1)^2_{\ce}$ representations, this corresponds to describing the massless module as~$(\varrho_{\sL} \otimes\widetilde{\varrho}_{\sL})^{\oplus2}$. As discussed in detail in ref.~\cite{Borsato:2014hja} and as we briefly recalled, equivalent alternative description are $(\varrho_{\sR} \otimes\widetilde{\varrho}_{\sR})^{\oplus2}$, $(\varrho_{\sL} \otimes\widetilde{\varrho}_{\sL})\oplus (\varrho_{\sR} \otimes\widetilde{\varrho}_{\sR})$ or $(\varrho_{\sR} \otimes\widetilde{\varrho}_{\sR})\oplus(\varrho_{\sL} \otimes\widetilde{\varrho}_{\sL})$; all these representations are isomorphic in the massless limit at $\k=0$. Here we expect the same to hold, as \textit{a priori} there is no reason to prefer any of these choices to describe massless modes.

Let us perform the redefinition
\begin{equation}
\ket{\chi^{a}}\to\frac{a^{\sL}_p}{b^{\sR}_p}\ket{\chi^{a}},
\qquad
\ket{\widetilde{\chi}^{a}}\to-\frac{b^{\sR}_p}{a^{\sL}_p}\ket{\widetilde{\chi}^{a}}.
\end{equation}
Using~\eqref{eq:massless-rep-shortening2} in the defining relations~\eqref{eq:repr-massless} we find that, as a result, the representation $(\varrho_{\sL} \otimes\widetilde{\varrho}_{\sL})^{\oplus2}$ is indeed isomorphic to $(\varrho_{\sR} \otimes\widetilde{\varrho}_{\sR})^{\oplus2}$.
We can then obtain the mixed cases $(\varrho_{\sL} \otimes\widetilde{\varrho}_{\sL})\oplus (\varrho_{\sR} \otimes\widetilde{\varrho}_{\sR})$ or $(\varrho_{\sR} \otimes\widetilde{\varrho}_{\sR})\oplus(\varrho_{\sL} \otimes\widetilde{\varrho}_{\sL})$ by performing the rescaling only on $\ket{\chi^2},\ket{\widetilde{\chi}^2}$ or $\ket{\chi^1},\ket{\widetilde{\chi}^1}$ respectively. It is also interesting to note that the rescaling coefficient is just a sign:
\begin{equation}
\frac{a^{\sL}_p}{b^{\sR}_p}=-\text{sgn}\Bigl[\sin\bigl(\frac{p}{2}\bigr)\Bigr].
\end{equation}

\subsubsection{Two-particle representations}
\label{sec:coproduct}
So far, we have described the action of the symmetries on the one-particle representations. In order to construct the S~matrix, we will also need to consider two-particle representations. These can be constructed by introducing a \emph{deformed coproduct}~\cite{Plefka:2006ze}, or  equivalently by appropriately picking the phase~$\xi$ in the one-particle representations~\cite{Arutyunov:2006yd}, \cf equation \eqref{eq:abparam}. A way to find such a coproduct is to require that the central charges~$\gen{C},\overline{\gen{C}}$ vanish on physical two-particle states~\cite{Arutyunov:2006yd}, so that they should be
\begin{equation}
\label{eq:C12}
\gen{C}_{(12)}=+\frac{i\h}{2}(e^{+i\gen{P}}-1),\qquad
\overline{\gen{C}}_{(12)}=-\frac{i\h}{2}(e^{-i\gen{P}}-1),
\end{equation}
where $\gen{P}$ is the \emph{total} worldsheet momentum,
\begin{equation}
\gen{P}\ket{p_1,\dotsc,p_n}=(p_1+\dotsb+p_n)\ket{p_1,\dotsc,p_n}\,.
\end{equation}
This then enforces, in the same way as in refs.~\cite{Borsato:2013qpa,Borsato:2014hja}, that the supercharges are%
\footnote{%
It is possible to pick different coproduct that are related to this one by a momentum-dependent change of the two-particle basis, as discussed in ref.~\cite{Borsato:2013qpa}.
}
\begin{equation}
\begin{gathered}
\gen{Q}^{\ \dot{a}}_{\sL}{}_{(12)}(p,q)=\gen{Q}^{\ \dot{a}}_{\sL}(p_1)\otimes \1 + e^{+\frac{i}{2}p}\Sigma\otimes\gen{Q}^{\ \dot{a}}_{\sL}(q)\,,\\
\gen{Q}_{\sR\dot{a}}{}_{(12)}(p,q)=\gen{Q}_{\sR\dot{a}}(p)\otimes \1 + e^{+\frac{i}{2}p}\Sigma\otimes\gen{Q}_{\sR\dot{a}}(q)\,,\\
\overline{\gen{Q}}{}_{\sL\dot{a}}{}_{(12)}(p,q)=\overline{\gen{Q}}{}_{\sL\dot{a}}(p)\otimes \1 + e^{-\frac{i}{2}p}\Sigma\otimes\overline{\gen{Q}}{}_{\sL\dot{a}}(q)\,,\\
\overline{\gen{Q}}{}^{\ \dot{a}}_{\sR}{}_{(12)}(p,q)=\overline{\gen{Q}}{}^{\ \dot{a}}_{\sR}(p)\otimes \1 + e^{-\frac{i}{2}p}\Sigma\otimes\overline{\gen{Q}}{}^{\ \dot{a}}_{\sR}(q)\,,
\end{gathered}
\end{equation}
where~$\Sigma$ is the fermion-sign matrix taking values $+1$, $-1$ on bosons and fermions respectively. Consequently, on the central charges we have
\begin{equation}
\begin{gathered}
\gen{C}_{(12)}(p,q)=\gen{C}(p)\otimes\1+e^{+ip}\1\otimes\gen{C}(q)\,,\\
\overline{\gen{C}}{}_{(12)}(p,q)=\overline{\gen{C}}(p)\otimes\1+e^{-ip}\1\otimes\overline{\gen{C}}(q)\,,
\end{gathered}
\end{equation}
consistently with~\eqref{eq:C12}, and finally
\begin{equation}
\label{eq:trivialcoproduct}
\gen{H}_{(12)}=\gen{H}\otimes\1+\1\otimes\gen{H}\,,
\qquad
\gen{M}_{(12)}=\gen{M}\otimes\1+\1\otimes\gen{M}\,.
\end{equation}
Similarly, the coproduct is trivial for the $\so(4)_2$ generators.

\subsubsection{A momentum-dependent mass?}
\label{sec:mass-comments}
It may appear unnatural that $\gen{M}$ depends on the momentum of the excitations, since in the algebra of superisometries it had the interpretation of an angular momentum. 
The resolution of this apparent contradiction is recalling that $\gen{M}$ is supposed to be identified with an isometry for a \emph{physical state}, \ie, on shell. This means that we should expect $\gen{M}$ to be integer-valued only when applied to states that satisfy the level-matching condition~\eqref{eq:level-matching}.
Let us rewrite~\eqref{eq:Mexact} in term of the integer level of the WZW term in the string action~$k = 2\pi\k \in\mathbb{Z}$. For a one-particle state we then have
\begin{equation}
\label{eq:Mwinding}
\gen{M}=m\1+\frac{k}{2\pi}\gen{P}\,.
\end{equation}
A physical state has worldsheet momentum~$2\pi w$, where $w\in\mathbb{Z}$ is the winding number. This shows that~$\gen{M}$ is integer on shell \emph{even for states with non-trivial winding}.

Note that linearity in the worldsheet momentum~$\gen{P}$ is crucial to extend this property to any physical \emph{multi-particle} state. In fact in the off-shell algebra the coproduct of~$\gen{M}$ \eqref{eq:trivialcoproduct} remains undeformed, so that its action on a multi-particle states is just additive. Any non-linear function in equation~\eqref{eq:Mwinding} would have prevented us from rewriting the eigenvalue of~$\gen{M}$ on a multiparticle state in terms of the \emph{total} worldsheet momentum, which is what is quantised on shell.%
\footnote{%
It is interesting to note that the winding number affects the ``mass'' of excitations, so that \eg when $k=w=1$ a right-moving excitation has the kinematics of a massless one. It would be interesting to understand if this has deeper implications, which may require analysing in more detail the complete bound-state spectrum of the theory.}

The quantisation of the angular momentum $\gen{M}$ explains why we have introduced the two coupling constants $\h$ and $\k$, even though they are both proportional to the string tension $\sqrt{\lambda}/2\pi$ to leading order at strong coupling. According to the above discussion, the quantisation of the momentum-dependent term in the mass follows from the fact that the WZW coupling is integer valued, a relation that should not get any quantum corrections.
The coupling $\h$, on the other hand, appears as an overall factor in front of the central charge $\gen{C}$ and is expected to receive corrections at higher orders in $1/\sqrt{\lambda}$.

It is also interesting to see how the momentum-dependence is compatible with the other symmetries of the theory. In ref.~\cite{Borsato:2014hja} it was argued that~$\gen{M}$ could not receive quantum correction without spoiling either  the~$\su(2)_{\circ}$ symmetry or crossing invariance. Let us see how that argument works in the present setting. Invariance under~$\su(2)_{\circ}$ dictates that $\gen{M}$ takes the same value on both~$\psu(1|1)^4_{\ce}$ massless modules. If we write $\gen{M}$ in a block-matrix form, with each block corresponding to a $\psu(1|1)^4_{\ce}$ module\footnote{Respectively, $\varrho_{\sL}\otimes\varrho_{\sL}$, $\varrho_{\sR}\otimes\varrho_{\sR}$ and the massless doublet $(\varrho_{\sL} \otimes\widetilde{\varrho}_{\sL})^{\oplus 2}$.}, we see that this indeed the case:
\begin{equation}
\gen{M}=\left(
\begin{array}{cccc}
+1+\k p & 0 & 0 & 0\\
0 & -1+\k p & 0 & 0\\
0 & 0 & \k p & 0 \\
0 & 0 & 0 & \k p
\end{array}\right).
\end{equation}
On the other hand, under a crossing transformation it should be possible to map every irreducible module to some other one for  which $\gen{M}$ has an opposite sign. We can see that at $\k=0$  this means sending right to  left movers, and massless modes to themselves. If we perform the crossing transformation at $\k>0$ we must account for the fact that $p$ flips sign, \ie
\begin{equation}
\gen{M}=\left(
\begin{array}{cccc}
+1-\k p & 0 & 0 & 0\\
0 & -1-\k p & 0 & 0\\
0 & 0 & -\k p & 0 \\
0 & 0 & 0 & -\k p
\end{array}\right),\qquad
{\text{at crossed }p}.
\end{equation}
This shows that indeed even at $\k>0$ one can implement crossing by swapping left and right movers and sending the massless modes to themselves. The condition for this to be possible is that the eigenvalue of~$\gen{M}$ in the massless sector is an odd function of~$\gen{P}$. This in particular rules out a constant correction to the mass.

These general considerations on the momentum-dependence of~$\gen{M}$ fit together nicely with our analysis of the $x^-$-dependence in the supercharges, which constrains the non-local coproduct to take the form discussed in section~\ref{sec:coproduct}. They are also consistent with the form of the dispersion relation found by studying semi-classical solutions~\cite{Hoare:2013lja, Ahn:2014tua, Babichenko:2014yaa} and with the analysis of the possible spectrum of bound states performed in ref.~\cite{Hoare:2013lja}.

\section{S matrix}
\label{sec:smat}
Our discussion of the S~matrix of fundamental particles for the mixed-flux backgrounds will be based on the one done in ref.~\cite{Borsato:2014hja} in the pure-R-R case. As we have seen, the particle content is the same in the two theories, and the symmetry representations of our case of interest are a deformation of the ones of~\cite{Borsato:2014hja}.

We define the S matrix as the operator~$\Smat_{(12)}(p,q)$ acting on the two-particle Hilbert space and relating in- and out-states as
\begin{equation}
\Smat_{(12)}(p,q)\;\ket{\mathcal{X}_p^{(\text{in})} \mathcal{Y}_q^{(\text{in})}}= \ket{\mathcal{Y}_q^{(\text{out})} \mathcal{X}_p^{(\text{out})}} ,
\end{equation}
where $\mathcal{X}_p^{(\text{in})}, \mathcal{Y}_q^{(\text{in})}$ are two arbitrary excitations and $\mathcal{X}_p^{(\text{out})}, \mathcal{Y}_q^{(\text{out})}$ are the product of their scattering---possibly a linear combination of several two-particle states.
For this S~matrix  to be physical and for the underlying theory to be integrable, several requirements should be satisfied. The most obvious is the invariance of~$\Smat$ under all symmetries of the theory
\begin{equation}\label{eq:S-mat-invariance}
\Smat_{(12)}(p,q)\;\gen{Q}_{(12)}(p,q)= \gen{Q}_{(12)}(q,p)\;\Smat_{(12)}(p,q).
\end{equation}
Here~$\gen{Q}_{(12)}(p,q)$ is any (super)charge of~$\mathcal{A}$, acting on a two-particle state. Note that we impose commutation with the \emph{off-shell} symmetries as~$\Smat_{(12)}(p,q)$ acts on particles that generally do not satisfy the level-matching condition. Next, we require \emph{braiding} and \emph{physical unitarity}, which read
\begin{equation}
\label{eq:Smat-symmetry}
\Smat_{(12)}(q,p)\;\Smat_{(12)}(p,q)= \1 ,
\qquad
\Bigl(\Smat_{(12)}(p,q)\Bigr)^{\dagger}\;\Smat_{(12)}(p,q)= \1 .
\end{equation}
We also impose the \emph{Yang-Baxter equation}  on the three-particle Hilbert space
\begin{equation}
\Smat_{(12)}(q,r)\;
\Smat_{(23)}(p,r)\;
\Smat_{(12)}(p,q)
=
\Smat_{(23)}(p,q)\;
\Smat_{(12)}(p,r)\;
\Smat_{(23)}(q,r),
\end{equation}
which ensures that factorised scattering can be consistently defined.
We will find that the Yang-Baxter equation \emph{automatically holds} for the $\psu(1|1)^{4}_{\ce}$ invariant S~matrices, signalling that this is a good candidate to be an integrable theory.
Lastly, there is the requirement of  invariance under the crossing transformation. We will come back to this in section~\ref{sec:crossing}.

We will start by briefly recalling the form of some invariant matrices which will then be useful to restrict the form of~$\Smat$ by means of~\eqref{eq:Smat-symmetry}. Since our charges take the same form as the ones in ref.~\cite{Borsato:2014hja} up to suitably redefining the Zhukovski variables~$x^{\pm}$, we expect the final result to be closely related to the one found there. We will  see that this is the case, even if there are some new features here. Imposing~\eqref{eq:Smat-symmetry} will fix the S~matrix up to some \emph{dressing factors}, which we will discuss in section~\ref{sec:crossing}.

\subsection{Invariant ~\texorpdfstring{$\su(1|1)^2_{\ce}$}{su(1|1)**2 c.e.} S matrices}
\label{sec:small-smat}
We start by considering operators that are invariant under $\su(1|1)^2_{\ce}$, which were first studied in ref.~\cite{Borsato:2012ud}. Since in this case the representations are much smaller, the resulting S~matrices are more manageable. Additionally, if we can define \eg a matrix~$\Smat^{\sL\sL}$ which commutes with all the generators of the two-particle~$\varrho_{\sL}$ representation, we are guaranteed that~$\Smat^{\sL\sL}\otimes\Smat^{\sL\sL}$ will commute with all generators of the two-particle~$\varrho_{\sL}\otimes\varrho_{\sL}$ one, which is one of the $\psu(1|1)^4_{\ce}$ representations which will be of interest to us. Clearly the same holds for all representations we need to consider.

\subsubsection{Same target-space-chirality scattering}
Let us start from the case where we have two excitations in the representation~$\varrho_{\sL}$. In~\cite{Borsato:2012ud} it was found that the invariant S~matrix takes the form
\begin{equation}\label{eq:Smat-LL-small}
\begin{aligned}
\mathcal{S}^{\sL\sL} \ket{\phi_p^{\sL} \phi_q^{\sL}} &= A_{pq}^{\sL\sL} \ket{\phi_q^{\sL} \phi_p^{\sL}},
\qquad
&\mathcal{S}^{\sL\sL} \ket{\phi_p^{\sL} \psi_q^{\sL}} &= B_{pq}^{\sL\sL} \ket{\psi_q^{\sL} \phi_p^{\sL}} +  C_{pq}^{\sL\sL} \ket{\phi_q^{\sL} \psi_p^{\sL}}, \\
%%%
\mathcal{S}^{\sL\sL} \ket{\psi_p^{\sL} \psi_q^{\sL}} &= F_{pq}^{\sL\sL} \ket{\psi_q^{\sL} \psi_p^{\sL}},\qquad
&\mathcal{S}^{\sL\sL} \ket{\psi_p^{\sL} \phi_q^{\sL}} &= D_{pq}^{\sL\sL} \ket{\phi_q^{\sL} \psi_p^{\sL}} +  E_{pq}^{\sL\sL} \ket{\psi_q^{\sL} \phi_p^{\sL}}.
\end{aligned}
\end{equation}
This is the case also for us, with the ratio of the S-matrix elements being a function of~$x^{\pm}_{\sL\,p}$ and $x^{\pm}_{\sL\,q}$, and of course the overall normalisation being arbitrary. We collect the expressions for these coefficients in appendix~\ref{app:smat-param}. It is interesting to note that those expressions depend on~$\h,\k$ and~$m$ only through the Zhukovski variables.

From~\eqref{eq:Smat-LL-small} we can immediately find the invariant S~matrix describing the scattering of \eg two particles that both are in the~$\widetilde{\varrho}_{\sL}$ representation. In fact, since the representations~${\varrho}_{\sL}$ and $\widetilde{\varrho}_{\sL}$ are related by a change of basis, we will have that
\begin{equation}
\label{eq:Smat-tilde-def}
\mathcal{S}^{\stL\stL}=\Pi^{g}\; \mathcal{O}^{-1}\;\Pi^{g}\;\mathcal{S}^{\sL\sL}\;\mathcal{O} ,
\end{equation}
where $\mathcal{O}=\mathcal{O}^{-1}=\sigma_1\otimes\sigma_1$ is the change-of-basis matrix and $\Pi^{g}$ is the graded permutation that accounts for the fermion signs. Up to suitably choosing~$\mathcal{O}$, this also yields~$\mathcal{S}^{\stL\sL}$ and~$\mathcal{S}^{\sL\stL}$.%
\footnote{%
The explicit form of the matrices $\mathcal{S}^{\stL\stL},\mathcal{S}^{\stL\sL}$ and~$\mathcal{S}^{\sL\stL}$ is also spelled out in ref.~\cite{Borsato:2014hja}.
}

The case of $\Smat^{\sR\sR}$, \ie both particles being in the representation~$\varrho_{\sR}$, is similar and in fact follows from the previous one by left-right symmetry. All we have to do is relabel everywhere~L$\to$R and introduce new scattering elements~$A^{\sR\sR}_{pq}, B^{\sR\sR}_{pq}$, etc. These will now depend on $x^{\pm}_{\sR\,p}$ and~$x^{\pm}_{\sR\,q}$. In a similar way, $\Smat^{\stR\sR}, \Smat^{\sR\stR}$ and $\Smat^{\stR\stR}$ can be easily found.

\subsubsection{Opposite target-space-chirality scattering}
Let us now consider the case where one particle transforms in~$\varrho_{\sL}$ and one transforms in~$\varrho_{\sR}$. At $\k=0$, such a set-up gives a scattering process of the form~\cite{Borsato:2012ud, Borsato:2013qpa}
\begin{equation}
\Smat \ket{\mathcal{X}^{\sL}_p\mathcal{Y}^{\sR}_q}= T_{pq} \ket{\mathcal{Y}^{\sR}_q\mathcal{X}^{\sL}_p} + R_{pq} \ket{\mathcal{Y}^{\sL}_q\mathcal{X}^{\sR}_p} , \end{equation}
where $T_{pq}$ is the transmission amplitude and $R_{pq}$ is the reflection one. Then, imposing LR-symmetry and unitarity requires either amplitude to vanish, and comparison with perturbative calculations sets $R_{pq}=0$.

On the other hand when $\k\neq 0$ we have that
\begin{equation}
\gen{H}\ket{\mathcal{X}^{\sL}_p\mathcal{Y}^{\sR}_q} \neq
\gen{H}\ket{\mathcal{X}^{\sR}_p\mathcal{Y}^{\sL}_q} ,
\end{equation}
which immediately sets~$R_{pq}=0$ when imposing~\eqref{eq:S-mat-invariance} for the Hamiltonian. This is an additional \textit{a posteriori} validation of the choice of a pure-transmission S~matrix originally made in ref.~\cite{Borsato:2012ud}.

We can therefore write down the matrix~$\Smat^{\sL\sR}$ as
\begin{equation}\label{eq:su(1|1)2-Smat-LRgrad1}
\begin{aligned}
\mathcal{S}^{\sL\sR} \ket{\phi^{\sL}_p \phi^{\sR}_q} &= A^{\sL\sR}_{pq} \ket{\phi^{\sR}_q \phi^{\sL}_p} + B^{\sL\sR}_{pq} \ket{\psi^{\sR}_q \psi^{\sL}_p}, \qquad 
&\mathcal{S}^{\sL\sR} \ket{\phi^{\sL}_p \psi^{\sR}_q} &= C^{\sL\sR}_{pq} \ket{\psi^{\sR}_q \phi^{\sL}_p} , \\
\mathcal{S}^{\sL\sR} \ket{\psi^{\sL}_p \psi^{\sR}_q} &= E^{\sL\sR}_{pq} \ket{\psi^{\sR}_q \psi^{\sL}_p}+F^{\sL\sR}_{pq} \ket{\phi^{\sR}_q \phi^{\sL}_p} ,  \qquad 
& \mathcal{S}^{\sL\sR} \ket{\psi^{\sL}_p \phi^{\sR}_q} &= D^{\sL\sR}_{pq} \ket{\phi^{\sR}_q \psi^{\sL}_p} .
\end{aligned}
\end{equation}
Here the S-matrix elements are functions of~$x^{\pm}_{\sL\,p}$ and~$x^{\pm}_{\sR\,q}$, see appendix~\ref{app:smat-param}.
Just as before, we can use changes of basis such as~\eqref{eq:Smat-tilde-def} to write down $\Smat^{\sL\stR},\Smat^{\stL\sR}$ and~$\Smat^{\stL\stR}$.

Finally, we can once more use LR symmetry to write down~$\Smat^{\sR\sL},\Smat^{\stR\sL},\Smat^{\sR\stL}$ and $\Smat^{\stR\stL}$. Due to the relabelling  L$\leftrightarrow$R, these will all depend on~$x^{\pm}_{\sR\,p}$ and~$x^{\pm}_{\sL\,q}$.

\subsubsection{Tensor-product structure}
As we have argued, the tensor product of any pair of S~matrices invariant under~$\su(1|1)^2_{\ce}$ will yield a $\psu(1|1)^4_{\ce}$-invariant S~matrix in a given representation. Let us consider a pair of particles, transforming in two $\psu(1|1)^4_{\ce}$ representations which we call $\varrho_{X_1}\otimes\varrho_{Y_1}$ and $\varrho_{X_2}\otimes\varrho_{Y_2}$ respectively, where $X_i$ and $Y_i$ could be L, R, $\tilde{\text{L}}$ or $\tilde{\text{R}}$. The $\psu(1|1)^4_{\ce}$-invariant S matrix will be given by the tensor product of two $\su(1|1)^2_{\ce}$  S~matrices.  In formulae,
\begin{equation}
\Smat_{\psu(1|1)^4}=\Smat^{X_1 X_2}_{\su(1|1)^2}\,\check{\otimes}\;\Smat^{Y_1 Y_2}_{\su(1|1)^2}.
\end{equation} 
Note that we have to account for signs arising from swapping fermionic excitations. To this end we define the \emph{graded} tensor product~$\check{\otimes}$, given by
\begin{equation}
\left( \mathcal{A}\,\check{\otimes}\,\mathcal{B} \right)_{MM',NN'}^{KK',LL'} = (-1)^{\epsilon_{M'}\epsilon_{N}+\epsilon_{L}\epsilon_{K'}} \ \mathcal{A}_{MN}^{KL} \  \mathcal{B}_{M'N'}^{K'L'} ,
\end{equation}
where~$\epsilon=0$ for bosons and $\epsilon=1$ for fermions.

\subsection{Matrix part of~\texorpdfstring{$\Smat$}{S}}
Let us decompose the S~matrix in different sectors, depending on the mass of the incoming particles, which clearly will be conserved during the scattering. We denote by $\Smat^{\bullet\bullet}$ the sector of massive excitations, $\Smat^{\circ\circ}$ the one of massless excitations, and by $\Smat^{\bullet\circ},\Smat^{\circ\bullet}$ the S-matrix blocks scattering particles of mixed~mass.

\subsubsection{Massive sector}
We can decompose the massive sector depending on whether the incoming particles have left or right target-space chirality. According to the discussion in the previous section, the scattering in each block should be given by the graded tensor product of two $\su(1|1)^2_{\ce}$-invariant S~matrices, and indeed
\begin{equation}
\Smat^{\bullet\bullet} = 
\left(
\begin{array}{cc}
\sigma^{\bullet\bullet}_{\sL\sL}\; \Smat^{\sL\sL}\,\check{\otimes}\,\Smat^{\sL\sL} & 
\widetilde{\sigma}^{\bullet\bullet}_{\sR\sL}\; \Smat^{\sR\sL}\,\check{\otimes}\,\Smat^{\sR\sL}\\\\
\widetilde{\sigma}^{\bullet\bullet}_{\sL\sR}\;\Smat^{\sL\sR}\,\check{\otimes}\,\Smat^{\sL\sR} & 
\sigma^{\bullet\bullet}_{\sR\sR}\;\Smat^{\sR\sR}\,\check{\otimes}\,\Smat^{\sR\sR}\\
\end{array}
\right).
\end{equation}
Note here that we are writing down four undetermined factors, rather than the two (same-chirality and opposite-chirality) that we would have in the pure-R-R case. Still, these are related pairwise by left-right symmetry, so we take
\begin{equation}
\begin{gathered}
\sigma^{\bullet\bullet}_{\sL\sL}(p,q)=\sigma^{\bullet\bullet}(x^{\pm}_{p\,\sL},x^{\pm}_{q\,\sL}),
\qquad
\sigma^{\bullet\bullet}_{\sR\sR}(p,q)=\sigma^{\bullet\bullet}(x^{\pm}_{p\,\sR},x^{\pm}_{q\,\sR}),\\
\widetilde{\sigma}^{\bullet\bullet}_{\sL\sR}(p,q)=\widetilde{\sigma}^{\bullet\bullet}(x^{\pm}_{p\,\sL},x^{\pm}_{q\,\sR}),
\qquad
\widetilde{\sigma}^{\bullet\bullet}_{\sR\sL}(p,q)=\widetilde{\sigma}^{\bullet\bullet}(x^{\pm}_{p\,\sR},x^{\pm}_{q\,\sL}),
\end{gathered}
\end{equation}
where~$\sigma^{\bullet\bullet}$ and~$\widetilde{\sigma}^{\bullet\bullet}$ are two appropriately defined functions.

\subsubsection{Massless sector}
In the massless sector we have two irreducible representations of $\psu(1|1)^4_{\ce}$ that form a doublet under~$\su(2)_{\circ}$. For this reason the S~matrix here is the tensor product of a $\psu(1|1)^4_{\ce}$-invariant one with an $\su(2)$-invariant pre-factor:
\begin{equation}
\Smat^{\circ\circ}=\sigma^{\circ\circ}\;\Smat_{\su(2)}\otimes \Bigl(\Smat^{\sL\sL} \,\check{\otimes}\, \Smat^{\stL\stL}\Bigr),
\end{equation}
where
\begin{equation}
\Smat_{\su(2)}=\frac{1}{1+\varsigma_{pq}} \bigl(\1+\varsigma_{pq}\Pi\bigr).
\end{equation}
In fact, the Yang-Baxter equation implies that $\Smat_{\su(2)}$ should be precisely the S~matrix of the Heisenberg model, \ie
\begin{equation}
\varsigma(p,q)=i\,(w_p-w_q) ,
\end{equation}
where $w_p$ is an appropriate rapidity.

This all follows closely what was found in refs.~\cite{Borsato:2014exa,Borsato:2014hja}. However, a few differences emerge at~$\k\neq 0$. Firstly, since now~$x^+\neq 1/x^-$ at $m=0$, the kinematics is richer. Consequently, scattering processes that accidentally had the same amplitude at $\k =m=0$ may now differ. For example
\begin{equation}
\Bra{\chi_q^{c}\chi_p^{d}}\Smat \Ket{\chi_p^{a}\chi_q^{b}}=
%e^{i(q-p)}\left(\frac{x^{+}_{\sL\,p}- x^{-}_{\sL\,q}}{x^{-}_{\sL\,p}- x^{+}_{\sL\,q}}\right)^2\,
\Bra{\widetilde{\chi}_q^{c}\widetilde{\chi}_p^{d}}\Smat \Ket{\widetilde{\chi}_p^{a}\widetilde{\chi}_q^{b}}
\qquad\text{only at }\k=0 ,
\end{equation}
while at $\k\neq 0$ the ratio of the two amplitudes is given by~$(A^{\sL\sL}_{pq}/F^{\sL\sL}_{pq})^2$. Additionally, both the dressing factor~$\sigma^{\circ\circ}$ and the rapidity~$w_p$ might have a more complicated form in the mixed-flux case. 

\subsubsection{Mixed-mass sector}
Let us now consider the scattering of a massive particle with a massless one. On symmetry grounds we can write
\begin{equation}
\Smat^{\bullet\circ} =
\Bigl[\sigma^{\bullet\circ}_{\sL} (\Smat^{\sL\sL}\otimes\Smat^{\sL\stL})^{\oplus 2}\Bigr]
\oplus
\Bigl[\sigma^{\bullet\circ}_{\sR}(\Smat^{\sR\sL}\otimes\Smat^{\sR\stL})^{\oplus 2}\Bigr] ,
\end{equation}
where the subscript indices L, R on the dressing factors refer to the target-space chirality of the massive excitation.
The presence of two copies of the S matrix inside each square bracket is due the fact that the massless $\psu(1|1)^4_{\ce}$ modules are doublets under~$\su(2)_\circ$. Once again, we have two dressing factors, which should be related to one another by replacing~$x^\pm_{p\,\sL}\leftrightarrow x^\pm_{p\,\sR}$ everywhere, \ie
\begin{equation}
\label{eq:mixed-ansatz}
\sigma^{\bullet\circ}_{\sL}(p,q)= \sigma^{\bullet\circ}(x_{\sL\,p}^\pm,x_{\sL\,q}^\pm),
\qquad
\sigma^{\bullet\circ}_{\sR}(p,q)= \sigma^{\bullet\circ}(x_{\sR\,p}^\pm,x_{\sR\,q}^\pm)= \sigma^{\bullet\circ}\bigl(x_{\sR\,p}^\pm,\frac{1}{x_{\sL\,q}^\mp}\bigr),
\end{equation}
where in the last equation we used that  $x^{\pm}_{\sR}=1/x^{\mp}_{\sL}$ at $m=0$. The scattering elements and hence the expression of $\sigma^{\bullet\circ}$ should not depend on whether we represent the massless particles as left- or right-movers, which is a constraint on the form of the dressing factor.

In a similar way, we can also can write
\begin{equation}
\Smat^{\circ\bullet} =
\Bigl[\sigma^{\circ\bullet}_{\sL} (\Smat^{\sL\sL}\otimes\Smat^{\stL\sL})^{\oplus 2}\Bigr]
\oplus
\Bigl[\sigma^{\circ\bullet}_{\sR}(\Smat^{\sL\sR}\otimes\Smat^{\stL\sR})^{\oplus 2}\Bigr] ,
\end{equation}
with the same caveats for the dressing factors as above.

\subsection{Dressing factors}
\label{sec:crossing}
The linear symmetries that we used in the previous subsection cannot constrain the scalar factors. On the other hand, braiding and physical unitarity, and crossing symmetry will impose new constraints. Before discussing those, let us fix the normalisation of each block of the S~matrix.

\subsubsection{Normalisations}
The normalisation each S-matrix block can be read off from the elements listed below. In the massive sector we have chosen
\begin{equation}
\begin{aligned}
\bra{Y^{\sL}_q \, Y^{\sL}_p} \mathcal{S} \ket{Y^{\sL}_p \, Y^{\sL}_q} & = \frac{x^+_{\sL\, p}}{x^-_{\sL\, p}} \, \frac{x^-_{\sL\, q}}{x^+_{\sL\, q}} \, \frac{x^-_{\sL\, p} - x^+_{\sL\, q}}{x^+_{\sL\, p} - x^-_{\sL\, q}} \, \frac{1-\frac{1}{x^-_{\sL\, p} x^+_{\sL\, q}}}{1-\frac{1}{x^+_{\sL\, p} x^-_{\sL\, q}}} \, \frac{1}{\left(\sigma^{\bullet\bullet}_{\sL\sL\,pq} \right)^2 }, \\
%%%
\bra{Y^{\sR}_q \, Y^{\sL}_p} \mathcal{S} \ket{Y^{\sL}_p \, Y^{\sR}_q} & = \frac{x^+_{\sL\, p}}{x^-_{\sL\, p}} \, \frac{x^-_{\sR\, q}}{x^+_{\sR\, q}} \, \frac{1-\frac{1}{x^+_{\sL\, p} x^-_{\sR\, q}}}{1-\frac{1}{x^+_{\sL\, p} x^+_{\sR\, q}}}  \, \frac{1-\frac{1}{x^-_{\sL\, p} x^+_{\sR\, q}}}{1-\frac{1}{x^-_{\sL\, p} x^-_{\sR\, q}}} \, \frac{1}{\left(\tilde{\sigma}^{\bullet\bullet}_{\sL\sR\,pq} \right)^2 }, \end{aligned}
\end{equation}
with two more equations following by LR symmetry when we exchange everywhere~L$\leftrightarrow$R.
In the massless sector  we set
\begin{equation}
\begin{aligned}
\label{eq:unitarity-varsigma}
\bra{T^{\dot{a}a}_q \, T^{\dot{a}a}_p} \mathcal{S} \ket{T^{\dot{a}a}_p \, T^{\dot{a}a}_q} & =  \frac{1}{\left(\sigma^{\circ\circ}_{pq} \right)^2 }.
\end{aligned}
\end{equation}
Finally, in the mixed-mass sector we choose
\footnote{%
The normalisation chosen here takes a different form to the one of~\cite{Borsato:2014hja} but reduces to it at~$\k =0$. It is chosen to simplify the form of the constraints imposed by unitarity.
}
\begin{equation}
\begin{aligned}
\bra{T^{\dot{a}a}_q \, Y^{\sL}_p} \mathcal{S} \ket{Y^{\sL}_p \, T^{\dot{a}a}_q} 
&=
 \left( \frac{x^+_{\sL\,p} - x^+_{\sL\,q}}{x^+_{\sL\,p} - x^-_{\sL\,q}}  \, \frac{x^-_{\sL\,p} - x^-_{\sL\,q}}{x^-_{\sL\,p} - x^+_{\sL\,q}} \right)^{1/2} \, \frac{1}{\left(\sigma^{\bullet\circ}_{\sL\sL\,pq} \right)^2 }, \\
\bra{Y^{\sL}_q \, T^{\dot{a}a}_p} \mathcal{S} \ket{T^{\dot{a}a}_p \, Y^{\sL}_q} 
&=
 \left( \frac{x^+_{\sL\,p} - x^+_{\sL\,q}}{x^+_{\sL\,p} - x^-_{\sL\,q}}  \, \frac{x^-_{\sL\,p} - x^-_{\sL\,q}}{x^-_{\sL\,p} - x^+_{\sL\,q}} \right)^{1/2} \, \frac{1}{\left(\sigma^{\circ\bullet}_{\sL\sL\,pq} \right)^2 },
\end{aligned}
\end{equation}
and using  LR-symmetry and the relation~\eqref{eq:magic-relation} this implies
\begin{equation}
\begin{aligned}
\bra{T^{\dot{a}a}_q \, Y^{\sR}_p} \mathcal{S} \ket{Y^{\sR}_p \, T^{\dot{a}a}_q} 
&=
 \left( \frac{1-\frac{1}{x^+_{\sR\,p} x^-_{\sL\,q}}}{1-\frac{1}{x^+_{\sR\,p} x^+_{\sL\,q}}}  \, \frac{1-\frac{1}{x^-_{\sR\,p} x^+_{\sL\,q}}}{1-\frac{1}{x^-_{\sR\,p} x^-_{\sL\,q}}} \right)^{1/2} \, \frac{1}{\left(\sigma^{\bullet\circ}_{\sR\sL\,pq} \right)^2 }, \\
\bra{Y^{\sR}_q \, T^{\dot{a}a}_p} \mathcal{S} \ket{T^{\dot{a}a}_p \, Y^{\sR}_q} 
&=
 \left( \frac{1-\frac{1}{x^+_{\sL\,p} x^-_{\sR\,q}}}{1-\frac{1}{x^+_{\sL\,p} x^+_{\sR\,q}}}  \, \frac{1-\frac{1}{x^-_{\sL\,p} x^+_{\sR\,q}}}{1-\frac{1}{x^-_{\sL\,p} x^-_{\sR\,q}}} \right)^{1/2} \, \frac{1}{\left(\sigma^{\circ\bullet}_{\sL\sR\,pq} \right)^2 }.
\end{aligned}
\end{equation}

\subsubsection{Unitarity}
Owing to our choice of normalisation, the requirements of braiding an physical unitarity take a simple form 
\begin{equation}
\begin{aligned}
\sigma^{\bullet\bullet}_{qp}=\frac{1}{\sigma^{\bullet\bullet}_{pq}}=\bigl(\sigma^{\bullet\bullet}_{pq}\bigr)^* ,
\qquad
\tilde{\sigma}^{\bullet\bullet}_{qp}=\frac{1}{\tilde{\sigma}^{\bullet\bullet}_{pq}}=\bigl(\tilde{\sigma}^{\bullet\bullet}_{pq}\bigr)^* ,
\qquad
\sigma^{\circ\circ}_{qp}=\frac{1}{\sigma^{\circ\circ}_{pq}}=\bigl(\sigma^{\circ\circ}_{pq}\bigr)^* ,\\
\sigma^{\bullet\circ}_{qp}=\frac{1}{\sigma^{\circ\bullet}_{pq}}=\bigl(\sigma^{\bullet\circ}_{pq}\bigr)^* ,
\qquad
\sigma^{\circ\bullet}_{qp}=\frac{1}{\sigma^{\bullet\circ}_{pq}}=\bigl(\sigma^{\circ\bullet}_{pq}\bigr)^* ,
\qquad
\varsigma_{qp}=-\varsigma_{pq}=\bigl(\varsigma_{pq}\bigr)^* ,
\end{aligned}
\end{equation}
where~$*$ denotes complex conjugation.

\subsubsection{Crossing symmetry}
The crossing transformation acts by flipping the sign of momentum and making the energy negative:
\begin{equation}
\label{eq:general-crossing}
p\to \bar{p}=-p ,
\qquad
E(p)\to E(\bar{p})=-E(-p).
\end{equation}
Note that in the latter relation we also flip  the sign of~$p$ in~$E(p)$. That sign is irrelevant in a parity-invariant theory, but it does affect the sign of the linear terms in the momenta in our case.

In general, it is convenient to describe the crossing transformation by introducing a rapidity variable that uniformises the dispersion relation. In relativistic theories, this can be done using a hyperbolic parametrisation, while for $\AdS_5\times\Sphere^5$ strings one can use elliptic functions to describe a rapidity torus~\cite{Janik:2006dc}. In both scenarios, crossing then amounts to an imaginary shift of the rapidity.
It is less clear how to uniformise the dispersion relation here. On the other hand, we can realise~\eqref{eq:general-crossing} in terms of the Zhukovski variables by setting
\begin{equation}
\label{eq:crossingmap}
x^{\pm}_{\sL}(\bar{p})=\frac{1}{x^{\pm}_{\sR}(p)} ,
\qquad
x^{\pm}_{\sR}(\bar{p})=\frac{1}{x^{\pm}_{\sL}(p)}.
\end{equation}
The supercharges are not meromorphic in~$x^\pm$, so that we have to resolve a square-root ambiguity when performing crossing. We do this by setting
\begin{equation}
\eta^{\sL}(\bar{p})= \frac{i}{x^{+}_{\sR}(p)}\eta^{\sR}(p) ,
\qquad
\eta^{\sR}(\bar{p})= \frac{i}{x^{+}_{\sL}(p)}\eta^{\sL}(p).
\end{equation}

In order to write down the crossing equation, we need to define a charge conjugation matrix, which in our case will be the same as the one of~\cite{Borsato:2014hja}. Let us pick a basis
\begin{equation}
( Y^{\sL}, \eta^{\sL 1}, \eta^{\sL 2}, Z^{\sL} ) \oplus ( Y^{\sR}, \eta^{\sR 1}, \eta^{\sR 2}, Z^{\sR} ) \oplus ( T^{11}, T^{21}, T^{12}, T^{22} ) \oplus ( \widetilde{\chi}^1, \chi^1, \widetilde{\chi}^2, \chi^2 ).
\end{equation}
The the charge conjugation matrix is
\begin{equation}
\label{eq:chargeconj}
\newcommand{\0}{\color{black!40}0}
  \renewcommand{\arraystretch}{1.1}
  \setlength{\arraycolsep}{3pt}
  \mathscr{C}_p=\!\left(\!
    \mbox{\footnotesize$
      \begin{array}{cccc|cccc}
        \0 & \0 & \0 & \0 & 1 & \0 & \0 & \0 \\
        \0 & \0 & \0 & \0 & \0 & \0 & -i & \0 \\
        \0 & \0 & \0 & \0 & \0 & i & \0 & \0 \\
        \0 & \0 & \0 & \0 & \0 & \0 & \0 & 1 \\
        \hline
        1 & \0 & \0 & \0 & \0 & \0 & \0 & \0 \\
        \0 & \0 & i & \0 & \0 & \0 & \0 & \0 \\
        \0 & -i & \0 & \0 & \0 & \0 & \0 & \0 \\
        \0 & \0 & \0 & 1 & \0 & \0 & \0 & \0 \\
      \end{array}$}\!
  \right) 
	\oplus
	\!\left(\!
    \mbox{\footnotesize$
      \begin{array}{cccc|cccc}
        \0 & \0 & \0 & 1 & \0 & \0 & \0 & \0 \\
        \0 & \0 & -1 & \0 & \0 & \0 & \0 & \0 \\
        \0 & -1 & \0 & \0 & \0 & \0 & \0 & \0 \\
        1 & \0 & \0 & \0 & \0 & \0 & \0 & \0 \\
        \hline
        \0 & \0 & \0 & \0 & \0 & \0 & \0 & -i c_p \\
        \0 & \0 & \0 & \0 & \0 & \0 & i c_p & \0 \\
        \0 & \0 & \0 & \0 & \0 & i c_p & \0 & \0 \\
        \0 & \0 & \0 & \0 & -i c_p & \0 & \0 & \0 \\
      \end{array}$}\!
  \right),
\end{equation}
where we note that the dependence on~$p$ in the massless sector comes through
\begin{equation}
c_p=\frac{a_{\sL}(p)}{b_{\sR}(p)}=-\text{sgn}\Bigl[\sin \frac{p}{2}\Bigr] ,
\end{equation}
where the last equality uses that $m=0$. This is reassuring as it indicates that if we treat separately left- and right-movers on the worldsheet, $\mathscr{C}$ is indeed a constant matrix. Note that, even if we are taking the massless modes to be in the left representation, we see the coefficients of the right representation appearing. This is not surprising as crossing exchanges the two kinematics.

We can now write down the crossing equation%
\footnote{%
There is one more crossing equation where crossing is performed in the second variable, which is equivalent to the one given here by unitarity.
}
\begin{equation}
\begin{aligned}
\mathscr{C}_p \otimes \1 \cdot  \mathbf{S}^{\text{t}_1}(\bar{p},q) \cdot  \mathscr{C}^{-1}_p \otimes \1 \cdot \mathbf{S}(p,q) &= \1 \otimes \1,
\end{aligned}
\end{equation}
where we have introduced the short-hand~$\gen{S}=\Pi\,\Smat$ defined in terms of the permutation~$\Pi$, and~$\text{t}_j$ denotes transposition in the $j$th space.
From this we can read off the constraints on the dressing factors.

In the massive sector we find
\begin{equation}
\begin{aligned}
\sigma^{\bullet\bullet}_{\sL\sL}(p,q)^2 \ \tilde{\sigma}^{\bullet\bullet}_{\sR\sL}(\bar{p},q)^2 &= \left( \frac{x^-_{\sL\,q}}{x^+_{\sL\,q}} \right)^2 \frac{(x^-_{\sL\,p}-x^+_{\sL\,q})^2}{(x^-_{\sL\,p}-x^-_{\sL\,q})(x^+_{\sL\,p}-x^+_{\sL\,q})} \frac{1-\frac{1}{x^-_{\sL\,p}x^+_{\sL\,q}}}{1-\frac{1}{x^+_{\sL\,p}x^-_{\sL\,q}}} ,
 \\
\sigma^{\bullet\bullet}_{\sL\sR}(p,q)^2 \ \tilde{\sigma}^{\bullet\bullet}_{\sR\sR}(\bar{p},q)^2 &= \left( \frac{x^-_{\sR\,q}}{x^+_{\sR\,q}} \right)^2 \frac{\left(1-\frac{1}{x^+_{\sL\,p}x^+_{\sR\,q}}\right)\left(1-\frac{1}{x^-_{\sL\,p}x^-_{\sR\,q}}\right)}{\left(1-\frac{1}{x^+_{\sL\,p}x^-_{\sR\,q}}\right)^2} \frac{x^-_{\sL\,p}-x^+_{\sR\,q}}{x^+_{\sL\,p}-x^-_{\sR\,q}}.
\end{aligned}
\end{equation}
Note that the Zhukovski variables carry the appropriate flavours with respect to the phases. This is guaranteed by the fact that both the transformation~\eqref{eq:crossingmap} and the charge conjugation matrix~\eqref{eq:chargeconj} swap left with right in the massive sector. Two additional equations can be written down by LR symmetry, exhausting the constraints of crossing in this sector.

In the massless sector we have
\begin{equation}
\begin{aligned}
\sigma^{\circ\circ}(p,q)^2 \ \sigma^{\circ\circ}(\bar{p},q)^2 &= \frac{\varsigma_{pq}-1}{\varsigma_{pq}} \, \frac{x^-_{\sL\,p}-x^+_{\sL\,q}}{x^+_{\sL\,p}-x^+_{\sL\,q}} \frac{x^+_{\sL\,p}-x^-_{\sL\,q}}{x^-_{\sL\,p}-x^-_{\sL\,q}},
\qquad
\varsigma_{\bar{p}q} &= \varsigma_{pq}-1.
\end{aligned}
\end{equation}
It is interesting to note that
\begin{equation}
\frac{x^-_{\sL\,p}-x^+_{\sL\,q}}{x^+_{\sL\,p}-x^+_{\sL\,q}} \frac{x^+_{\sL\,p}-x^-_{\sL\,q}}{x^-_{\sL\,p}-x^-_{\sL\,q}}=\frac{x^-_{\sR\,p}-x^+_{\sR\,q}}{x^+_{\sR\,p}-x^+_{\sR\,q}} \frac{x^+_{\sR\,p}-x^-_{\sR\,q}}{x^-_{\sR\,p}-x^-_{\sR\,q}}
\qquad
\text{at}\quad  m=0,
\end{equation}
so that the crossing equation for the massless phases does not depend on whether we decided to represent the massless modes as left or right particles.

Finally, in the mixed-mass sector we have
\begin{equation}
\begin{aligned}
\sigma^{\bullet \circ}_{\sL}(p,q)^2 \  \sigma^{\bullet \circ}_{\sR}(\bar{p},q)^2 &=
\frac{x^-_{\sL\,p}-x^+_{\sL\,q}}{x^+_{\sL\,p}-x^+_{\sL\,q}} \frac{x^+_{\sL\,p}-x^-_{\sL\,q}}{x^-_{\sL\,p}-x^-_{\sL\,q}} =
\sigma^{\circ\bullet}_{\sL}(q,p)^2 \  \sigma^{\circ\bullet}_{\sL}(\bar{q},p)^2 ,
\end{aligned}
\end{equation}
where in order to write down this equation only in term of \emph{left} massless Zhukovski variables  we have used~\eqref{eq:magic-relation}. The same formula, together with LR symmetry, yields
\begin{equation}
\begin{aligned}
\sigma^{\bullet \circ}_{\sR}(p,q)^2 \  \sigma^{\bullet \circ}_{\sL}(\bar{p},q)^2 &=
\frac{1-\frac{1}{x^+_{\sR\,p}x^+_{\sL\,q}}}{1-\frac{1}{x^+_{\sR\,p}x^-_{\sL\,q}}}
\frac{1-\frac{1}{x^-_{\sR\,p}x^-_{\sL\,q}}}{1-\frac{1}{x^-_{\sR\,p}x^+_{\sL\,q}}} =
\sigma^{\circ\bullet}_{\sR}(q,p)^2 \  \sigma^{\circ\bullet}_{\sR}(\bar{q},p)^2.
\end{aligned}
\end{equation}
Note also that the form of this crossing equation is compatible with~\eqref{eq:mixed-ansatz}.

\section{Discussion and outlook}
\label{sec:conclusion}

In this paper we have determined the complete all-loop worldsheet S matrix of Type IIB string theory on $\AdS_3\times\Sphere^3\times\Torus^4 $ with mixed R-R and NS-NS three-form flux, up to the dressing factors. We further wrote down the crossing relations that these dressing factors have to satisfy. In constructing this S matrix we relied on the off-shell symmetry algebra $\mathcal{A}$ of the gauge-fixed theory and its representations. This method, initially advocated in the context of $\AdS_5\times \Sphere^5$~\cite{Arutyunov:2006ak}, has recently been shown to be particularly well-suited in the study of \emph{massless}, as well as massive, excitations and allowed for the construction of the non-perturbative S matrix for the pure R-R flux $\AdS_3\times\Sphere^3\times\Torus^4$ theory~\cite{Borsato:2014exa,Borsato:2014hja}. Our present work demonstrates the versatility of this approach and provides strong evidence that, by using these methods, one will be able to tackle other classes of backgrounds such  as  $\AdS_3\times\Sphere^3\times\Sphere^3 \times\Sphere^1$~\cite{Abbott:2012dd, Babichenko:2009dk, Beccaria:2012kb, Beccaria:2012pm,Abbott:2013ixa,Sundin:2012gc, Borsato:2012ud,Borsato:2012ss}, or the less-supersymmetric backgrounds discussed in the context of integrability in refs.~\cite{Sorokin:2011rr,Wulff:2014kja,Murugan:2012mf,Abbott:2013kka,Hoare:2014kma}. It would be particularly interesting to investigate these backgrounds as well as to solve the crossing relations proposed in this paper and we hope to return to this subject in the near future.

The analysis carried out here provides strong evidence for the validity of the $\fl$-dependent dispersion relation proposed in ref.~\cite{Hoare:2013lja}. We find in particular that this form of  dispersion relation is satisfied by both massive and massless modes. 

The S-matrix construction presented here elucidates a feature pure-R-R S matrix. There, for massive particles, it was found that symmetries and unitarity leave two choices for the all-loop S~matrix: one where the target space chirality is always transmitted, and one where it is always reflected~\cite{Borsato:2012ud,Borsato:2013qpa}. Only the former option satisfies the Yang-Baxter equation and is compatible with tree-level perturbative calculations~\cite{Rughoonauth:2012qd}. Interestingly, in the presence of NS-NS fluxes we find that the pure-reflection S~matrix is immediately ruled out by symmetries, further motivating the choice of a reflectionless S~matrix in refs.~\cite{Borsato:2012ud,Borsato:2013qpa}.

The S matrix presented in this paper provides a two-parameter family ($\lambda$ and $\fl$) of quantum integrable models. It is likely that this parameter space can be further enhanced by the study of so-called $\eta$-deformations~\cite{Beisert:2008tw,Delduc:2013qra,Arutyunov:2013ega,Kawaguchi:2014qwa,Arutynov:2014ota}. This large parameter space, and the presence of novel massless modes deserves a detailed investigation in the context of integrability and may provide us with new inputs into the relationship between integrability and holography~\cite{Pittelli:2014ria}.

From our previous work~\cite{Borsato:2014exa,Borsato:2014hja} and the present paper we are lead to conclude that the solution of the spectral problem in $\AdS_3/\CFT_2$ is likely to be within reach using integrable methods. In particular it would be important to understand the mirror Thermodynamical Bethe Ansatz and Quantum Spectral Curve for these backgrounds~\cite{Ambjorn:2005wa,Arutyunov:2007tc, Arutyunov:2009zu, Gromov:2009tv,Bombardelli:2009ns,Arutyunov:2009ur, Cavaglia:2010nm, Gromov:2013pga,Gromov:2014caa}.
Given this, it would be interesting to investigate other aspects of this class of dualities, such as scattering amplitudes, Wilson loops or entanglement entropy using integrable methods. It would also be important to connect the results presented here to the higher-spin holography of $\AdS_3$ backgrounds investigated in recent papers such as~\cite{Gaberdiel:2013vva,Ahn:2013oya,Gaberdiel:2014yla, Creutzig:2014ula,Gaberdiel:2014cha}.

Finally, our work suggests that connections between the integrable approach and other studies of the $\AdS_3/\CFT_2$ correspondence, that deserve to be explored more fully. Two links naturally suggest themselves. Firstly, the mixed-flux theory investigated in the present paper was analysed some time ago in the hybrid formalism~\cite{Berkovits:1999im} and it would be interesting to establish connections between those results and the work presented in this paper. For example, might one be able to see integrable structure in the framework of ref.~\cite{Berkovits:1999im}? Secondly, type IIB string theory on $\AdS_3\times\Sphere^3\times\Torus^4$ with only NS-NS flux was investigated in detail from the point of view of a WZW theory in refs.~\cite{Maldacena:2000hw,Maldacena:2000kv,Maldacena:2001km}, where worldsheet CFT methods were used to powerful effect. It would be intriguing to see if there is a way to take a $\fl\rightarrow 1$ limit of our integrable structure in a controlled way.
Uncovering the role of integrability in the pure NS-NS theory, as recently investigated in ref.~\cite{Hernandez:2014eta}, might provide new connections between integrable and worldsheet CFT approaches.

\section*{Acknowledgements}
We would like to thank Gleb Arutyunov, Sergey Frolov, Ben Hoare, Alessandro Torrielli, Arkady Tseytlin, Dan Waldram and Kostya Zarembo for helpful discussions, and Ben Hoare and Arkady Tseytlin for their comments on the manuscripts.
We are very grateful to Riccardo Borsato for his collaboration on earlier related work, critical reading and invaluable comments on the manuscript and for many useful discussions.
T.L. is supported by an STFC studentship.
O.O.S.'s  work was supported by the ERC Advanced grant No.~290456,
 ``Gauge theory -- string theory duality''.
A.S.'s work is funded by the People Programme (Marie Curie Actions) of the European Union, Grant Agreement No.~317089 (GATIS). A.S. also acknowledges the hospitality at APCTP where part of this work was
done.
B.S.\@ acknowledges funding support from an STFC Consolidated Grant  ``Theoretical Physics at City University''
ST/J00037X/1.

%%%%%%%%%%%%%%%%%%%%%%%%%%%%%%%%%%%%%%%%%%%%%%%%
%%%%%%%%%%%%%%%%%%%%%%%%%%%%%%%%%%%%%%%%%%%%%%%%
\appendix

\section{Conventions}
\label{app:conventions}
In this appendix we give our conventions for indices. We denote worldsheet coordinates by $\alpha,\beta, \dotsc = \tau,\sigma$; spacetime coordinates by indices $m,n,\dotsc = 0,\dotsc 9$; and $\so(1,9)$ tangent coordinates by $A,B,\dotsc = 0,\dotsc 9$. Indices $I,J,\dotsc = 1,2$ denote the two sets of spacetime spinors.

We also use indices referring to representations of the algebra $\so(4)_1 \times \so(4)_2$, as described in ref.~\cite{Borsato:2014hja}, where $\so(4)_1$ corresponds to rotations along the $\AdS_3 \times \Sphere^3$ directions transverse to the light-cone directions $t$ and $\phi$ and $\so(4)_2$ corresponds to rotations along $\Torus^4$. We use indices $\underline{a},\underline{b},\dotsc = 1,2$ and $\underline{\dot{a}},\underline{\dot{b}},\dotsc = 1,2$ for the two Weyl spinors of $\so(4)_1$; and indices $a,b,\dotsc = 1,2$ and $\dot{a},\dot{b},\dotsc = 1,2$ for the two Weyl spinors of $\so(4)_2$. We use indices $\underline{i},\underline{j},\dotsc = 1,\dotsc,4$ for the vector of $\so(4)_1$. We use the same indices for the transverse coordinates of $\AdS_3$ and $\Sphere^3$ themselves ($z_{\underline{i}}$ and $y_{\underline{i}}$ respectively) with the understanding that $z_3=z_4=y_1=y_2=0$.

We raise and lower spinor indices with epsilon symbols normalised as
\begin{equation}
\epsilon^{12}=-\epsilon_{12}=+1\ .
\end{equation}
We also occasionally write $\epsilon^{\underline{i}\underline{j}}$, by this we will always mean an expression of the following form
\begin{equation}
\epsilon^{\underline{i}\underline{j}}z_{\underline{i}} \partial_\alpha z_{\underline{j}} = z_1 \partial_\alpha z_2 - z_2 \partial_\alpha z_1\ ,\quad \epsilon^{\underline{i}\underline{j}}y_{\underline{i}} \partial_\alpha y_{\underline{j}} = y_3 \partial_\alpha y_4 - y_4 \partial_\alpha y_3\ .
\end{equation}
Similarly in our conventions
\begin{equation}
\dot{z} \cdot \pri{z} = \dot{z}_{\underline{i}} \pri{z}^{\underline{i}} = \dot{z}_1 \pri{z}_1 + \dot{z}_2 \pri{z}_2\ , \quad \dot{y} \cdot \pri{y} = \dot{y}_{\underline{i}} \pri{y}^{\underline{i}} =\dot{y}_3 \pri{y}_3 + \dot{y}_4 \pri{y}_4 \ .
\end{equation}

At various times we make use of $\so(4)$ gamma matrices. These are understood to be embedded inside 10d gamma matrices in the way described in ref.~\cite{Borsato:2014hja}. Here we collect only those explicit choices for these gamma matrices from ref.~\cite{Borsato:2014hja} that are needed in this paper. We have matrices $(\gamma^{\underline{i}})^{\underline{a}}{}_{\underline{\dot{b}}}$, $(\tilde{\gamma}^{\underline{i}})^{\underline{\dot{a}}}{}_{\underline{b}}$, $(\tau^i)^a{}_{\dot{b}}$ and $(\tilde{\tau}^i)^{\dot{a}}{}_b$ chosen to be
\begin{align}\label{eq:so4-gamma}
&\gamma^1=+\sigma_3\ ,\quad \gamma^2=-i\mathbbm{1}\ ,\quad \gamma^3=+\sigma_2\ ,\quad \gamma^4=+\sigma_1\ ,\quad \tilde{\gamma}^{\underline{i}}=+(\gamma^{\underline{i}})^\dagger\nonumber\\
&\tau^6=+\sigma_1\ ,\quad \tau^7=+\sigma_2\ ,\quad \tau^8=+\sigma_3\ ,\quad \tau^9=+i\mathbbm{1}\ ,\quad \tilde{\tau}^i=-(\tau^i)^\dagger\ .
\end{align}
We also define
\begin{equation}
(\gamma^{\underline{i}\underline{j}})^{\underline{a}}_{\ \underline{b}}=\frac{1}{2}(\gamma^{\underline{i}}\tilde{\gamma}^{\underline{j}}-\gamma^{\underline{j}}\tilde{\gamma}^{\underline{i}})^{\underline{a}}_{\ \underline{b}}\ ,\quad
(\tilde{\gamma}^{\underline{i}\underline{j}})^{\underline{\dot{a}}}_{\ \underline{\dot{b}}}=\frac{1}{2}(\tilde{\gamma}^{\underline{i}}\gamma^{\underline{j}}-\tilde{\gamma}^{\underline{j}}\gamma^{\underline{i}})^{\underline{\dot{a}}}_{\ \underline{\dot{b}}}\ .
\end{equation}

\section{Killing spinors}
\label{app:killing-spinors}
In this appendix we present a derivation of the solution of the mixed-flux Killing spinor equations~\eqref{eq:AdS3-S3-T4-Killing-spinor-eq}. It turns out to be helpful to re-write these equations in terms of  two independent spinors $\epsilon_1$ and $\epsilon_2$ by introducing
\begin{equation}
  \tilde{\epsilon}_1 = \sqrt{\frac{1+\flt}{2}} \, \epsilon_1 - \sqrt{\frac{1-\flt}{2}} \, \epsilon_2 , \qquad
  \tilde{\epsilon}_2 = \sqrt{\frac{1+\flt}{2}} \, \epsilon_2 + \sqrt{\frac{1-\flt}{2}} \, \epsilon_1 .
\end{equation}
After some simple algebra the Killing spinor equations can then be written as
\begin{equation}\label{eq:killing-spinor-eqs-1}
  \Bigl( \bigl( \partial_m + \frac{1}{4} \slashed{\omega}_m \bigr)\delta_{IJ} + \bigl( \frac{\fl}{8}\slashed{H}_m + \frac{\flt}{48} \slashed{F} \slashed{E}_m\bigr)\sigma^3_{IJ} \Bigr) \epsilon_J 
 =   \Bigl( \frac{\fl}{48} \slashed{F} \slashed{E}_m - \frac{\flt}{8}\slashed{H}_m \Bigr)\sigma^1_{IJ} \epsilon_J 
\end{equation}
Now we insert the form of the fluxes (\cf equations~\eqref{eq:Hslash} and~\eqref{eq:Fslash}) and use the fact that the spinors $\epsilon_I$ are anti-chiral with respect to $\Gamma^{012345}$~\cite{Borsato:2014hja}. From this we find that the right-hand side of equation~\eqref{eq:killing-spinor-eqs-1} vanishes. Furthermore we can combine the terms coming from each flux on the left-hand side into a single term proportional to the R-R flux. Altogether we can rewrite equation~\eqref{eq:killing-spinor-eqs-1} as
\begin{equation}
\Bigl( \bigl( \partial_m + \frac{1}{4} \slashed{\omega}_m \bigr)\delta_{IJ} + \frac{\sigma^3_{IJ}}{48\flt} \slashed{F} \slashed{E}_m \Bigr) \epsilon_J = 0\ .
\end{equation}
Since the R-R flux here is related to the flux of the pure R-R background by an overall rescaling by $\flt$, this is precisely the equation satisfied by the Killing spinors of the pure R-R background~\cite{Borsato:2014hja}.

Recall, that the $\fl=0$ Killing spinors are given by
\begin{equation}
  \varepsilon^1
  =
  \hat{M} \varepsilon_0^1,
  ,\qquad\qquad
  \varepsilon^2
  =
  \check{M} \varepsilon_0^2,
\end{equation}
where $\varepsilon_0^I$ are constant 9+1 dimensional Majorana-Weyl spinors%\footnote{Our spinor and gamma matrix conventions are given in appendix~\ref{app:conventions}.}
, which further satisfy
\begin{equation}
  \frac{1}{2} ( 1 + \Gamma^{012345} ) \varepsilon^I = 
    \frac{1}{2} ( 1 + \Gamma^{012345} ) \varepsilon_0^I  = 0.
\end{equation}
The matrices  $\hat{M}$ and $\check{M}$ are given by
\begin{equation}
  \hat{M} = M_0 M_t , \qquad
  \check{M} = M_0^{-1} M_t^{-1} ,
\end{equation}
where
\begin{equation}\label{eq:M0-def}
  \begin{aligned}
    M_0 &= \frac{1}{\sqrt{\bigl( 1 - \frac{z_1^2 + z_2^2}{4} \bigr) \bigl( 1 + \frac{y_3^2 + y_4^2}{4} \bigr)}}
    \Bigl( \mathds{1} - \frac{1}{2} z_{\underline{i}} \Gamma^{\underline{i}} \Gamma^{012} \Bigr)
    \Bigl( \mathds{1} - \frac{1}{2} y_{\underline{i}} \Gamma^{\underline{i}} \Gamma^{345} \Bigr) ,
    \\
    M_0^{-1} &= \frac{1}{\sqrt{\bigl( 1 - \frac{z_1^2 + z_2^2}{4} \bigr) \bigl( 1 + \frac{y_3^2 + y_4^2}{4} \bigr)}}
    \Bigl( \mathds{1} + \frac{1}{2} z_{\underline{i}} \Gamma^{\underline{i}} \Gamma^{012} \Bigr)
    \Bigl( \mathds{1} + \frac{1}{2} y_{\underline{i}} \Gamma^{\underline{i}} \Gamma^{345} \Bigr) ,
  \end{aligned}
\end{equation}
and
\begin{equation}\label{eq:Mt-def}
  M_t = e^{-\frac{1}{2} ( t \, \Gamma^{12} + \phi \, \Gamma^{34} )} , \qquad
  M_t^{-1} = e^{+\frac{1}{2} ( t \, \Gamma^{12} + \phi \, \Gamma^{34} )} .
\end{equation}
In summary, the Killing spinors of the mixed-flux background, $\tilde{\epsilon}_I$, are given by a $\fl$-dependent linear combination of the $\fl=0$ Killing spinors,  $\epsilon_I$, as given in equation~\eqref{eq:killing-spinor-lin-combs}.

\section{Explicit expressions for the Lagrangian}
\label{app:lagrangian}

In this appendix we write down the explicit form of the Lagrangian to quadratic order in fermions which are redefined in each of the two ways described in section~\ref{sec:GS-action}. We will make use of the following definitions for contraction of the vielbeins $E_m^A$ over $\AdS_3 \times \Sphere^3$ and $\Torus^4$ separately
\begin{equation}
  \slashed{\bar{E}}_m = \sum_{A=0}^5 E_m^A \Gamma_A , \qquad
  \slashed{\dot{E}}_m = \sum_{A=6}^9 E_m^A \Gamma_A \ ,
\end{equation}
and we will then define two different sets of rotated vielbeins for the two different sets of field redefinitions.

First, using the redefinition of the fermions coming from the combination of equations~\eqref{eq:fermion-IJ-rot} and~\eqref{eq:theta-vartheta-def}, we define the rotated vielbeins $\hat{K}_m^A$ and $\check{K}_m^A$ via
\begin{equation}
  \slashed{\hat{K}} = \hat{M}^{-1} \slashed{\bar{E}} \hat{M} , \qquad
  \slashed{\check{K}} =  \check{M}^{-1} \slashed{\bar{E}} \check{M} .
\end{equation}
With these rotations, $\lagr_{\text{kin}}$ is given by
\begin{equation}
\label{eq:pre-gauge-action-kin}
  \begin{aligned}
    \lagr_{\text{kin}}
    =
    -i\gamma^{\alpha\beta} \Bigl[
    &\bar{\vartheta}^-_1 \slashed{\hat{K}}_\alpha \partial_\beta \vartheta^-_1 
    + 2\bar{\vartheta}^+_1 \slashed{\dot{E}}_\alpha \partial_\beta \vartheta^-_1
    + \bar{\vartheta}^+_1 \slashed{\hat{K}}_\alpha \partial_\beta \vartheta^+_1 \\
    + & \bar{\vartheta}^-_2 \slashed{\check{K}}_\alpha \partial_\beta \vartheta^-_2
    + 2\bar{\vartheta}^+_2 \slashed{\dot{E}}_\alpha \partial_\beta \vartheta^-_2
    + \bar{\vartheta}^+_2 \slashed{\check{K}}_\alpha \partial_\beta \vartheta^+_2 \\
    - &\frac{\flt^2}{2} \sigma^3_{IJ}\bar{\vartheta}^+_I \Gamma^{012} \vartheta^+_J E_\alpha^A E_\beta^B \eta_{AB} \\
    + &\frac{\fl \flt}{2} \bar{\vartheta}^+_2 \check{M}^{-1} \hat{M} \Gamma^{012} \vartheta^+_1 E_\alpha^A E_\beta^B \eta_{AB} \\
    + &\frac{\fl \flt}{2} \bar{\vartheta}^+_1 \hat{M}^{-1} \check{M} \Gamma^{012} \vartheta^+_2 E_\alpha^A E_\beta^B \eta_{AB}
    \bigr].
  \end{aligned}
\end{equation}
and $\lagr_{\text{WZ}}$ is given by
\begin{equation}
\label{eq:pre-gauge-action-WZ}
  \begin{aligned}
    \lagr_{\text{WZ}} = +i\epsilon^{\alpha\beta} \Bigl[
    \flt & ( 
    \bar{\vartheta}^-_1 \hat{M}^{-1} \check{M} \slashed{\check{K}}_\alpha \partial_\beta \vartheta^-_2
    + \bar{\vartheta}^+_1 \hat{M}^{-1} \check{M} \slashed{\check{K}}_\alpha \partial_\beta \vartheta^+_2
    ) \\
    + \flt & ( 
    \bar{\vartheta}^-_2 \check{M}^{-1} \hat{M} \slashed{\hat{K}}_\alpha \partial_\beta \vartheta^-_1
    + \bar{\vartheta}^+_2 \check{M}^{-1} \hat{M} \slashed{\hat{K}}_\alpha \partial_\beta \vartheta^+_1
    ) \\
    + \flt & ( 
    \bar{\vartheta}^-_1 \hat{M}^{-1} \check{M} \slashed{\dot{E}}_\alpha \partial_\beta \vartheta^+_2
    + \bar{\vartheta}^+_1 \hat{M}^{-1} \check{M} \slashed{\dot{E}}_\alpha \partial_\beta \vartheta^-_2
    ) \\
    + \flt & ( 
    \bar{\vartheta}^-_2 \check{M}^{-1} \hat{M} \slashed{\dot{E}}_\alpha \partial_\beta \vartheta^+_1
    + \bar{\vartheta}^+_2 \check{M}^{-1} \hat{M} \slashed{\dot{E}}_\alpha \partial_\beta \vartheta^-_1
    ) \\
    + \frac{\flt}{2} &
    \bar{\vartheta}^+_1 \hat{M}^{-1} \check{M} ( \slashed{\check{K}}_\alpha \slashed{\check{K}}_\beta + \slashed{\dot{E}}_\alpha \slashed{\dot{E}}_\beta ) \Gamma^{012} \vartheta^+_2 
    \\
    + \frac{\flt}{2} &
    \bar{\vartheta}^+_2 \check{M}^{-1} \hat{M} ( \slashed{\hat{K}}_\alpha \slashed{\hat{K}}_\beta + \slashed{\dot{E}}_\alpha \slashed{\dot{E}}_\beta ) \Gamma^{012} \vartheta^+_1 
    \\
    + \fl & ( \bar{\vartheta}^-_1 \slashed{\hat{K}}_\alpha \partial_\beta \vartheta^-_1 + 2\bar{\vartheta}^+_1 \slashed{\dot{E}}_\alpha \partial_\beta \vartheta^-_1 + \bar{\vartheta}^+_1 \slashed{\hat{K}}_\alpha \partial_\beta \vartheta^+_1 ) \\
    - \fl & ( \bar{\vartheta}^-_2 \slashed{\check{K}}_\alpha \partial_\beta \vartheta^-_2 + 2\bar{\vartheta}^+_2 \slashed{\dot{E}}_\alpha \partial_\beta \vartheta^-_2 + \bar{\vartheta}^+_2 \slashed{\check{K}}_\alpha \partial_\beta \vartheta^+_2 )
    \Bigr].
  \end{aligned}
\end{equation}

For a redefinition of the fermions coming from the combination of equations~\eqref{eq:fermion-IJ-rot} and~\eqref{eq:theta-vartheta-def}, we define rotated vielbeins $\hat{E}_m^A$ and $\check{E}_m^A$ via
\begin{equation}
  \slashed{\hat{E}} = M_0^{-1} \slashed{\bar{E}} M_0 , \qquad
  \slashed{\check{E}} =  M_0 \slashed{\bar{E}} M_0^{-1} .
\end{equation}
We fix the light-cone kappa gauge~\eqref{eq:bmn-lc-kappa-gauge} and define light-cone coordinates
\begin{equation}
 E^{\pm} = \frac{1}{2} ( E^5 \pm E^0 ) , \qquad
 x^{\pm} = \frac{1}{2} ( \phi \pm t ) .
\end{equation}
$\lagr_{\text{kin}}$ is then given by
\begin{equation}\label{eq:post-gauge-action-kin}
  \begin{aligned}
    \lagr_{\text{kin}}=-2i\gamma^{\alpha\beta}\bigg[&\bar{\eta}_1\hat{E}_\alpha^+\Gamma^-\left(\partial_\beta\eta_1+\Gamma^{12}\eta_1\partial_\beta x^+\right)+\bar{\eta}_2\check{E}_\alpha^+\Gamma^-\left(\partial_\beta\eta_2-\Gamma^{12}\eta_2\partial_\beta x^+\right) \\
    +&\bar{\chi}_1\hat{E}_\alpha^+\Gamma^-\left(\partial_\beta\chi_1-\Gamma^{12}\chi_1\partial_\beta x^-\right)+\bar{\chi}_2\check{E}_\alpha^+\Gamma^-\left(\partial_\beta\chi_2+\Gamma^{12}\chi_2\partial_\beta x^-\right) \\
    -&\frac{\flt^2}{4}\sigma^3_{IJ}\bar{\chi}_I\Gamma^{012}\chi_J
    E_\alpha^AE_\beta^B\eta_{AB} \\
    +&\frac{\fl\flt}{4}\left(
      \bar{\chi}_1M_0^{-2}\Gamma^{012}\chi_2+\bar{\chi}_2M_0^2\Gamma^{012}\chi_1
    \right)
    E_\alpha^AE_\beta^B\eta_{AB}
    % \nonumber\\
    % +&\frac{\fl\sqrt{1-\fl^2}}{4}
    % \bar{\chi}_2M_0^2\Gamma^{012}\chi_1
    % E_\alpha^AE_\beta^B\eta_{AB}
    \bigg]
  \end{aligned}
\end{equation}
and $\lagr_{\text{WZ}}$ by
\begin{equation}\label{eq:post-gauge-action-WZ}
  \begin{aligned}\!\!\!
    \lagr_{\text{WZ}} = 2i\epsilon^{\alpha\beta}\bigg[
    &\frac{\flt}{4}\bar{\chi}_1\left(\slashed{\hat{E}}_\alpha\slashed{\hat{E}}_\beta+\slashed{\dot{E}}_\alpha\slashed{\dot{E}}_\beta\right)M_0^{-2}\Gamma^{012}\chi_2
    - \frac{\flt}{4}\bar{\chi}_2\left(\slashed{\check{E}}_\alpha\slashed{\check{E}}_\beta+\slashed{\dot{E}}_\alpha\slashed{\dot{E}}_\beta\right)M_0^2\Gamma^{012}\mathrlap{\chi_1} \\
    +&\frac{\flt}{2} \bar{\chi}_1\slashed{\check{E}}_\alpha M_0^{-2}\left(\partial_\beta\chi_2+\Gamma^{12}\chi_2\partial_\beta x^-\right)
    + \frac{\flt}{2} \bar{\chi}_2\slashed{\hat{E}}_\alpha M_0^2\left(\partial_\beta\chi_1-\Gamma^{12}\chi_1\partial_\beta x^-\right) \\
    +&\flt\bar{\chi}_1\slashed{\dot{E}}_\alpha M_0^{-2}\left(\partial_\beta\eta_2-\Gamma^{12}\eta_2\partial_\beta x^+\right)
    + \flt\bar{\chi}_2\slashed{\dot{E}}_\alpha M_0^2\left(\partial_\beta\eta_1+\Gamma^{12}\eta_1\partial_\beta x^+\right) \\
    +&\frac{\flt}{2} \bar{\eta}_1\slashed{\hat{E}}_\alpha M_0^{-2}\left(\partial_\beta\eta_2-\Gamma^{12}\eta_2\partial_\beta x^+\right)
    + \frac{\flt}{2} \bar{\eta}_2\slashed{\check{E}}_\alpha M_0^2\left(\partial_\beta\eta_1+\Gamma^{12}\eta_1\partial_\beta x^+\right) \\
    +&\fl \bar{\eta}_1\hat{E}_\alpha^+\Gamma^-\left(\partial_\beta\eta_1+\Gamma^{12}\eta_1\partial_\beta x^+\right)
    -\fl \bar{\eta}_2\check{E}_\alpha^+\Gamma^-\left(\partial_\beta\eta_2-\Gamma^{12}\eta_2\partial_\beta x^+\right) \\
    +&\fl \bar{\chi}_1\hat{E}_\alpha^+\Gamma^-\left(\partial_\beta\chi_1-\Gamma^{12}\chi_1\partial_\beta x^-\right)
    - \fl \bar{\chi}_2\check{E}_\alpha^+\Gamma^-\left(\partial_\beta \chi_2+\Gamma^{12}\chi_2\partial_\beta x^-\right)
    \bigg].
  \end{aligned}
\end{equation}

\section{Supercurrents and bosonic Hamiltonian}
\label{app:quartic-currents}

In this appendix we give the bosonic Hamiltonian to quartic order, as well as the supercurrents to cubic order in bosons and first order in fermions. Some of these expressions are fairly lengthy. To make them a bit more manageable we have left out the indices when there is a simple contraction between two fields. The quartic order bosonic Hamiltonian is given by
\begin{equation}\label{eq:quartic-hamiltonian}
  \begin{aligned}
    \mathcal{H} = & 
    +\frac{1}{2} \bigl(
    p_z^2 + p_y^2 + p_x^2
    + \pri{z}^2 + \pri{y}^2 + \pri{x}^2
    + z^2 + y^2
    - 2 \fl \epsilon^{\underline{ij}} ( z_{\underline{i}} \pri{z}_{\underline{j}} + y_{\underline{i}} \pri{y}_{\underline{j}} )
    \bigr)
    \\ &
    +\frac{1}{4} \bigl( p_z^2 + p_y^2 + p_x^2 + \pri{z}^2 + \pri{y}^2 + \pri{x}^2 \bigr) 
    \bigl( (z^2 - y^2) + \fl \epsilon^{\underline{ij}} ( \pri{z}_{\underline{i}} z_{\underline{j}} - \pri{y}_{\underline{i}} y_{\underline{j}} )
    \\ &
    + \frac{1}{4} z^2 \bigl( \pri{z}^2 - p_z^2 + \fl \epsilon^{\underline{ij}} ( \pri{z}_{\underline{i}} z_{\underline{j}} + \pri{y}_{\underline{i}} y_{\underline{j}} \bigr)
    - \frac{1}{4} y^2 \bigl( \pri{y}^2 - p_y^2 + \fl \epsilon^{\underline{ij}} ( \pri{z}_{\underline{i}} z_{\underline{j}} + \pri{y}_{\underline{i}} y_{\underline{j}} \bigr)
    \\ &
    - \frac{\fl}{2} \epsilon^{\underline{i}\underline{j}} \bigl( p_{z^{\underline{i}}} z_{\underline{j}} + p_{y^{\underline{i}}} y_{\underline{j}} \bigr)
    \bigl( p_z \cdot \pri{z} + p_y \cdot \pri{y} + p_x \cdot \pri{x} \bigr).
  \end{aligned}
\end{equation}
For the currents we give the expressions separately for the massless, massive and mixed massless-massive parts, so that the full currents are given as the sum
\begin{equation}
j_I^\alpha=j_{I,\mathrm{massless}}^\alpha+j_{I,\mathrm{massive}}^\alpha+j_{I,\mathrm{mixed}}^\alpha .
\end{equation}
The massless currents are given by
\begin{equation}
  \begin{aligned}
    j^{\tau}_{1,\text{massless}}
    &= i \gamma^{34} e^{+x^- \gamma^{34}} \Bigl(
    + \dot{x}^i \tilde{\tau}_i \chi_1 - \flt \pri{x}^i \tilde{\tau}_i \chi_2 - \fl \pri{x}^i \tilde{\tau}_i \chi_1
    \Bigr) ,
    \\
    j^{\tau}_{2,\text{massless}}
    &= i \gamma^{34} e^{-x^- \gamma^{34}} \Bigl(
    + \dot{x}^i \tilde{\tau}_i \chi_2 - \flt \pri{x}^i \tilde{\tau}_i \chi_1 + \fl \pri{x}^i \tilde{\tau}_i \chi_2
    \Bigr) ,
    \\
    j^{\sigma}_{1,\text{massless}}
    &= i \gamma^{34} e^{+x^- \gamma^{34}} \Bigl(
    - \pri{x}^i \tilde{\tau}_i \chi_1 + \flt \dot{x}^i \tilde{\tau}_i \chi_2 + \fl \dot{x}^i \tilde{\tau}_i \chi_1
    \Bigr) ,
    \\
    j^{\sigma}_{2,\text{massless}}
    &= i \gamma^{34} e^{-x^- \gamma^{34}} \Bigl(
    - \pri{x}^i \tilde{\tau}_i \chi_2 + \flt \dot{x}^i \tilde{\tau}_i \chi_1 - \fl \dot{x}^i \tilde{\tau}_i \chi_2
    \Bigr) .
  \end{aligned}
\end{equation}
The massive currents are given by
\begin{align} \nonumber
  j^{\tau}_{1,\text{massive}}
  = i e^{+x^- \gamma^{34}} \Bigl(
  + & ( \dot{z}^{\underline{i}} - \dot{y}^{\underline{i}} ) \gamma_{\underline{i}} \eta_1 
  +  ( z^{\underline{i}} - y^{\underline{i}} ) \gamma_{\underline{i}} \tilde{\gamma}^{34} \eta_1 
  - ( \pri{z}^{\underline{i}} - \pri{y}^{\underline{i}} ) \gamma_{\underline{i}} ( \flt \eta_2 + \fl \eta_1 )
  \\ \nonumber
  - & \tfrac{1}{2} \bigl( 
  (z^2 - y^2)(\dot{z}^{\underline{i}} - \dot{y}^{\underline{i}}) 
  - \tfrac{3}{2} (z^2 \dot{z}^{\underline{i}} + y^2 \dot{y}^{\underline{i}}) 
  + ( z \cdot \dot{z} z^{\underline{i}} + y \cdot \dot{y} y^{\underline{i}} )
  \bigr) \gamma_{\underline{i}} \eta_1
  \\ \nonumber
  - & \tfrac{1}{4} \bigl(
  ( \dot{z}^2 + \pri{z}^2 + \dot{y}^2 + \pri{y}^2 ) ( z^{\underline{i}} + y^{\underline{i}} )
  - (y^2 z^{\underline{i}} + z^2 y^{\underline{i}})
  \bigr) \gamma_{\underline{i}} \tilde{\gamma}^{34} \eta_1
  \\ \nonumber
  - & \tfrac{\flt}{2} \bigl( 
  (z^2 - y^2)(\pri{z}^{\underline{i}} - \pri{y}^{\underline{i}}) 
  - \tfrac{1}{2} (z^2 \pri{z}^{\underline{i}} + y^2 \pri{y}^{\underline{i}}) 
  \\ \nonumber &\qquad 
  + 2 ( z \cdot \pri{z} - y \cdot \pri{y} ) ( z^{\underline{i}} - y^{\underline{i}} ) 
  - ( z \cdot \pri{z} z^{\underline{i}} + y \cdot \pri{y} y^{\underline{i}} )
  \bigr) \gamma_{\underline{i}} \eta_2
  \\ \nonumber
  - & \tfrac{\flt}{2} ( \dot{z} \cdot \pri{z} + \dot{y} \cdot \pri{y} ) ( z^{\underline{i}} + y^{\underline{i}} ) \gamma_{\underline{i}} \tilde{\gamma}^{34} \eta_2
  \\ \nonumber
  - & \tfrac{\fl}{4} \bigl( 
  ( z^2 \pri{z}^{\underline{k}} + y^2 \pri{y}^{\underline{k}} )
  + 2 \epsilon_{\underline{ij}} ( z^{\underline{i}} \dot{z}^{\underline{j}} - y^{\underline{i}} \dot{y}^{\underline{j}} ) ( \pri{z}^{\underline{k}} - \pri{y}^{\underline{k}} )
  \\ \nonumber &\qquad
  + 2\epsilon_{\underline{ij}} ( y^{\underline{i}} \pri{y}^{\underline{j}} \dot{z}^{\underline{k}} + z^{\underline{i}} \pri{z}^{\underline{j}} \dot{y}^{\underline{k}} ) 
  \bigr) \gamma_{\underline{k}} \eta_1
  \\ \nonumber
  - & \tfrac{\fl}{2} \bigl(
  ( z \cdot \dot{z} \pri{z}^{\underline{k}} + y \cdot \dot{y} \pri{y}^{\underline{k}} )
  - ( \dot{y} \cdot \pri{y} z^{\underline{k}} + \dot{z} \cdot \pri{z} y^{\underline{k}} ) 
  \\ &\qquad
  - \epsilon_{\underline{ij}} ( y^{\underline{i}} \pri{y}^{\underline{j}} z^{\underline{k}} + z^{\underline{i}} \pri{z}^{\underline{j}} y^{\underline{k}} ) 
  \bigr) \gamma_{\underline{k}} \tilde{\gamma}^{34} \eta_1
  \Bigr) ,
\end{align}
\begin{align} \nonumber
  j^{\tau}_{2,\text{massive}}
  = i e^{-x^- \gamma^{34}} \Bigl(
  + & ( \dot{z}^{\underline{i}} - \dot{y}^{\underline{i}} ) \gamma_{\underline{i}} \eta_2 
  -  ( z^{\underline{i}} - y^{\underline{i}} ) \gamma_{\underline{i}} \tilde{\gamma}^{34} \eta_2 
  - ( \pri{z}^{\underline{i}} - \pri{y}^{\underline{i}} ) \gamma_{\underline{i}} ( \flt \eta_1 - \fl \eta_2 )
  \\ \nonumber
  - \tfrac{1}{2} & \bigl( 
  (z^2 - y^2)(\dot{z}^{\underline{i}} - \dot{y}^{\underline{i}}) 
  - \tfrac{3}{2} (z^2 \dot{z}^{\underline{i}} + y^2 \dot{y}^{\underline{i}}) 
  + ( z \cdot \dot{z} z^{\underline{i}} + y \cdot \dot{y} y^{\underline{i}} )
  \bigr) \gamma_{\underline{i}} \eta_2
  \\ \nonumber
  - \tfrac{1}{4} & \bigl(
  ( \dot{z}^2 + \pri{z}^2 + \dot{y}^2 + \pri{y}^2 ) ( z^{\underline{i}} + y^{\underline{i}} )
  + (y^2 z^{\underline{i}} + z^2 y^{\underline{i}})
  \bigr) \gamma_{\underline{i}} \tilde{\gamma}^{34} \eta_2
  \\ \nonumber
  - \tfrac{\flt}{2} & \bigl( 
  (z^2 - y^2)(\pri{z}^{\underline{i}} - \pri{y}^{\underline{i}}) 
  - \tfrac{1}{2} (z^2 \pri{z}^{\underline{i}} + y^2 \pri{y}^{\underline{i}}) 
  \\ \nonumber &\qquad 
  + 2 ( z \cdot \pri{z} - y \cdot \pri{y} ) ( z^{\underline{i}} - y^{\underline{i}} ) 
  - ( z \cdot \pri{z} z^{\underline{i}} + y \cdot \pri{y} y^{\underline{i}} )
  \bigr) \gamma_{\underline{i}} \eta_1
  \\ \nonumber
  - \tfrac{\flt}{2} & ( \dot{z} \cdot \pri{z} + \dot{y} \cdot \pri{y} ) ( z^{\underline{i}} + y^{\underline{i}} ) \gamma_{\underline{i}} \tilde{\gamma}^{34} \eta_1
  \\ \nonumber
  + \tfrac{\fl}{4} & \bigl(
  ( z^2 \pri{z}^{\underline{k}} + y^2 \pri{y}^{\underline{k}} ) 
  - 2 \epsilon_{\underline{ij}} ( z^{\underline{i}} \dot{z}^{\underline{j}} - y^{\underline{i}} \dot{y}^{\underline{j}} ) ( \pri{z}^{\underline{k}} - \pri{y}^{\underline{k}} )
  \\ \nonumber & \qquad
  - 2 \epsilon_{\underline{ij}} ( y^{\underline{i}} \pri{y}^{\underline{j}} \dot{z}^{\underline{k}} + z^{\underline{i}} \pri{z}^{\underline{j}} \dot{y}^{\underline{k}} ) 
  \bigr) \gamma_{\underline{k}} \eta_2
  \\ \nonumber
  - \tfrac{\fl}{2} & \bigl(
  ( z \cdot \dot{z} \pri{z}^{\underline{k}} + y \cdot \dot{y} \pri{y}^{\underline{k}} ) 
  - ( \dot{y} \cdot \pri{y} z^{\underline{k}} + \dot{z} \cdot \pri{z} y^{\underline{k}} ) 
  \\ & \qquad
  - \epsilon_{\underline{ij}} ( y^{\underline{i}} \pri{y}^{\underline{j}} z^{\underline{k}} + z^{\underline{i}} \pri{z}^{\underline{j}} y^{\underline{k}} ) 
  \bigr) \gamma_{\underline{k}} \tilde{\gamma}^{34} \eta_2
  \Bigr) ,
\end{align}
\begin{align} \nonumber
  j^{\sigma}_{1,\text{massive}}
  = i e^{+x^- \gamma^{34}} \Bigl(
  - & ( \pri{z}^{\underline{i}} - \pri{y}^{\underline{i}} ) \gamma_{\underline{i}} \eta_1 
  -  ( z^{\underline{i}} - y^{\underline{i}} ) \gamma_{\underline{i}} \tilde{\gamma}^{34} \eta_1 
  + ( \dot{z}^{\underline{i}} - \dot{y}^{\underline{i}} ) \gamma_{\underline{i}} ( \flt \eta_2 + \fl \eta_1)
  \\ \nonumber
  - \tfrac{1}{2} & \bigl( 
  (z^2 - y^2)(\pri{z}^{\underline{i}} - \pri{y}^{\underline{i}}) 
  + \tfrac{3}{2} (z^2 \pri{z}^{\underline{i}} + y^2 \pri{y}^{\underline{i}}) 
  - ( z \cdot \pri{z} z^{\underline{i}} + y \cdot \pri{y} y^{\underline{i}} )
  \bigr) \gamma_{\underline{i}} \eta_1
  \\ \nonumber
  - \tfrac{1}{2} & ( \dot{z} \cdot \pri{z} + \dot{y} \cdot \pri{y} ) ( z^{\underline{i}} + y^{\underline{i}} ) \gamma_{\underline{i}} \tilde{\gamma}^{34} \eta_1
  \\ \nonumber
  + \tfrac{\flt}{4} & \bigl( 
  (z^2 - y^2)(\dot{z}^{\underline{i}} - \dot{y}^{\underline{i}}) 
  - (z^2 \dot{z}^{\underline{i}} + y^2 \dot{y}^{\underline{i}}) 
  \\ \nonumber &\qquad 
  + 2 ( z \cdot \dot{z} - y \cdot \dot{y} ) ( z^{\underline{i}} - y^{\underline{i}} ) 
  - 2 ( y \cdot \dot{y} z^{\underline{i}} + z \cdot \dot{z} y^{\underline{i}} )
  \bigr) \gamma_{\underline{i}} \eta_2
  \\ \nonumber
  - \tfrac{\flt}{4} & \bigl( 
  ( \dot{z}^2 + \pri{z}^2 + z^2 + \dot{y}^2 + \pri{y}^2 + y^2 ) ( z^{\underline{i}} + y^{\underline{i}} )
  - 3 ( z^2 z^{\underline{i}} + y^2 y^{\underline{i}} )
  \bigr) \gamma_{\underline{i}} \tilde{\gamma}^{34} \eta_2
  \\ \nonumber
  + \tfrac{\fl}{4} & \bigl(
  ( z^2 \dot{z}^{\underline{k}} + y^2 \dot{y}^{\underline{k}} )
  - 2 \epsilon_{\underline{ij}} ( z^{\underline{i}} \dot{z}^{\underline{j}} - y^{\underline{i}} \dot{y}^{\underline{j}} ) ( \dot{z}^{\underline{k}} - \dot{y}^{\underline{k}} ) 
  \\ \nonumber & \qquad
  + 2 \epsilon_{\underline{ij}} ( z^{\underline{i}} \pri{z}^{\underline{j}} - y^{\underline{i}} \pri{y}^{\underline{j}} ) ( \pri{z}^{\underline{k}} - \pri{y}^{\underline{k}} ) 
  \bigl) \gamma_{\underline{k}} \eta_1
  \\ \nonumber
  + \tfrac{\fl}{4} & \bigl(
  ( \dot{z}^2 + \pri{z}^2 - z^2 + \dot{y}^2 + \pri{y}^2 - y^2 ) ( z^{\underline{k}} + y^{\underline{k}} )
  + 3 ( z^2 z^{\underline{k}} + y^2 y^{\underline{k}} ) 
  \\ & \qquad
  + 2 \epsilon_{\underline{ij}} ( y^{\underline{i}} \dot{y}^{\underline{j}} z^{\underline{k}} + z^{\underline{i}} \dot{z}^{\underline{j}} y^{\underline{k}} ) 
  \bigr) \gamma_{\underline{k}} \tilde{\gamma}^{34} \eta_1
  \Bigr) ,
\end{align}
\begin{align} \nonumber
  j^{\sigma}_{2,\text{massive}}
  = i e^{-x^- \gamma^{34}} \Bigl(
  - & ( \pri{z}^{\underline{i}} - \pri{y}^{\underline{i}} ) \gamma_{\underline{i}} \eta_2 
  +  ( z^{\underline{i}} - y^{\underline{i}} ) \gamma_{\underline{i}} \tilde{\gamma}^{34} \eta_2 
  + ( \dot{z}^{\underline{i}} - \dot{y}^{\underline{i}} ) \gamma_{\underline{i}} ( \flt \eta_1 - \fl \eta_2)
  \\ \nonumber
  - \tfrac{1}{2} & \bigl( 
  (z^2 - y^2)(\pri{z}^{\underline{i}} - \pri{y}^{\underline{i}}) 
  + \tfrac{3}{2} (z^2 \pri{z}^{\underline{i}} + y^2 \pri{y}^{\underline{i}}) 
  - ( z \cdot \pri{z} z^{\underline{i}} + y \cdot \pri{y} y^{\underline{i}} )
  \bigr) \gamma_{\underline{i}} \eta_2
  \\ \nonumber
  + \tfrac{1}{2} & ( \dot{z} \cdot \pri{z} + \dot{y} \cdot \pri{y} ) ( z^{\underline{i}} + y^{\underline{i}} ) \gamma_{\underline{i}} \tilde{\gamma}^{34} \eta_2
  \\ \nonumber
  + \tfrac{\flt}{4} & \bigl( 
  (z^2 - y^2)(\dot{z}^{\underline{i}} - \dot{y}^{\underline{i}}) 
  - (z^2 \dot{z}^{\underline{i}} + y^2 \dot{y}^{\underline{i}}) 
  \\ \nonumber &\qquad 
  + 2 ( z \cdot \dot{z} - y \cdot \dot{y} ) ( z^{\underline{i}} - y^{\underline{i}} ) 
  - 2 ( y \cdot \dot{y} z^{\underline{i}} + z \cdot \dot{z} y^{\underline{i}} )
  \bigr) \gamma_{\underline{i}} \eta_1
  \\ \nonumber
  - \tfrac{\flt}{4} & \bigl( 
  ( \dot{z}^2 + \pri{z}^2 + z^2 + \dot{y}^2 + \pri{y}^2 + y^2 ) ( z^{\underline{i}} + y^{\underline{i}} )
  + 3 ( z^2 z^{\underline{i}} + y^2 y^{\underline{i}} )
  \bigr) \gamma_{\underline{i}} \tilde{\gamma}^{34} \eta_1
  \\ \nonumber
  - \tfrac{\fl}{4} & \bigl(
  ( z^2 \dot{z}^{\underline{k}} + y^2 \dot{y}^{\underline{k}} )
  +2 \epsilon_{\underline{ij}} ( z^{\underline{i}} \dot{z}^{\underline{j}} - y^{\underline{i}} \dot{y}^{\underline{j}} ) ( \dot{z}^{\underline{k}} - \dot{y}^{\underline{k}} ) 
  \\ \nonumber & \qquad
  -2 \epsilon_{\underline{ij}} ( z^{\underline{i}} \pri{z}^{\underline{j}} - y^{\underline{i}} \pri{y}^{\underline{j}} ) ( \pri{z}^{\underline{k}} - \pri{y}^{\underline{k}} ) 
  \bigr) \gamma_{\underline{k}} \eta_2
  \\ \nonumber
  + \tfrac{\fl}{4} & \bigl(
  ( \dot{z}^2 + \pri{z}^2 - z^2 + \dot{y}^2 + \pri{y}^2 - y^2 ) ( z^{\underline{k}} + y^{\underline{k}} )
  + 3 ( z^2 z^{\underline{k}} + y^2 y^{\underline{k}} ) 
  \\ & \qquad
  - 2 \epsilon_{\underline{ij}} ( y^{\underline{i}} \dot{y}^{\underline{j}} z^{\underline{k}} + z^{\underline{i}} \dot{z}^{\underline{j}} y^{\underline{k}} ) 
  \bigr) \gamma_{\underline{k}} \tilde{\gamma}^{34} \eta_2
  \Bigr) .
\end{align}
The mixed currents are given by
\begin{equation}
  \begin{aligned}
    j^{\tau}_{1,\text{mixed}}
    = i e^{+x^- \gamma^{34}} \Bigl(
    - \tfrac{1}{2} & (z^2 - y^2) ( \dot{x}^i \gamma^{34} \tilde{\tau}_i \chi_1 + \flt \, \pri{x}^i \gamma^{34} \tilde{\tau}_i \chi_2 ) 
    + \flt \, z^{\underline{i}} y^{\underline{j}} \pri{x}^k \gamma^{34} \gamma_{\underline{ij}} \tilde{\tau}_k \chi_2
    \\
    + \tfrac{1}{4} & ( \dot{x}^2 + \pri{x}^2 ) ( z^{\underline{i}} + y^{\underline{i}} ) \gamma_i \tilde{\gamma}^{34} \eta_1
    + \tfrac{\flt}{2} ( \dot{x} \cdot \pri{x} ) ( z^{\underline{i}} + y^{\underline{i}} ) \gamma_{\underline{i}} \tilde{\gamma}^{34} \eta_2
    \\
    + \tfrac{\fl}{2} & \bigl( \epsilon_{{\underline{ij}}} ( z^{\underline{i}} \pri{z}^{\underline{j}} - y^{\underline{i}} \pri{y}^{\underline{j}} ) \dot{x}^k - \epsilon_{\underline{ij}} ( z^{\underline{i}} \dot{z}^{\underline{j}} - y^{\underline{i}} \dot{y}^{\underline{j}} ) \pri{x}^k \bigr) \gamma^{34} \tilde{\tau}_k \chi_1
    \\ 
    - \tfrac{\fl}{2} & ( \dot{x} \cdot \pri{x} ) ( z^{\underline{i}} + y^{\underline{i}} ) \gamma_{\underline{i}} \tilde{\gamma}^{34} \eta_1
    \Bigr) ,
  \end{aligned}
\end{equation}
\begin{equation}
  \begin{aligned}
    j^{\tau}_{2,\text{mixed}}
    = i e^{-x^- \gamma^{34}} \Bigl(
    - \tfrac{1}{2} & (z^2 - y^2) ( \dot{x}^i \gamma^{34} \tilde{\tau}_i \chi_2 + \flt \, \pri{x}^i \gamma^{34} \tilde{\tau}_i \chi_1 ) 
    + \flt \, z^{\underline{i}} y^{\underline{j}} \pri{x}^k \gamma^{34} \gamma_{\underline{ij}} \tilde{\tau}_k \chi_1
    \\
    + \tfrac{1}{4} & ( \dot{x}^2 + \pri{x}^2 ) ( z^{\underline{i}} + y^{\underline{i}} ) \gamma_{\underline{i}} \tilde{\gamma}^{34} \eta_2
    - \tfrac{\flt}{2} ( \dot{x} \cdot \pri{x} ) ( z^{\underline{i}} + y^{\underline{i}} ) \gamma_{\underline{i}} \tilde{\gamma}^{34} \eta_1
    \\
    + \tfrac{\fl}{2} & \bigl( \epsilon_{\underline{ij}} ( z^{\underline{i}} \pri{z}^{\underline{j}} - y^{\underline{i}} \pri{y}^{\underline{j}} ) \dot{x}^k - \epsilon_{\underline{i}} ( z^{\underline{i}} \dot{z}^{\underline{j}} - y^{\underline{i}} \dot{y}^{\underline{j}} ) \pri{x}^k \bigr) \gamma^{34} \tilde{\tau}_k \chi_2
    \\
    - \tfrac{\fl}{2} & ( \dot{x} \cdot \pri{x} ) ( z^{\underline{i}} + y^{\underline{i}} ) \gamma_{\underline{i}} \tilde{\gamma}^{34} \eta_2
    \Bigr) ,
  \end{aligned}
\end{equation}
\begin{equation}
  \begin{aligned}
    j^{\sigma}_{1,\text{mixed}}
    = i e^{+x^- \gamma^{34}} \Bigl(
    - \tfrac{1}{2} & (z^2 - y^2) ( \pri{x}^i \gamma^{34} \tilde{\tau}_i \chi_1 - \flt \, \dot{x}^i \gamma^{34} \tilde{\tau}_i \chi_2 ) 
    - \flt \, z^{\underline{i}} y^{\underline{j}} \dot{x}^k \gamma^{34} \gamma_{\underline{ij}} \tilde{\tau}_k \chi_2
    \\
    - \tfrac{\flt}{4} & ( \dot{x}^2 + \pri{x}^2 ) ( z^{\underline{i}} + y^{\underline{i}} ) \gamma_{\underline{i}} \tilde{\gamma}^{34} \eta_2
    - \tfrac{1}{2} ( \dot{x} \cdot \pri{x} ) ( z^{\underline{i}} + y^{\underline{i}} ) \gamma_{\underline{i}} \tilde{\gamma}^{34} \eta_1
    \\
    - \tfrac{\fl}{2} & \bigl( \epsilon_{\underline{ij}} ( z^{\underline{i}} \dot{z}^{\underline{j}} - y^{\underline{i}} \dot{y}^{\underline{j}} ) \dot{x}^k - \epsilon_{\underline{ij}} ( z^{\underline{i}} \pri{z}^{\underline{j}} - y^{\underline{i}} \pri{y}^{\underline{j}} ) \pri{x}^k \bigr) \gamma^{34} \tilde{\tau}_k \chi_1
    \\
    + \tfrac{\fl}{4} & ( \dot{x}^2 + \pri{x}^2 ) ( z^{\underline{i}} + y^{\underline{i}} ) \gamma_{\underline{i}} \tilde{\gamma}^{34} \eta_1
    \Bigr) ,
  \end{aligned}
\end{equation}
\begin{equation}
  \begin{aligned}
    j^{\sigma}_{2,\text{mixed}}
    = i e^{-x^- \gamma^{34}} \Bigl(
    - \tfrac{1}{2} & (z^2 - y^2) ( \pri{x}^i  \gamma^{34} \tilde{\tau}_i \chi_2 - \flt \, \dot{x}^i \gamma^{34} \tilde{\tau}_i \chi_1 ) 
    - \flt \, z^{\underline{i}} y^{\underline{j}} \dot{x}^k \gamma^{34} \gamma_{\underline{ij}} \tilde{\tau}_k \chi_1
    \\
    + \tfrac{\flt}{4} & ( \dot{x}^2 + \pri{x}^2 ) ( z^{\underline{i}} + y^{\underline{i}} ) \gamma_{\underline{i}} \tilde{\gamma}^{34} \eta_1
    + \tfrac{1}{2} ( \dot{x} \cdot \pri{x} ) ( z^{\underline{i}} + y^{\underline{i}} ) \gamma_{\underline{i}} \tilde{\gamma}^{34} \eta_2
    \\
    - \tfrac{\fl}{2} & \bigl( \epsilon_{\underline{ij}} ( z^{\underline{i}} \dot{z}^{\underline{j}} - y^{\underline{i}} \dot{y}^{\underline{j}} ) \dot{x}^k - \epsilon_{\underline{ij}} ( z^{\underline{i}} \pri{z}^{\underline{j}} - y^{\underline{i}} \pri{y}^{\underline{j}} ) \pri{x}^k \bigr) \gamma^{34} \tilde{\tau}_k \chi_2
    \\
    + \tfrac{\fl}{4} & ( \dot{x}^2 + \pri{x}^2 ) ( z^{\underline{i}} + y^{\underline{i}} ) \gamma_{\underline{i}} \tilde{\gamma}^{34} \eta_2
    \Bigr) .
  \end{aligned}
\end{equation}

\section{Poisson brackets}
\label{app:poisson}

In this appendix we present the Poisson bracket expressions for fermions that are used in computing $\mathcal{A}$ in section~\ref{sec:currents-and-algebra}. The fermionic Poisson bracket is symmetric and has the non-vanishing elements
\begin{equation}
  \begin{aligned}
    \acommPB{(\eta_1)^{\underline{\dot{a}}\dot{a}}}{(\eta_1)^{\underline{\dot{b}}\dot{b}}} 
    &= - \frac{i}{4} ( 1 + A_1 )  \, \epsilon^{\underline{\dot{a}\dot{b}}}  \, \epsilon^{\dot{a}\dot{b}} , 
    &
    \acommPB{(\eta_1)^{\underline{\dot{a}}\dot{a}}}{(\eta_2)^{\underline{\dot{b}}\dot{b}}} 
    &= - \frac{i}{4} (A_2)^{\underline{\dot{a}}}{}_{\underline{\dot{c}}}  \, \epsilon^{\underline{\dot{c}\dot{b}}}  \, \epsilon^{\dot{a}\dot{b}} , 
    \\
    \acommPB{(\eta_2)^{\underline{\dot{a}}\dot{a}}}{(\eta_2)^{\underline{\dot{b}}\dot{b}}} 
    &= - \frac{i}{4} ( 1 - A_1 )  \, \epsilon^{\underline{\dot{a}\dot{b}}}  \, \epsilon^{\dot{a}\dot{b}} , 
    &
    \acommPB{(\chi_1)^{\underline{a}a}}{(\chi_2)^{\underline{b}b}} 
    &= - \frac{i}{4} ( A_3 )^{\underline{a}}{}_{\underline{c}}  \, \epsilon^{\underline{cb}}  \, \epsilon^{ab} , 
    \\
    \acommPB{(\chi_1)^{\underline{a}a}}{(\chi_1)^{\underline{b}b}} 
    &= - \frac{i}{4} ( 1 + A_1 )  \, \epsilon^{\underline{ab}}  \, \epsilon^{ab} , 
    &
    \acommPB{(\eta_1)^{\underline{\dot{a}}\dot{a}}}{(\chi_2)^{\underline{b}b}} 
    &= - \frac{i}{4} ( A_4 )^{\underline{\dot{a}}\dot{a}}{}_{\underline{c}c}  \, \epsilon^{\underline{cb}}  \, \epsilon^{cb} , 
    \\
    \acommPB{(\chi_2)^{\underline{a}a}}{(\chi_2)^{\underline{b}b}} 
    &= - \frac{i}{4} ( 1 - A_1 )  \, \epsilon^{\underline{ab}}  \, \epsilon^{ab} , 
    &\ \ 
    \acommPB{(\eta_2)^{\underline{\dot{a}}\dot{a}}}{(\chi_1)^{\underline{b}b}} 
    &= + \frac{i}{4} ( A_4 )^{\underline{\dot{a}}\dot{a}}{}_{\underline{c}c}  \, \epsilon^{\underline{cb}}  \, \epsilon^{cb} ,
  \end{aligned}
\end{equation}
where the coefficients $A_i$ are given to quadratic order by
\begin{equation}
  \begin{gathered}
    A_1 = -\tfrac{1}{2} \epsilon^{\underline{ij}} ( z_{\underline{i}} \dot{z}_{\underline{j}} - y_{\underline{i}} \dot{y}_{\underline{j}} ) , \\
    A_2 = -\tfrac{\flt}{2} \tilde{\gamma}^{34} ( z \cdot \pri{z} - y \cdot \pri{y} ) 
    + \tfrac{\flt}{2} \tilde{\gamma}^{34} \tilde{\gamma}^{\underline{ij}} ( z_{\underline{i}} \pri{y}_{\underline{j}} + \pri{z}_{\underline{i}} y_{\underline{j}} ) , \\
    A_3 = +\tfrac{\flt}{2} \gamma^{34} ( z \cdot \pri{z} + y \cdot \pri{y} ) 
    + \tfrac{\flt}{2} \gamma^{34} \gamma^{\underline{ij}} ( z_{\underline{i}} \pri{y}_{\underline{j}} - \pri{z}_{\underline{i}} y_{\underline{j}} ) , \\
    A_4 = +\tfrac{\flt}{2} \tilde{\gamma}^{\underline{i}} \tilde{\tau}^k ( z_{\underline{i}} - y_{\underline{i}} ) \pri{x}_k .
  \end{gathered}
\end{equation}
We note that these Poisson brackets are modified in a very simple way from the pure R-R Poisson brackets: the terms mixing the fermions $\eta_1$ and $\chi_1$ with $\eta_2$ and $\chi_2$ are rescaled by a factor $\flt$.

\section{Tensor-product representations}
\label{app:tensorprod}
The superalgebra $\su(1|1)^2_{\ce}$ features the supercharges $\gen{Q}_{\sL},\gen{Q}_{\sR}$ and their conjugates $\overline{\gen{Q}}_{\sL}$,$\overline{\gen{Q}}_{\sR}$ satisfying the relations~\eqref{eq:cealgebra-small}.
The supercharges
\begin{equation}
  \begin{gathered}
    \gen{Q}_{\smallL}^{\ 1} = \gen{Q}_{\smallL} \otimes \1 , \qquad
    \gen{Q}_{\smallL}^{\ 2} = \1 \otimes \gen{Q}_{\smallL}, \qquad
    \gen{Q}_{\smallR 1} = \gen{Q}_{\smallR} \otimes \1 , \qquad
    \gen{Q}_{\smallR 2} = \1 \otimes \gen{Q}_{\smallR} ,\\
    \overline{\gen{Q}}{}_{\smallL 1} = \overline{\gen{Q}}{}_{\smallL} \otimes \1 ,\, \qquad
    \overline{\gen{Q}}{}_{\smallL 2} = \1 \otimes \overline{\gen{Q}}{}_{\smallL} ,
    \qquad
    \overline{\gen{Q}}{}_{\smallR}^{\ 1} = \overline{\gen{Q}}{}_{\smallR} \otimes \1 , \qquad
    \overline{\gen{Q}}{}_{\smallR}^{\ 2} = \1 \otimes \overline{\gen{Q}}{}_{\smallR} ,
  \end{gathered}
\end{equation}
satisfy the algebra~$\psu(1|1)^4_{\ce}$ of equation~\eqref{eq:cealgebra}, and carry an (anti)fundamental~$\su(2)_{\bullet}$ index.

We can construct bi-fundamental representations of $\psu(1|1)^4_{\ce}$ out of pairs of fundamental representations  of $\su(1|1)^2_{\ce}$ of section~\ref{sec:repr-smallalgebra}. Consider the representation~$\varrho_{\sL}\otimes\varrho_{\sL}$, and identify
\begin{equation}
    Y^{\sL} = \phi^{\sL} \otimes \phi^{\sL} , \qquad
    \eta^{\sL 1} = \psi^{\sL} \otimes \phi^{\sL} ,\;  \qquad
    \eta^{\sL 2} = \phi^{\sL} \otimes \psi^{\sL} ,\;  \qquad
    Z^{\sL} = \psi^{\sL} \otimes \psi^{\sL} .
\end{equation}
This yields precisely the left representation~\eqref{eq:repr-massive-L}. Similarly, the right representation is found from $\varrho_{\sR}\otimes\varrho_{\sR}$ by setting
\begin{equation}
    Y^{\sR} = {\phi}^{\sR} \otimes {\phi}^{\sR} , \qquad
    \eta^{\sR}_{\ 1} = {\psi}^{\sR} \otimes {\phi}^{\sR} , \qquad
    \eta^{\sR}_{\ 2} = {\phi}^{\sR} \otimes {\psi}^{\sR} , \qquad
    Z^{\sR} = {\psi}^{\sR} \otimes {\psi}^{\sR} .
\end{equation}
Finally, for the massless modes we use the representation~$(\varrho_{\sL} \otimes\widetilde{\varrho}_{\sL})^{\oplus 2}$ and identify
\begin{equation}
    T^{1a} = \bigl(\psi^{\sL} \otimes \tilde{\psi}^{\sL}\bigr)^{a} ,\quad
    \widetilde{\chi}^{a} = \bigl(\psi^{\sL} \otimes \tilde{\phi}^{\sL}\bigr)^{a} , \quad
    \chi^{a} = \bigl(\phi^{\sL} \otimes \tilde{\psi}^{\sL}\bigr)^{a} , \quad
     T^{2a} = \bigl(\phi^{\sL} \otimes \tilde{\phi}^{\sL}\bigr)^{a} ,
\end{equation}
where the index~$a$ labels the two irreducible $\psu(1|1)^4_{\ce}$ modules.

\section{Oscillator representation of the charges}
\label{app:charges:osc}
The mixed-flux currents at quadratic order in fields are given in equation~\eqref{eq:quadratic-currents}. Let us write the fermions in components as follows:
\begin{equation}
\begin{aligned}
&(\eta_1)^{\underline{\dot{a}}\dot{b}}=\left(
\begin{array}{cc}
-e^{+i\pi/4}\bar{\eta}_{\sL 2} & -e^{+i\pi/4}\bar{\eta}_{\sL 1} \\
\phantom{+}e^{-i\pi/4}\eta_{\sL}^{\ 1} & -e^{-i\pi/4}\eta_{\sL}^{\ 2}
\end{array}
\right) ,\quad
&&(\eta_2)^{\underline{\dot{a}}\dot{b}}=\left(
\begin{array}{cc}
-e^{-i\pi/4}\eta_{\sR 2} & -e^{-i\pi/4}\eta_{\sR 1} \\
\phantom{+}e^{+i\pi/4}\bar{\eta}_{\sR}^{\ 1} & -e^{+i\pi/4}\bar{\eta}_{\sR}^{\ 2}
\end{array}
\right) ,\\
&(\chi_1)^{\underline{a}b}=\left(
\begin{array}{cc}
-e^{+i\pi/4}\bar{\chi}_{+2} & \phantom{+}e^{+i\pi/4}\bar{\chi}_{+1} \\
-e^{-i\pi/4}\chi_+^{\ 1} & -e^{-i\pi/4}\chi_+^{\ 2}
\end{array}
\right) ,\quad
&&(\chi_2)^{\underline{a}b}=\left(
\begin{array}{cc}
\phantom{+}e^{-i\pi/4}\chi_-^{\ 1} & \phantom{+}e^{-i\pi/4}\chi_-^{\ 2} \\
-e^{+i\pi/4}\bar{\chi}_{-2} & \phantom{+}e^{+i\pi/4}\bar{\chi}_{-1}
\end{array}\right).
\end{aligned}
\end{equation}
%The fermion appear in the currents with both indices raised as written here, and the gamma matrices have indices $\gamma^{\underline{a}}_{\ \underline{\dot{b}}}$, $\tilde{\tau}^{\dot{a}}_{\ b}$, so the currents carry indices $(j_I)^{\underline{a}\dot{a}}$.

We introduce complex bosonic coordinates 
\begin{equation}
\begin{gathered}
Z=-z_2+iz_1 ,\quad \bar{Z}=-z_2-iz_1 ,
\qquad
Y=-y_3-iy_4 ,\quad \bar{Y}=-y_3+iy_4 , \\
X^{11}=-x_6+ix_7=(X^{22})^\dagger ,\qquad X^{12}=x_8-ix_9=-(X^{21})^\dagger ,
\end{gathered}
\end{equation}
with conjugate momenta
\begin{equation}
\begin{gathered}
P_Z=2\dot{Z} , \quad P_{\bar{Z}}=2\dot{\bar{Z}} ,
\qquad
P_Z=2\dot{Z} , \quad P_{\bar{Y}}=2\dot{\bar{Y}} ,  \\
P_{11}=P_{22}^\dagger=2\dot{X}^{22} , \qquad P_{12}=-P_{21}^\dagger=-2\dot{X}^{21} .
\end{gathered}
\end{equation}
These satisfy the commutation relations
\begin{equation}
\begin{gathered}
[Z(\sigma_1),P_{\bar{Z}}(\sigma_2)]=[\bar{Z}(\sigma_1),P_Z(\sigma_2)]=i\delta(\sigma_1-\sigma_2) ,\\
[Y(\sigma_1),P_{\bar{Y}}(\sigma_2)]=[\bar{Y}(\sigma_1),P_Y(\sigma_2)]=i\delta(\sigma_1-\sigma_2) ,\\
[X^{\dot{a}a}(\sigma_1),P_{\dot{b}b}(\sigma_2)]=i\delta^{\dot{a}}_{\dot{b}}\delta^a_b\delta(\sigma_1-\sigma_2) ,
\end{gathered}
\end{equation}
and the anti-commutation relations 
\begin{equation}
\begin{gathered}
\{\bar{\eta}_{\sL\dot{a}}(\sigma_1),\eta_{\sL}^{\ \dot{b}}(\sigma_2)\}=\{\bar{\eta}_{\sR}^{\ \dot{b}}(\sigma_1),\eta_{\sR\dot{a}}(\sigma_2)\}=\delta_{\dot{a}}^{\dot{b}}\delta(\sigma_1-\sigma_2) ,\\
\{\bar{\chi}_{+a}(\sigma_1),\chi_+^b(\sigma_2)\}=\{\bar{\chi}_{-a}(\sigma_1),\chi_-^b(\sigma_2)\}=\delta_a^b\delta(\sigma_1-\sigma_2).
\end{gathered}
\end{equation}

Let us associate the components of the currents $(j_I^\tau)^{\underline{a}\dot{a}}$ to charges $\gen{Q}_{\sL}^{\ \dot{a}}$ and $\gen{Q}_{\sR\dot{a}}$ which carry a natural $\su(2)$ structure. Explicitly,
\begin{equation}
\begin{aligned}
\gen{Q}_{\sL}^{\ 1}&=-\int\de\sigma\;(j_1^\tau)^{21},
\qquad\qquad
\gen{Q}_{\sL}^{\ 2}=+\int\de\;\sigma(j_1^\tau)^{22}, \\
\gen{Q}_{\sR 1}&=+\int\de\sigma\;(j_2^\tau)^{12},
\qquad\qquad
\gen{Q}_{\sR 2}=+\int\de\sigma\;(j_2^\tau)^{11} .
\end{aligned}
\end{equation}
The Hermitian conjugates of these,
\begin{equation}
\overline{\gen{Q}}_{\sL\dot{a}}= (\gen{Q}_{\sL}^{\ \dot{a}})^\dagger ,
\qquad \overline{\gen{Q}}_{\sR}^{\ \dot{a}} =(\gen{Q}_{\sR\dot{a}})^\dagger ,
\end{equation}
are then given in terms of the remaining components of the currents by:
\begin{equation}
\begin{aligned}
\overline{\gen{Q}}_{\sL 1}&= +\int\de\sigma\;(j_1^\tau)^{12},\qquad\qquad
\overline{\gen{Q}}_{\sL 2}= +\int\de\sigma\;(j_1^\tau)^{11} ,\\
\overline{\gen{Q}}_{\sR}^{\ 1}&= -\int\de\sigma\;(j_2^\tau)^{21},\qquad\qquad
\overline{\gen{Q}}_{\sR}^{\ 2}= +\int\de\sigma\;(j_2^\tau)^{22} .
\end{aligned}
\end{equation}

The explicit form of these charges, at quadratic order in the fields are:
\begin{equation}
\begin{aligned}
\gen{Q}_{\sL}^{\ \dot{a}}=e^{-i\pi/4} \int\de\sigma\bigg[
&\frac{1}{2}P_Z\eta_{\sL}^{\ \dot{a}}-Z'(i\flt\,\bar{\eta}_{\sR}^{\ \dot{a}}
+\fl\,\eta_{\sL}^{\ \dot{a}})+iZ\eta_{\sL}^{\ \dot{a}}\\
&-\epsilon^{\dot{a}\dot{b}}
\left(\frac{i}{2}P_{\bar{Y}}\bar{\eta}_{\sL\dot{b}} -\bar{Y}'(\flt\,\eta_{\sR\dot{b}}+i\fl\,\bar{\eta}_{\sL\dot{b}}) +\bar{Y}\bar{\eta}_{\sL\dot{b}}\right)\\
&\qquad\quad-\frac{1}{2}\epsilon^{\dot{a}\dot{b}} P_{\dot{b}a}\chi_+^{\ a}- (X^{\dot{a}a})'(i\flt\,\bar{\chi}_{-a} +\fl\,\epsilon_{ab}\chi_+^{\ b})\bigg] ,\\
\gen{Q}_{\sR\dot{a}}=e^{-i\pi/4}\int\mathrm{d}\sigma\bigg[
&\frac{1}{2}P_{\bar{Z}}\eta_{\sR\dot{a}} -\bar{Z}'(i\flt\,\bar{\eta}_{\sL\dot{a}}-\fl\,\eta_{\sR\dot{a}})+i\bar{Z}\eta_{\sR\dot{a}}\\
&+\epsilon_{\dot{a}\dot{b}}\left(\frac{i}{2}P_Y\bar{\eta}_{\sR}^{\ \dot{b}}-Y'(\flt\,\eta_{\sL}^{\ \dot{b}} -i\fl\,\bar{\eta}_{\sR}^{\ \dot{b}}) +Y\bar{\eta}_{\sR}^{\ \dot{b}}\right)\\
&\qquad\quad+\frac{1}{2}P_{\dot{a}a}\chi_-^{\ a}- \epsilon_{\dot{a}\dot{b}}(X^{\dot{b}a})'(i\flt\,\bar{\chi}_{+a}+\fl\,\epsilon_{ab}\chi_-^{\ b})\bigg].
\end{aligned}
\end{equation}

We will now introduce creation and annihilation operators for both bosons and fermions and use them to rewrite the supercharges. For the massive bosons we have
\begin{equation}
\begin{aligned}
a_{\sL z}(p)&=\frac{1}{\sqrt{2\pi}}\int\frac{\de\sigma}{\sqrt{\omega_p^{\sL}}}\left(\omega_p^{\sL}\bar{Z}+\frac{i}{2}P_{\bar{Z}}\right)e^{-ip\sigma} ,
\\
a_{\sR z}(p)&=\frac{1}{\sqrt{2\pi}}\int\frac{\de\sigma}{\sqrt{\omega_p^{\sR}}} \left(\omega_p^{\sR}Z+\frac{i}{2}P_{Z}\right)e^{-ip\sigma},
\\
a_{\sL y}(p)&=\frac{1}{\sqrt{2\pi}}\int\frac{\de\sigma}{\sqrt{\omega_p^{\sL}}} \left(\omega_p^{\sL}\bar{Y}+\frac{i}{2}P_{\bar{Y}}\right)e^{-ip\sigma},
\\
a_{\sR y}(p)&=\frac{1}{\sqrt{2\pi}}\int\frac{\de\sigma}{\sqrt{\omega_p^{\sR}}} \left(\omega_p^{\sR}Y+\frac{i}{2}P_{Y}\right)e^{-ip\sigma},
\end{aligned}
\end{equation}
with creation operators given by complex conjugation. The annihilation operators for the massive fermions are
\begin{align}
d_{\sL\dot{a}}(p)&=+\frac{e^{+i\pi/4}}{\sqrt{2\pi}}\int\frac{\de\sigma}{\sqrt{\omega_p^{\sL}}} \epsilon_{\dot{a}\dot{b}}(f_p^{\sL}\eta_{\sL}^{\ \dot{b}}-ig_p^{\sL}\bar{\eta}_{\sR}^{\ \dot{b}})e^{-ip\sigma},\nonumber\\
d_{\sR}^{\ \dot{a}}(p)&=-\frac{e^{+i\pi/4}}{\sqrt{2\pi}}\int\frac{\de\sigma}{\sqrt{\omega_p^{\sR}}} \epsilon^{\dot{a}\dot{b}}(f_p^{\sR}\eta_{\sR\dot{b}} -ig_p^{\sR}\bar{\eta}_{\sL\dot{b}})e^{-ip\sigma}.
\end{align}
The corresponding creation operators are found by taking the complex conjugate and raising or lowering the $\su(2)$ indices as appropriate. The representation parameters $f^{\sL}_p,f^{\sR}_p,g^{\sL}_p, g^{\sR}_p$ and the dispersion $\omega^{\sL}_p,\omega^{\sR}_p$ are given in section \ref{sec:repr-paraman-nfs}.
In the same way, for the massless bosons we define 
\begin{equation}
a_{\dot{a}a}(p)=-\frac{1}{\sqrt{2\pi}}\int\frac{\mathrm{d}\sigma}{\tilde{\omega}_p}\left(\tilde{\omega}_pX_{\dot{a}a}+\frac{i}{2}P_{\dot{a}a}\right)e^{-ip\sigma},
\end{equation}
where $X_{\dot{a}a}= \left(X^{\dot{a}a}\right)^\dagger$. For the massless fermions we define
\begin{align}
\tilde{d}_a(p)&=\frac{e^{-i\pi/4}}{\sqrt{2\pi}}\int\frac{\mathrm{d}\sigma}{\tilde{\omega}_p} \left(\tilde{f}_p\bar{\chi}_{+a} -i\tilde{g}_p\epsilon_{ab}\chi_-^{\ b} \right)e^{-ip\sigma},\nonumber\\
d_a(p)&=\frac{e^{+i\pi/4}}{\sqrt{2\pi}}\int\frac{\mathrm{d}\sigma}{\tilde{\omega}_p}\left(\tilde{f}_p \epsilon_{ab}\chi_+^{\ b} - i\tilde{g}_p\bar{\chi}_{-b}\right) e^{-ip\sigma} .
\end{align}
All these operators satisfy canonical (anti)commutation relations.
%
%\begin{align}
%[a_{Lz}^\dagger(p_1),a_{Lz}(p_2)]=[a_{Rz}^\dagger(p_1),a_{Rz}(p_2)]&=\delta(p_1-p_2),\nonumber\\
%[a_{Ly}^\dagger(p_1),a_{Ly}(p_2)]=[a_{Ry}^\dagger(p_1),a_{Ry}(p_2)]&=\delta(p_1-p_2),\nonumber\\
%[a^{\dot{b}b}\*^\dagger(p_1),a_{\dot{a}a}(p_2)]&=\delta_{\dot{a}}^{\dot{b}}\delta_a^b\delta(p_1-p_2),\nonumber\\
%\{d_L^{\dot{a}}\*^\dagger(p_1),d_{L\dot{b}}(p_2)\}=\{d_{R\dot{b}}^\dagger(p_1),d_R^{\dot{a}}(p_2)\}&=\delta_{\dot{b}}^{\dot{a}}\delta(p_1-p_2),\nonumber\\
%\{\tilde{d}^a\*^\dagger(p_1),\tilde{d}_b(p_2)\}=\{d^a\*^\dagger(p_1),d_b(p_2)\}&=\delta_b^a\delta(p_1-p_2)
%\end{align}

In terms of these creation and annihilation operators, the supercharges become:
\begin{equation}
\begin{aligned}
\gen{Q}_{\sL}^{\ \dot{a}}=
\int\de p\bigg[
(d_{\sL}^{\ \dot{a}}\*^\dagger a_{\sL  y}+\epsilon^{\dot{a}\dot{b}}a_{\sL z}^\dagger d_{\sL\dot{b}})f_p^{\sL}&+(a_{\sR y}^\dagger d_{\sR}^{\ \dot{a}}+\epsilon^{\dot{a}\dot{b}} d_{\sR\dot{b}}^\dagger a_{\sR z})g_p^{\sR}\\
&\quad +(\epsilon^{\dot{a}\dot{b}}\tilde{d}^a\*^\dagger a_{\dot{b}a}+a^{\dot{a}a}\*^\dagger d_a)\tilde{f}_p\bigg] ,\\
\gen{Q}_{\sR\dot{a}}=\int\de p\bigg[
(d_{\sR\dot{a}}^\dagger a_{\sR y} -\epsilon_{\dot{a}\dot{b}} a_{\sR z}^\dagger d_{\sR}^{\ \dot{b}})f_p^{\sR} &+(a_{\sL y}^\dagger d_{\sL\dot{a}}- \epsilon_{\dot{a}\dot{b}} d_{\sL}^{\ \dot{b}}\*^\dagger a_{\sL z})g_p^{\sL}\\
&\quad +(d^a\*^\dagger a_{\dot{a}a}- \epsilon_{\dot{a}\dot{b}} a^{\dot{b}a}\*^\dagger\tilde{d}_a) \tilde{g}_p\bigg].
\end{aligned}
\end{equation}

\section{Parametrisation of the S-matrix elements}
\label{app:smat-param}
In this appendix we define the coefficients of the $\su(1|1)^2_{\ce}$ S-matrix elements introduced in section~\ref{sec:small-smat}. We have
\begin{equation}
\begin{aligned}
A^{\sL\sL}_{pq} &= 1, &
\quad
B^{\sL\sL}_{pq} &= \phantom{-}\left( \frac{x^-_{\sL\,p}}{x^+_{\sL\,p}}\right)^{1/2} \frac{x^+_{\sL\,p}-x^+_{\sL\,q}}{x^-_{\sL\,p}-x^+_{\sL\,q}}, \\
C^{\sL\sL}_{pq} &= \left( \frac{x^-_{\sL\,p}}{x^+_{\sL\,p}} \frac{x^+_{\sL\,q}}{x^-_{\sL\,q}}\right)^{1/2} \frac{x^-_{\sL\,q}-x^+_{\sL\,q}}{x^-_{\sL\,p}-x^+_{\sL\,q}} \frac{\eta^{\sL}_{p}}{\eta^{\sL}_{q}}, 
\quad &
D^{\sL\sL}_{pq} &= \phantom{-}\left(\frac{x^+_{\sL\,q}}{x^-_{\sL\,q}}\right)^{1/2}  \frac{x^-_{\sL\,p}-x^-_{\sL\,q}}{x^-_{\sL\,p}-x^+_{\sL\,q}}, \\
E^{\sL\sL}_{pq} &= \frac{x^-_{\sL\,p}-x^+_{\sL\,p}}{x^-_{\sL\,p}-x^+_{\sL\,q}} \frac{\eta^{\sL}_{q}}{\eta^{\sL}_{p}}, 
\quad &
F^{\sL\sL}_{pq} &= - \left(\frac{x^-_{\sL\,p}}{x^+_{\sL\,p}} \frac{x^+_{\sL\,q}}{x^-_{\sL\,q}}\right)^{1/2} \frac{x^+_{\sL\,p}-x^-_{\sL\,q}}{x^-_{\sL\,p}-x^+_{\sL\,q}},
\end{aligned}
\end{equation}
and
\begin{equation}
\begin{aligned}
 A^{\sL\sR}_{pq} &= \sqrt{\frac{x^+_{\sL\,p}}{x^-_{\sL\,p}}}\,
  \frac{1-\frac{1}{x^+_{\sL\,p} x^-_{\sR\,q}}}{1-\frac{1}{x^-_{\sL\,p} x^-_{\sR\,q}}}, 
 \quad &
 C^{\sL\sR}_{pq} &= \phantom{-}1, \\
B^{\sL\sR}_{pq} &= -\frac{2i}{\h} \, \sqrt{\frac{x^-_{\sL\,p}}{x^+_{\sL\,p}}\frac{x^+_{\sR\,q}}{x^-_{\sR\,q}}}\,
 \frac{\eta^{\sL}_{p}\eta^{\sR}_{q}}{ x^-_{\sL\,p} x^+_{\sR\,q}} \frac{1}{1-\frac{1}{x^-_{\sL\,p} x^-_{\sR\,q}}},  
\quad &
D^{\sL\sR}_{pq} &=\phantom{-}\sqrt{\frac{x^+_{\sL\,p}}{x^-_{\sL\,p}}\frac{x^+_{\sR\,q}}{x^-_{\sR\,q}}}\,
 \frac{1-\frac{1}{x^+_{\sL\,p} x^+_{\sR\,q}}}{1-\frac{1}{x^-_{\sL\,p} x^-_{\sR\,q}}}, \\
F^{\sL\sR}_{pq} &= \frac{2i}{\h} \sqrt{\frac{x^+_{\sL\,p}}{x^-_{\sL\,p}}\frac{x^+_{\sR\,q}}{x^-_{\sR\,q}}}\,
  \frac{\eta^{\sL}_{p}\eta^{\sR}_{q}}{ x^+_{\sL\,p} x^+_{\sR\,q}} \frac{1}{1-\frac{1}{x^-_{\sL\,p} x^-_{\sR\,q}}},
\quad &
 E^{\sL\sR}_{pq} &= - \sqrt{\frac{x^+_{\sR\,q}}{x^-_{\sR\,q}}}\,
  \frac{1-\frac{1}{x^-_{\sL\,p} x^+_{\sR\,q}}}{1-\frac{1}{x^-_{\sL\,p} x^-_{\sR\,q}}}.
\end{aligned}
\end{equation}
Note that the appearance of~$\h$ in these last formulae is an artefact of our definition of~$\eta_p$. Once we express this in terms of~$x^{\pm}$, there is no explicit dependence on~$\h$.

The remaining S-matrix elements follow by left-right symmetry.

\bibliographystyle{nb}
\bibliography{ads3-s3-t4-mixed-flux}

\end{document}